\documentclass[11pt]{article}
\usepackage{geometry}
\usepackage[parfill]{parskip}
\usepackage{setspace}

\usepackage{graphicx}
\usepackage{amsmath,amsthm,amssymb,mathrsfs,bbm}
\usepackage[colorlinks, linkcolor= blue, urlcolor = blue, citecolor = blue]{hyperref}
\usepackage[multiple]{footmisc}

\usepackage{newpxtext, newpxmath}

\usepackage[width=0.8\textwidth]{caption}
\usepackage{booktabs}
\usepackage{multirow}
\usepackage[flushleft]{threeparttable}

\usepackage{float}
\usepackage{subcaption}

\usepackage{natbib,threeparttable,dcolumn,multicol,lscape,bigstrut,booktabs,multirow,colortbl,array}
\usepackage{url}

\usepackage{abstract}
\usepackage[titletoc,title]{appendix}

\title{Education Opportunities for Rural Areas: \\Evidence from China’s Higher Education Expansion\footnote{We are grateful to Travis Baseler, Lisa Kahn, Ronni Pavan, John Singleton, and Kegon Tan for their continuous guidance and support. We thank Jinfu He, Zibin Huang, Joanna Venator, and seminar participants at Rochester Applied Reading Group, Rochester Student Seminar, AEFP 2022 Annual conference, CES 2022 Annual Conference, and Economics Graduate Student Conference 2022 at WUSTL. We thank Jie Li and Mingyang Lv at RCRE, who provided invaluable assistance on the NFPS data set. All errors are ours.}}
\author{Ande~Shen\thanks{PhD student, Department of Economics, University of Rochester. Email: \href{mailto:ande.shen@rochester.edu}{ande.shen@rochester.edu}.} \quad Jiwei~Zhou\thanks{PhD student, Department of Economics, University of Rochester. Email: \href{mailto:jzhou51@ur.rochester.edu}{jzhou51@ur.rochester.edu}.}}
\date{\today \\ \textit{(click \href{https://www.dropbox.com/scl/fi/nny34woif2kknrokgg32v/HEE_Ande_Shen.pdf?rlkey=wfc3s66h1yh14i892f2w8i04j&dl=0}{here} for the latest version)}}

\begin{document}
\maketitle
\begin{abstract}
\normalsize
\noindent This paper explores the causal impact of education opportunities on rural areas by exploiting the higher education expansion (HEE) in China in 1999. By utilizing the detailed census data, the cohort-based difference-in-differences design indicates that the HEE increased college attendance and encouraged more people to attend senior high schools and that the effect is more significant in rural areas. Then we apply a similar approach to a novel panel data set of rural villages and households to examine the effect of education opportunities on rural areas. We find contrasting impacts on income and life quality between villages and households. Villages in provinces with higher HEE magnitudes underwent a drop in the average income and worse living facilities. On the contrary, households sending out migrants after the HEE experienced an increase in their per capita income. The phenomenon where villages experienced a ``brain drain'' and households with migrants gained after the HEE is explained by the fact that education could serve as a way to overcome the barrier of rural-urban migration. Our findings highlight the opposed impacts of education opportunities on rural development and household welfare in rural areas.

\bigskip


\end{abstract}

\newpage
\section{Introduction}\label{introduction}

A large literature on growth and development has documented a significant gap of productivity and living standards between urban and rural areas (\cite{gollin2014agricultural}; \cite{Lagakos2020-gc}). It's also widely accepted that education is a leading instrument for accelerating human capital accumulation, promoting economic development, and boosting long-term growth, especially for less-developed regions to climb out of poverty (\cite{lucas1988mechanics}; \cite{bloom2006higher}). Thus, providing more education opportunities could be an efficient way to narrow down the rural-urban gap.

If the urban and rural areas were completely isolated, the benefit from providing more education opportunities could be fully realized since those well-educated people are preserved in rural areas, resulting in a ``brain gain.'' However, education can also serve as an approach to break the isolation. As people with higher education attainment are more competitive in the labor market, they are more likely to move out of rural areas due to the attraction of higher wages and better amenities in urban areas, leading to a ``brain drain.'' Specifically, given the large gap between rural and urban areas and the internal migration restriction, we would like to know how the increased education opportunities could affect rural areas.

This paper examines this impact in the context of China. China's setting fits this purpose for several reasons. First, China's urban-rural dualism and the \textit{hukou} system (household registration) help to identify the response of rural people and urban people to education opportunities, respectively.\footnote{The initial reason why China implemented this particular registration system to keep most of the population on farms is because of the low level of agricultural productivity and the need to ensure food provision for cities, which were deemed essential for industrialization \citep{meng2012labor}.} Second, to study the brain-drain effect and the migration channel due to education, we need to rule out other potential causes of migration. The \textit{hukou} constraint in China can alleviate this concern since it substantially restricts people's migration but places much fewer restrictions on people with higher education.\footnote{It was very difficult to switch \textit{hukou} status for rural people \citep{fan2007china}. People without urban \textit{hukou} cannot get full access to the public good in urban areas, including public schools, medical services, social security, and social benefits (\cite{Zhao1997-rs};  \cite{Congressional-Executive_Commission_on_China2005-oq};  \cite{meng2012labor}; \cite{ngai2019china}; \cite{wu2020welfare}).} Third, China's higher education system experienced a dramatic and unanticipated change in 1999, leading to a massive increase in higher education opportunities. Finally, the place-based feature of China's higher education admission system creates rich variations across regions, allowing for causally identifying the effect of education opportunities.

Within a relatively long period, it is very hard for many rural residents in China to access the higher education (i.e., attend colleges and universities) as there has been fierce competition for seats due to the low admission rate of the National College Entrance Examination (also called Gaokao, henceforth, NCEE). Since its resumption in 1977, NCEE has become a standard way for Chinese students to attend colleges and obtain higher education degrees. Due to the limited education resources, the admission rate of NCEE and the number of admitted students had been remaining at very low levels for over 20 years.\footnote{For example, only 25\%-35\% of NCEE takers can get admitted into colleges between the late 1980s and late 1990s, and only 0.1\% of the population can attend colleges.}

However, as a result of the market demand for college graduates, without any anticipation by the public, the Chinese government relaxed the cap on NCEE admission in 1999 by increasing the national average admission rate from 34\% in 1998 to 56\% in 1999, and this pattern continued through the subsequent years. This higher education expansion (henceforth, HEE) dramatically increases education opportunities and provides more access to higher education for rural residents.

Since China's higher education admission features a place-based admission quota system, the HEE magnitudes varies substantially across provinces. To capture the differences in admission rates and population density across provinces, we use the change in the number of admitted students per capita as a measure for the HEE magnitudes. We use the level change after the HEE as the main measure, which is highly correlated with other measures.


By utilizing the cross-province variations of the HEE magnitudes and detailed census data, we first conduct a cohort-based difference-in-differences (DID) design to examine the HEE's impact on people's educational attainment.\footnote{We include only 24 provinces (out of 31 provinces in mainland China) in the empirical part for several reasons. First, we exclude from our sample the four province-level municipalities (Beijing, Tianjin, Shanghai, and Chongqing) because they have administrative priorities over other provinces. Second, we drop Tibet and Inner Mongolia because the village-level survey didn't cover these two provinces during our sample period. Finally, we exclude Hainan since the village-level survey only covers one village in this province.} Before the HEE, there was no significant difference in educational attainment between any cohort in provinces with higher and lower magnitudes. After that, cohorts in provinces with higher HEE magnitudes are 3\% more prone to attend colleges than their peers in provinces with a lower magnitude. The effect is pronounced for both rural and urban people, classified by their \textit{hukou} type. We also find that the HEE encouraged more rural people to attend high schools, though this spillover effect is moderate for urban people, possibly due to the fact that urban areas already had high college attendance rates. Finally, we dig into the labor market outcomes associated with educational attainment and find that rural people who are more affected by the HEE experienced a short-term earnings drop due to college attendance and that they are more likely to choose to work in the non-agricultural sector, indicating the HEE's impact on the structural transformation of rural areas.

The HEE brought more education opportunities for rural people, providing us with a chance to detect whether rural areas would benefit or suffer from the increment in education opportunities. On the one hand, the effect could be a brain gain for rural areas. As more rural people attended colleges and high schools, the human capital accumulation would be accelerated, spurring the economic growth. In contrast, the effect could be a brain drain. Taking those higher education opportunities enables individuals to acquire skills, be competitive in urban labor markets, and earn higher salaries, attracting them to move into urban areas, thus reducing the human capital in rural areas.

To answer this question empirically, we utilize a novel panel data set on rural villages and households (National Fixed Point Survey, NFPS) from the Chinese Ministry of Agriculture. The data set spans from 1995 to 2008.\footnote{The reason why we use both village panel and household panel is that the village panel includes more variables to measure the development of rural areas, and that the household panel is important for us to detect the migration channel through which education could affect household welfare.} The panel structure allows for the inclusion of village or household fixed effects to absorb time-invariant heterogeneity across villages or households as well as year fixed effects to absorb other changes that could affect all villages or households uniformly. Thus, we implement a difference-in-differences design to identify the causal impact of increased education opportunities on rural villages. We compare the change over time in outcomes of villages or households with varying exposure to the HEE. The identifying assumption behind our analysis is that outcomes of villages (or households) in more exposed areas would evolve in a similar pattern as others. We test this assumption by ruling out other confounding factors and checking for the parallel trends using event studies.

We grouped all the village-level outcomes into four categories: population, labor force, agricultural activities, and income and life quality. First, we find that villages more exposed to the HEE experienced a 6\% more decline in population and an 10\% more drop in the number of households after the HEE. Second, the HEE extracted labor force with better educational attainment from rural villages, even though there is no effect on the total labor force. The rural villages in provinces with higher HEE magnitudes underwent over 6\% decline in the number of workers with senior high school degrees or above because of college attendance and out-migration. The labor force in the non-agricultural sector experiences a higher decline in villages that are more exposed to the HEE.

Since the agricultural sector includes all the main business in rural villages and requires a lower level of education, then it could be the case that those labor force left in rural villages have comparative advantages in agricultural activities, which could increase the efficiency of agricultural production because they can access to more resources (e.g., land). Then we use the area of cropped land to measure the agricultural activities in rural villages.\footnote{This measure is better than the output of agricultural products because the latter one can be affected by the change of local weather and geographical conditions over time, which is not observed to us.} The estimation indicates the opposite. Rather than improving the efficiency of agricultural production, more education opportunities lead to the shrinkage of agricultural activities, though the magnitudes of the impacts are not significantly different from zero. Finally, we examine the impact on average income and life qualities in rural villages. We find that people in villages more exposed to the HEE have a 6\% lower income than those in less-exposed villages. We also see that fewer households can get access to advanced living facilities like tap water and televisions in those more-exposed villages. In addition, our event studies on all outcomes present insignificant impacts before 1999, which validates the parallel trends assumption of our identification. Altogether, the results reveal that more education opportunities lead to the brain drain of rural villages, impede the development of rural areas, and lower the life quality of rural people who stayed behind.

Finally, we switch to the household panel data to further disentangle the impact on different households.  The impact of education opportunities might vary across households since some households may not be able to take those opportunities due to the selectivity of the college entrance examination. Probing the household-level impact would help us further distinguish the direct effect of education opportunities from its spillover effect. Due to data limitations, we focus on three primary outcomes: household size (and the number of migrants), household labor force with at least senior high school education, and household income (including remittance). We find contrasting impacts on rural households when compared to rural villages. The HEE provides opportunities for some rural residents to receive higher education, migrate to urban areas, and
get higher salaries, which ultimately increases their households welfare. That is, we see the positive impact of education opportunities on rural households who can take those opportunities. However, for households that cannot take those high education opportunities, they will experience decreases in income and worse living facilities, consistent with the HEE’s impact on rural villages. Our results highlight the opposed impacts of education opportunities on rural development and household welfare in rural areas.

Our research contributes to several strands of the literature. First, our paper is related to the discussions on education and regional development \citep{chatterton2000response}. Previous studies in this strand mainly focus on the labor market implications of education opportunities (\cite{Wan2006-wd}; \cite{Li2014-ip}; \cite{Demurger2019-st}), the education returns and social inequalities (\cite{Duflo2001-qb}; \cite{Psacharopoulos2004-bf}; \cite{Xing2014-yp}; \cite{blossfeld2016models}), and productivity and innovation (\cite{Yao2019-vg}; \cite{Ma_undated-gk}; \cite{Rong2020-ad}). Little research explores the effect of increased education opportunities on rural villages through the channel of reallocating people from rural areas to urban areas. Our project fills this gap by utilizing the exogenous variation caused by China's HEE and the novel village-level panel data.

Second, this paper relates to the literature about migration and the brain drain effect of education on the sending places (\cite{Beine2001-re, Beine2008-zv}; \cite{Docquier2012-kj}, \cite{Todaro2021-bd}). Most research in this literature conducts cross-country analysis, for example, \cite{Todaro2021-bd} discussed the emigration of high-skill professionals and technicians from developing countries to the developed world. The impact of the loss of educated people in poor areas is less obvious. On the one hand, it could lead to positive feedback to the sending places. Migrants may bring back remittances, additional skills, and networks in urban areas to the sending places. Besides, the education premium observed by people who stayed behind could also encourage them to invest more in education, which would raise the human capital accumulation in sending places (\cite{McCulloch1977-wz}; \cite{Mountford1997-al}; \cite{Beine2001-re}). On the contrary, it is also widely recognized that this loss of educated people and the associated human capital escaping could be detrimental to the migrant source places \citep{Beine2008-zv}. China's \textit{hukou} system restricts free migrations across domestic regions, provides us with a similar scenario as the international migration from poor to rich countries in the brain drain framework. Our research establishes the causal evidence on the brain drain consequence of education opportunities through internal migration.

The rest of this paper proceeds as follows. Section \ref{sec_Background} provides institutional background on China's place-based admission system and the sharp expansion of higher education in 1999. We describe the data in Section \ref{sec_Data} and the empirical strategy in Section \ref{sec_EmpiricalStrategy}. Section \ref{sec_EmpiricalResults} presents the results, analyzes the mechanisms, and Section \ref{sec_Conclusion} concludes.

\section{Institutional Background}\label{sec_Background}

This section describes the institutional setting related to China's higher education (post-secondary education). First, the national college entrance examination (NCEE) is the only way for almost all Chinese students to get higher education. Second, the admission process of NCEE is highly place-based, where the number of admitted students varies heavily across provinces. Finally, due to the urgent need for high-skilled workers in industrial production, in the late 1990s, the central government expanded the higher education system by increasing the number of admitted students of the NCEE in all provinces.

\subsection{National College Entrance Examination}\label{sec_Background_ncee}
The Chinese government resumed the national college entrance examination in 1977 after it had been silent for around ten years. Before 1977, the admission process to colleges is recommendation-based. The resumption then transferred it to be examination-based. Consequently, students are admitted to colleges based on their academic achievements, regardless of their \textit{hukou} status. Afterwards, NCEE becomes a standard way for almost all Chinese students to attend colleges and get higher education. 

\begin{figure}[htbp]
    \centering
	\subfloat[Average admission]{
		\includegraphics[width=.48\textwidth]{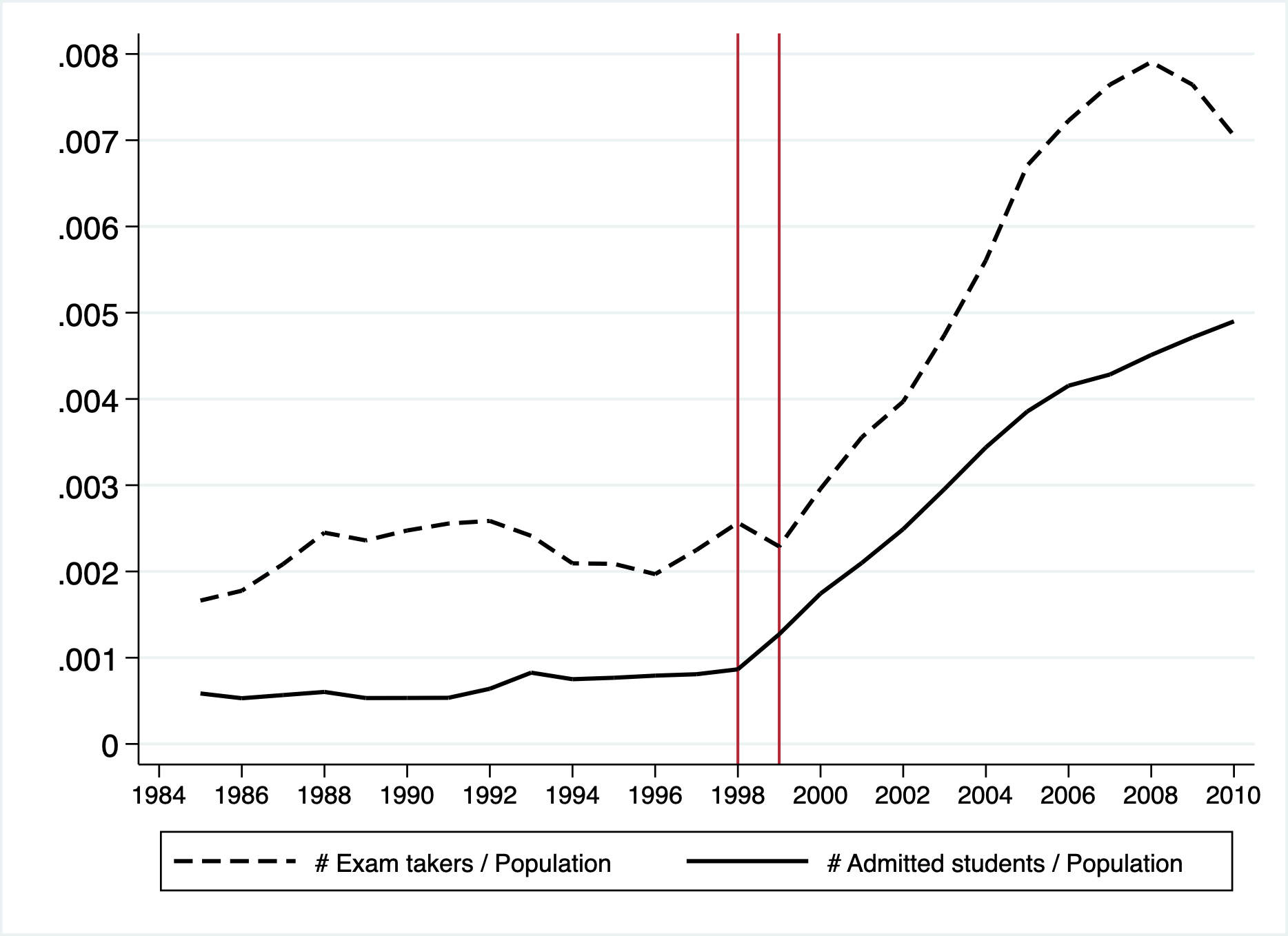}\label{fig_nation_avg_adm}}
	\subfloat[Overall admission]{
		\includegraphics[width=.48\textwidth]{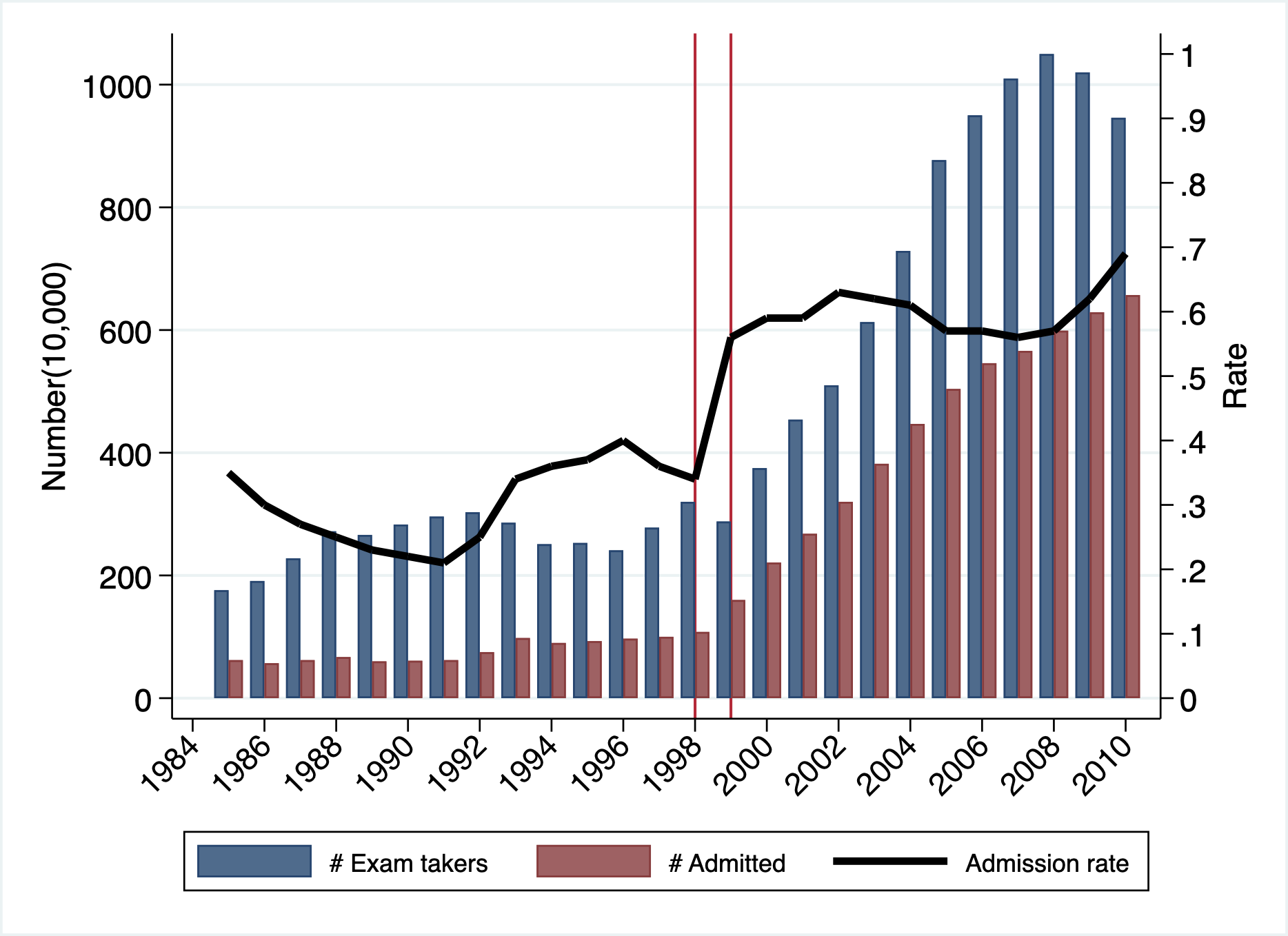}\label{fig_nation_overall_adm}}
	\caption{The national pattern of NCEE admission}\label{fig_nation_adm}
	\captionsetup{singlelinecheck=false}
    \caption*{\footnotesize \textit{Notes}: Panel (a) shows the change of NCEE participation rate and NCEE admission rate for the whole population. In Panel (b), the two bars refer to the number of exam takers and the number of admitted students in each NCEE from 1985 to 2010. The bold broken line shows the admission rate, which is the ratio between the two bars in each year. The two vertical lines in both graphs indicate the higher education expansion from 1998 to 1999.}
\end{figure}

Even though the number of admitted students is huge, the competition is fierce due to the large number of exam takers. In 1977, the total number of candidate students for NCEE was as many as 5.7 million. Although the Ministry of Education eventually expanded enrollment, adding 63,000 more to the admission quota, the admission ratio is just 4.8\% with only 272,971 students being admitted. After that, the admission rate (the ratio between the number of admitted students and the number of exam takers) becomes higher, but it is still fairly low compared to the availability of higher education in the Western world. Figure \ref{fig_nation_adm} shows the national pattern of NCEE admission from 1985 to 2010. In Figure \ref{fig_nation_avg_adm}, the NCEE participation rate (the number of exam takers per capita) was lower than 0.3\% before 1998 and the admission rate was even lower, every year just one person out of 1,000 people can become a college student. Figure \ref{fig_nation_overall_adm} demonstrates the number of exam takers and the number of admitted students in each NCEE, as well as the admission rate. Before 1998, on average around 3 million students took the annual NCEE and only 30\% of them can get admitted to higher education.\footnote{If we classify the higher education institutions into different tiers, the admission rate to better colleges (for example, Project-985 colleges) will be much lower. More details about college tiers can be found in \cite{yang2021place}.} Moreover, the gross enrollment ratio, which refers to the percentage of the 18–22 age group enrolling full time in higher education, is an important indicator to measure the development of higher education. Before 1997, it had been consistently lower than 7\%, which is far lower than the 15\% criterion for mass higher education \citep{Wan2006-wd}.

We also present the national admission pattern by people's \textit{hukou} status in Figure \ref{fig_nation_adm_ru}. During 1996 and 1998, more exam takers were from rural areas, while urban students were more likely to pass NCEE and get admitted to colleges. In other words, students in rural areas faced more fierce competition to get to college.

Since there is a high demand for well-educated workers in the job market, NCEE's success has generated substantial returns to its winners, especially those hailing from rural China. First, people with college degrees are more likely to earn a higher salary. Second, due to the \textit{hukou} system and the migration restrictions in China in the 1990s, it's extremely hard for people with rural \textit{hukou} to get a decent job in urban areas, so their salary is much lower than people with urban \textit{hukou}. In contrast, if they pass the NCEE and get a college degree, it is much easier for them to get a better job with higher wage in cities and maintain a permanent residence (i.e., transform to urban \textit{hukou}) there (\cite{Zhao1997-sw}; \cite{Zhao1997-rs}). In this way, the high salary and \textit{hukou} privileges associated with NCEE are inherently inspiring for rural students to pursue higher education.




\subsection{Place-Based Admission System}\label{sec_Background_PlaceBasedAdm}

China's higher education admission features a place-based admission quota system \citep{kang2005institutional}. Each year, the Ministry of Education coordinates with the National Development and Reform Commission to determine how many students each college can admit in each province during the next NCEE, taking into account the number of exam takers, the admission rate in each province, and the capacity of each college \citep{Moe2016-ke}.

\begin{figure}[htbp]
    \centering
	\subfloat[Admission rate]{
		\includegraphics[width=.48\textwidth]{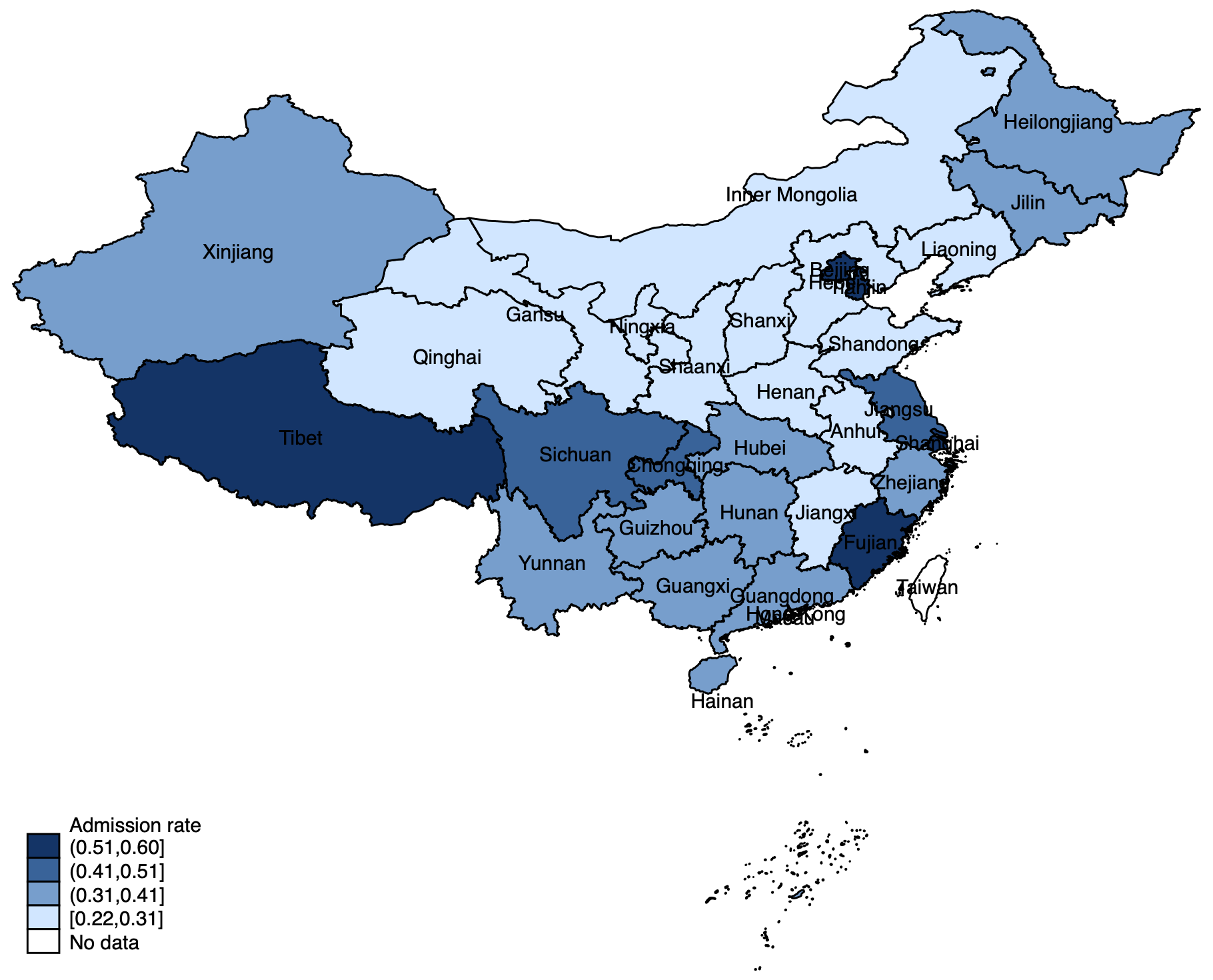}\label{fig_map_admrate98}}
	\subfloat[Average number of admitted students]{
		\includegraphics[width=.48\textwidth]{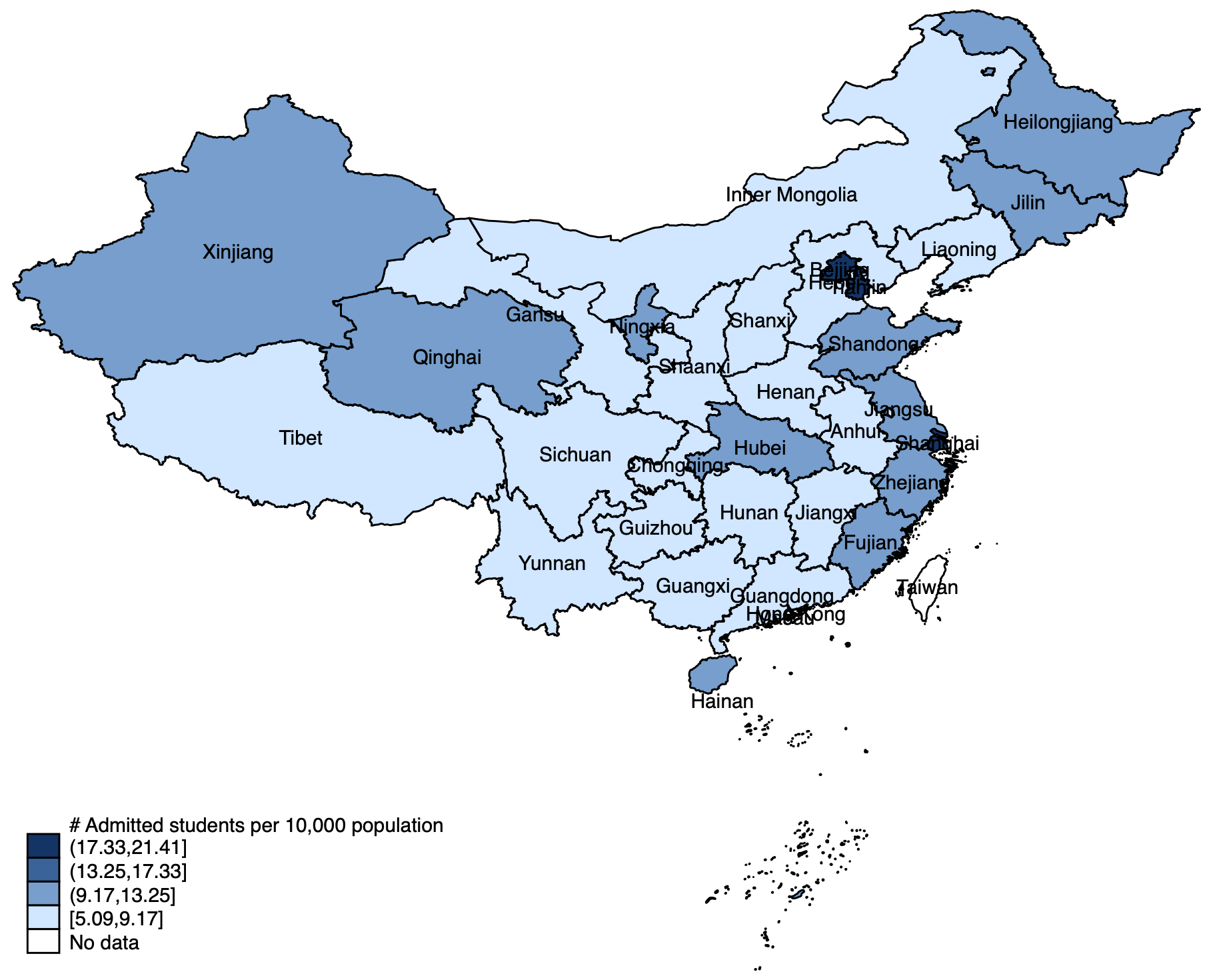}\label{fig_map_nadmpc98}}
	\caption{The NCEE admission pattern across provinces in 1998}\label{fig_province_adm}
	\captionsetup{singlelinecheck=false}
    \caption*{\footnotesize \textit{Notes}: This figure shows the cross-province variations of admission rate and average number of admitted students in 1998. The admission rate is the ratio between the number of admitted students and the number of NCEE takers, and the average number of admitted students is the number of admitted students divided by the population in that province.}
\end{figure}

In each year, about 95\% of the college candidates take NCEE in the province where their hukou is registered.\footnote{ Others can attend colleges by exemption (\textit{baosong}) due to exceptional or special talent or take other standardized entrance exams, such as those designed for adult education students.}\footnote{This location restriction is to alleviate the selective migrating to areas with less competition in NCEE.} After the NCEE, the education authority in each province operates the admission process under the predetermined admission quota scheme. All the exam takers are graded and ranked within their province, then the provincial admission office allocates a college to each student based on their scores, college preferences, and the available quota of each college in that province. Figure \ref{fig_province_adm} shows the admission pattern across provinces in 1998, we can see that both the admission rate (Figure \ref{fig_map_admrate98}) and the average number of admitted students (Figure \ref{fig_map_nadmpc98}) vary across provinces.

The availability of quotas is mainly determined by the number of colleges in each province. Almost all colleges admit students from more than one province, but local preference is highly pronounced when allocating admission quotas \citep{Kang2005-kt}. Even top colleges that are supposed to recruit students nationwide allocate more of their seats to their local province. For example, Sun Yat-Sen University, the top one university in Guangdong province, admitted 6,452 students nationwide in 2017, including 3,202 in Guangdong alone, accounting for 49.6 \% of its whole quota (\cite{Ma_undated-gk}; \cite{yang2021place}; \cite{Suyu2017-dt}). Figure \ref{fig_map_ncolpc} shows the number of colleges per capita in each province, which is consistent with the number of admitted students per capita, their correlation coefficient is 0.75. Therefore, provinces with more post-secondary institutions (colleges) can admit more students to higher education.



\subsection{Higher Education Expansion}\label{sec_Background_hee}
In response to the market demand for high-quality labor and the potential recession after the 1997 Asian financial crisis, the Chinese government dramatically expanded its higher education enrollment in 1999 \citep{Wan2006-wd}.

This policy was not announced to the public until June 1999, after that year's NCEE registration deadline \citep{Kang2000-uw}.\footnote{According to Baocheng Ji, the director of the Department of Development and Planning from the Ministry of Education, ``On the morning of June 2nd, the Minister told me that the State Council has decided to expand largely the scale of higher education admission this year. I was shocked since I thought it was too late to prepare for it'' \citep{Ji2009-cg}.} This fact is consistent with the total number of exam takers in 1999, as shown in Figure \ref{fig_nation_overall_adm}. In 1999, the number of NCEE takers decreased by about 10\%, while the number of admitted students increased by about 20\%, indicating that most students were unaware of the expansion and that the admission process did experience considerable changes.


The magnitude of the HEE has been remarkable and persistent since 1999. Overall, the admission rate (the ratio between the number of admitted students and the number of NCEE takers) surged from 34\% in 1998 to 56\% in 1999, mainly due to the increase in the number of admitted students from 1.08 million to 1.6 million. In the subsequent years, the rise in the number of admitted students persists, leading to the rise in the number of admitted students per capita, as indicated in Figure \ref{fig_nation_avg_adm}.
For example, in 2006, the number of admitted students has been more than 5 million, and the number of admitted students per capita has been four times higher than that in 1998 (Figure \ref{fig_nation_adm}).

\begin{figure}[htbp]
    \centering
	\subfloat[Admission rate]{
		\includegraphics[width=.48\textwidth]{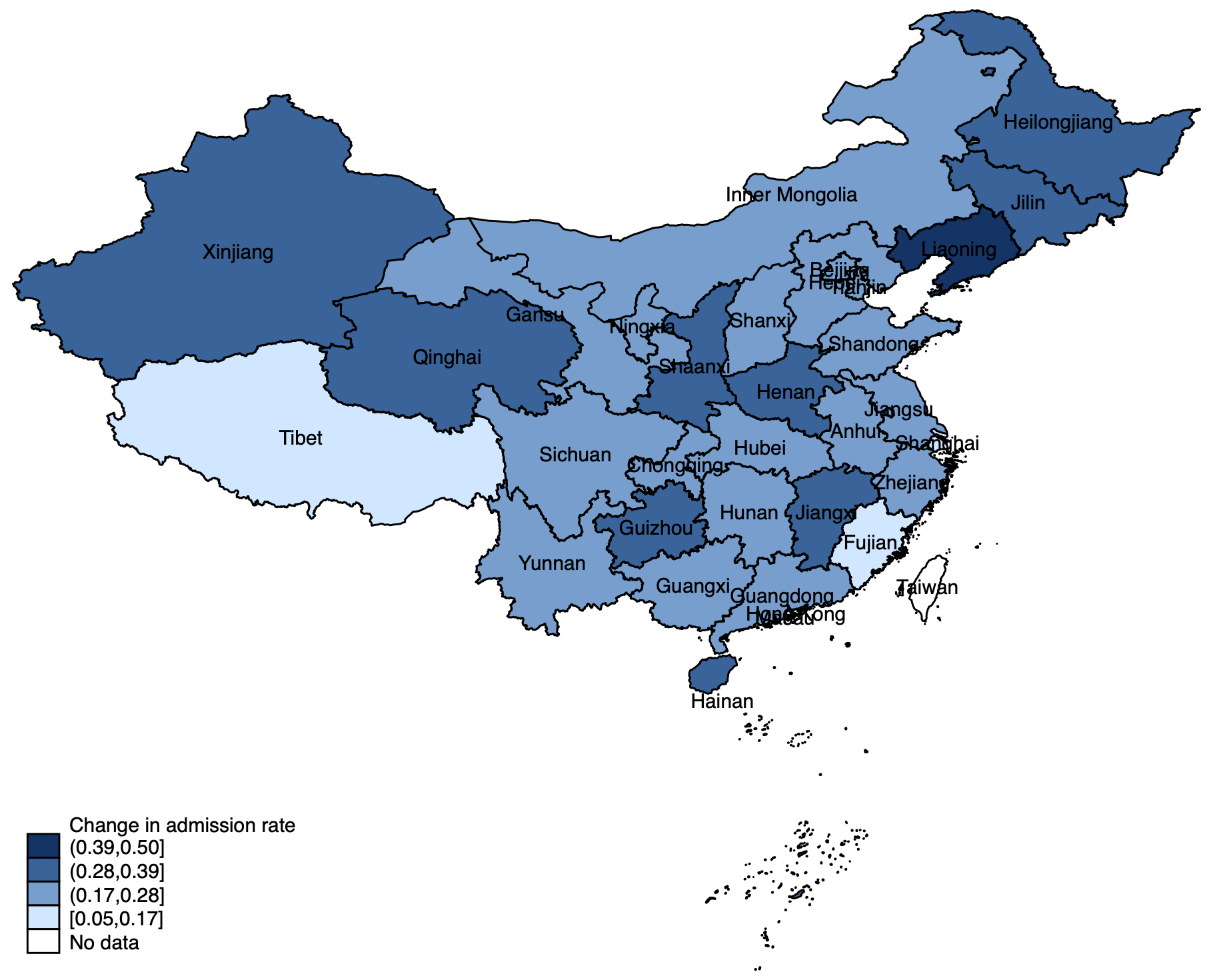}\label{fig_map_mean_admrate}}
	\subfloat[Average number of admitted students]{
		\includegraphics[width=.48\textwidth]{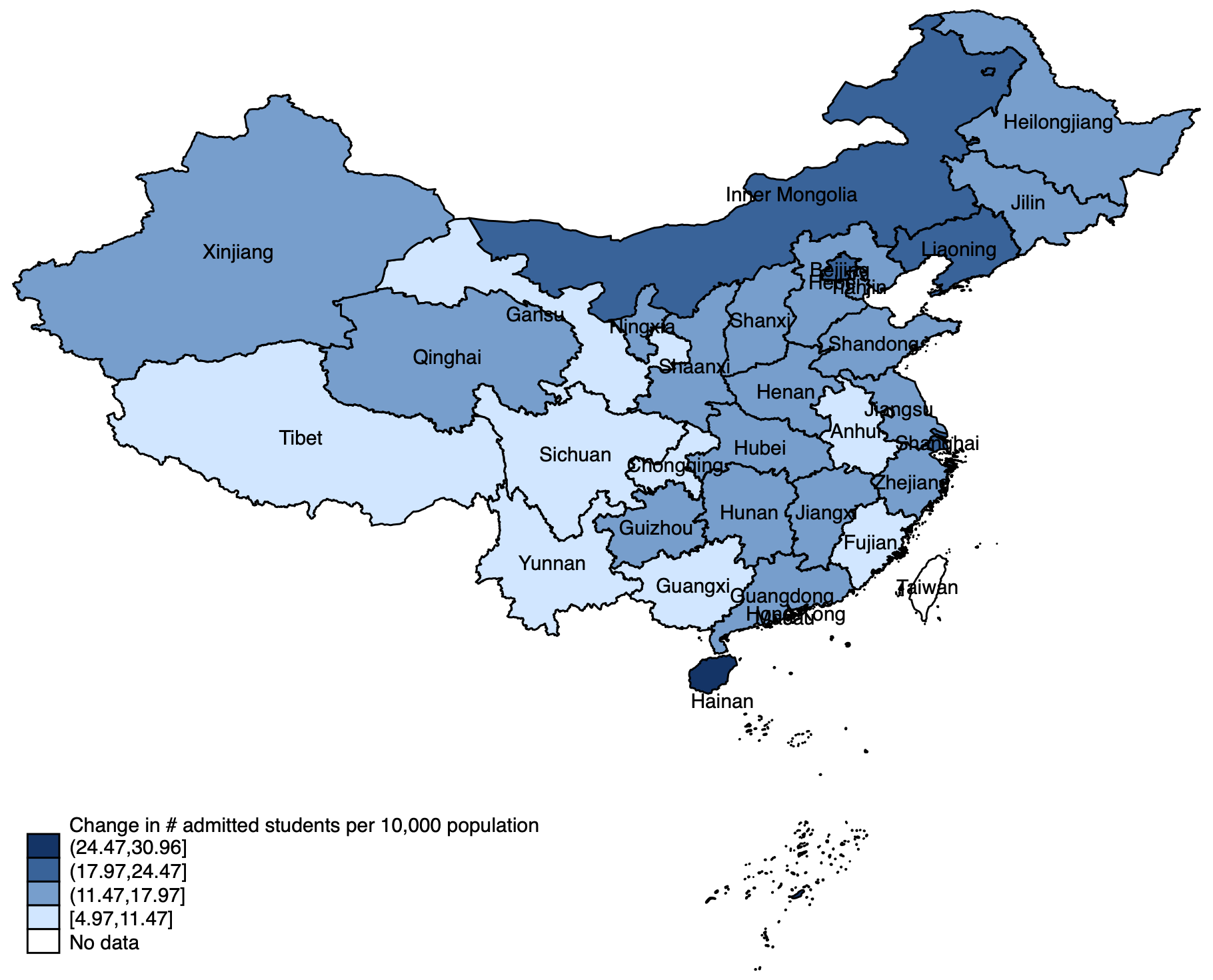}\label{fig_map_mean_nadmpc}}
	\caption{The change in the NCEE admission patterns across provinces}\label{fig_province_mean_adm}
	\captionsetup{singlelinecheck=false}
    \caption*{\footnotesize \textit{Notes}: This figure shows the change of provincial admission rate and average number of admitted students after 1999. The change is calculated as the difference between the level in 1998 and the average level during 1999 and 2003.}
\end{figure}

As mentioned above, the admission process is highly place-based, 
leading to the magnitude of the HEE varying a lot across provinces.\footnote{For example, Figure \ref{fig_province_admrate} shows the admission rates in four provinces over time. Both their admission rates in 1998 and the subsequent expansions differ significantly from each other.} Figure \ref{fig_province_mean_adm} presents the change in the admission rate and the average number of admitted students per 10,000 population across provinces from 1998 to 2003. 
We can observe significant variations in the changes. For example, the changes in the number of admitted students per 10,000 population are between 5 and 31, where the maximum is around 6 times as large as the minimum. We will use these province-level variations for our empirical analysis to implement causal inference.

\section{Data}\label{sec_Data}

Our analysis combines three primary data sets. In the first, we collect data on the admission and participation in NCEE, both of which are used to construct our measure of HEE. The second one is the 2000 and 2005 Census, which can be used to examine people's education enrollment. The last one is the National Fixed Point Survey (NFPS) conducted by the Ministry of Agriculture, which allows us to understand the effect of HEE in Rural China.

First, to measure the pattern of higher education expansion in China, we collect both national and provincial-level NCEE information from the National Education and Examination Yearbooks issued by the Ministry of Education. We mainly focus on the number of examinees and the number of admitted students. Data for the national-level NCEE admissions have been available since it was resumed in 1977. However, the provincial admission data can only go back to 1998. Missing province-year cells in the yearbook are either referenced from local official newspapers in that time or imputed by linear interpolation. Finally, we collect a balanced panel data set on admissions for all provinces from 1998 to 2012.

As a complementary, we also collect the number of students (in senior high schools or colleges) and the number of post-secondary institutions (colleges, universities, technical schools) in each province. Prior to 1998, the number of college students and higher education institutions were almost constant across years and provinces, indicating that it is valid to measure the magnitude of HEE by only using one pre-expansion year (i.e., year 1998).

Second, we use the 2005 and 2010 population censuses to evaluate people's education attainment choices due to HEE. The Census 2010 includes 1\% of the whole population, while Census 2005 covers 0.1\% population with weights allocated to individuals. Both of them are representative and provide detailed demographic information of individuals, including birthday, birth location, \textit{hukou} status, \textit{hukou} location, gender, ethnicity, education attainment, migration pattern, occupation (if any), and wage (only in Census 2005).

Next, to examine the impact of HEE on rural villages, we utilize the National Fixed Point Survey (NFPS), a panel survey conducted by the Research Center for the Rural Economy (RCRE) under the Ministry of Agriculture in China. The survey started in 1986 and continues until now. We use annual waves of data between 1995 and 2008 since the structure of the survey changed substantially in 1995.\footnote{NFPS is not available for years 1992 and 1994 since RCRE was unable to conduct the survey due to funding difficulties in 1992 and 1994.}\footnote{The content of the questionnaire of NFPS in the 1986-1992 version is substantially different from the later version.} We focus on those rural villages surveyed in NFPS, though they also include rural households. The data set used in our analysis covers around 250 rural villages across 24 provinces.\footnote{The dataset is balanced for 192 villages.}

RCRE selects those villages in NFPS for representativeness based on region, income, cropping pattern, population, and non-farm activities. \cite{benjamin2005evolution} provides a detailed description of the data and indicates that NFPS data are of good quality.

Over the period 1995 to 2008, we can observe both village-level information and household-level characteristics, but we use only the village-level panel due to the considerations from four perspectives. Firstly, the attrition of households. The survey runs in a rotation way, in each year, it includes around 10,000 households (in the 24 provinces we focus on) and replaces around 500 households by adding new households. However, we cannot distinguish whether it is due to the rotation or out-migration (or death of members) if some households are dropped out from the survey.\footnote{Though the out-migration of the whole household is rare in rural China because they would lose their land if all the families in the household moved out of their \textit{hukou} place \citep{Kinnan2018-bd}.} Secondly, age distribution. We cannot observe the age of each member in a household in NFPS before 2003. So we cannot observe which households are affected by the HEE by looking at whether there are schooling-age children in the households. Another way is to speculate the ages of individuals in each household from the data after 2003, but we cannot observe some people's age after 2003 if they had already moved out of villages due to the HEE or other reasons before 2003. Thirdly, education attainment. Before 2003, we only know the number of educated people within working ages who work in the village (the original question in the survey is ``In your household, how many people within working ages and working in the village have attended senior high school or above?"). Namely, we could only observe the education achievement of labors who stayed behind in villages, people who moved out are not counted in the data. Lastly, migration. The household panel only collects information on family members registered under the same \textit{hukou}, so people would not be counted if they moved out or changed their \textit{hukou} status. Also, what we could observe is the total time that each household spends working outside of the village, but we cannot know whether the one working outside is the one affected by the HEE. Given the question we want to answer with the NFPS data, the drawbacks of the household panel in these perspectives induce us to adopt the village panel in our main results.



In 2003, there was an adjustment to the villages surveyed in NFPS. After restricting the sample to cover only the 24 provinces, we observe a balanced panel for 239 villages from 1995 to 2003, but only 192 for 1995 to 2008. Due to the panel structure of the data being very useful for our empirical analysis and including more villages could increase the validity of statistical inferences, in addition to results based on the longer panel from 1995 to 2008, we also report the results using the panel from 1995 to 2003 in the appendix as robustness checks.

\begin{table}[h!]
\centering
\caption{Summary Statistics of NFPS Village Panel (1995-2008)}\label{tab_v_su}
\begin{tabular}{lccccc}
\hline \hline
& Mean & SD & Observations & Min & Max \\
\hline
\multicolumn{6}{l}{\textbf{Population}} \\
End-of-year population & 1,709.46 & 1,173.45 & 2,684 & 273 & 5,475 \\
End-of-year households & 448.63 & 321.01 & 2,683 & 73 & 1,398 \\
\hline
\multicolumn{6}{l}{\textbf{Labor force}} \\
Total & 902.12 & 680.76 & 2,466 & 152 & 3,104 \\
By education attainment: &  &  &  &  &  \\
\hspace{0.5cm}Senior school+ & 114.18 & 179.73 & 2,669 & 5 & 746 \\
\hspace{0.5cm}Junior school+ & 494.23 & 484.75 & 2,674 & 48 & 2,137 \\
\hspace{0.5cm}Primary school+ & 920.56 & 781.68 & 2,669 & 152 & 3,443 \\
\hspace{0.5cm}Illiteracy & 98.49 & 132.27 & 2,345 & 2 & 605 \\
By work type: &  &  &  &  &  \\
\hspace{0.5cm}Agriculture & 508.35 & 378.99 & 2,460 & 57 & 1,895 \\
\hspace{0.5cm}Non-agriculture & 340.22 & 443.47 & 2,171 & 6 & 1,781 \\ \hline
\multicolumn{6}{l}{\textbf{Area of cropped land}} \\
Total (unit: \textit{mu}) & 3,888.99 & 3,846.58 & 2,673 & 294 & 19,728 \\
\hspace{0.5cm}Grain crop & 2,872.83 & 2,898.46 & 2,656 & 90 & 16,590 \\
\hspace{0.5cm}Cash crop & 603.38 & 1,223.56 & 2,447 & 0 & 6,100 \\
\hline
\multicolumn{6}{l}{\textbf{Income and life quality}} \\
Income per capita (\textit{yuan}) & 2,202.23 & 2,016.14 & 2,602 & 510.58 & 8,575.66 \\
Num. of HHs with: &  &  &  &  &  \\
\hspace{0.5cm}Tap water & 251.19 & 313.48 & 2,413 & 0 & 1,267 \\
\hspace{0.5cm}Safe water & 323.42 & 306.08 & 2,377 & 0 & 1,272 \\
\hspace{0.5cm}Good housing & 171.05 & 276.34 & 2,586 & 0 & 1,313 \\
\hspace{0.5cm}Electricity & 438.23 & 324.27 & 2,671 & 2 & 1,411 \\
\hspace{0.5cm}TV & 397.23 & 299.47 & 2,661 & 5 & 1,340 \\
Num. of telephones & 163.81 & 285.14 & 2,672 & 0 & 1,298 \\ \hline \hline
\end{tabular}
\captionsetup{singlelinecheck=false}
\caption*{\footnotesize \textit{Notes}: This table reports the summary statistics for all villages in 24 provinces at village-year level during 1995 - 2008. We use \textit{mu} to measure the area of land, one \textit{mu} is equivalent to 4046.86 $m^2$. The income per capita is deflated to their levels in 1995 using provincial rural CPIs from \cite{Brandt2006-bn}. We measure income by the Chinese currency \textit{yuan}, and one \textit{yuan} is approximately 1/8.4 USD in 1995. Good housing refers to houses with reinforced concrete structures.}
\end{table}

Table \ref{tab_v_su} reports the summary statistics for villages in the NFPS data. We classify all the key outcomes into four groups: population, labor force, agricultural activities, and income and life quality. The number of population and households are calculated based on the \textit{hukou} registration. 

For labor force, Table \ref{tab_v_su} first reports the total number of rural labor force in each village. Except for students, military personnel, and the disabled, the labor force includes all rural residents between 15 and 64 years of age who have lived in villages for more than six months in a year and are not employed outside the village. The survey asks the education level and type for the left-behind laborers (i.e., laborers working in the village). This calculation is based on the sampling households within the village \citep{Moa_datacoll-aj}. The results show that more than half of the total rural labor force has only junior high school education. In addition, there are more workers engaged in the agricultural sector.

As for the land usage and living facilities, each village has their clear account recording these. We use the area of cropped land in each village to measure the agricultural activities. On average, more than 70\% (2703/3630) of the cropped land in rural villages is for grain crops, and just around 15\% of the cropped land is used for cash crops. We use average income and living facilities to describe life quality. The last part of Table \ref{tab_v_su} presents the outcomes. The income per capita is also calculated based on the sampled households and we deflate it to the level in year 1995 using provincial rural consumer price indices from \cite{Brandt2006-bn}. We measure income by the Chinese currency \textit{yuan}, where one \textit{yuan} is approximately 1/8.4 USD in 1995. To measure life quality in rural areas, we count the number of households with access to better living facilities (e.g., tap water, electricity).\footnote{In 2006, the Chinese government started a subsidy program to construct rural tap water system for all villages \citep{Feng2010Fiscal}. Before that, only rich villages can afford the construction cost of tap water, so the number of households using tap water could be an indicator of a village's wealth level.}\footnote{Safe water includes tap water and other water sources (e.g., water with filter process). The number of households using safe water could be an approximation for the health status in a village.}\footnote{The average price of a television was about 2,000 \textit{yuan} in 2003 \citep{Yang2008-tvprice}, which is close to the annual income per capita in rural villages in our sample.} We also quantify local amenities in rural villages by the number of telephones.

\begin{table}[h!]
\centering
\caption{Summary Statistics of NFPS Households Panel (1995-2008)}
\label{tab_hh_su}

\begin{tabular}{lccccc}
\hline\hline
 & Mean & SD & Observations & Min & Max \\ \hline
\textbf{Population} &  &  &  &  &  \\
\hspace{0.5cm}Num. residents & 4.536 & 1.646 & 48261 & 0 & 18 \\
\hspace{0.5cm}Num. migrants & 0.179 & 0.643 & 44746 & 0 & 13 \\ \hline
\textbf{Num. labor force} &  &  &  &  &  \\
\hspace{0.5cm}Total & 2.557 & 1.272 & 48261 & 0 & 13 \\
By education attainment &  &  &  &  &  \\
\hspace{0.5cm}Junior school+ & 1.074 & 1.021 & 48259 & 0 & 8 \\
\hspace{0.5cm}Senior school+ & 0.258 & 0.605 & 48254 & 0 & 11 \\ \hline
\textbf{Income \textit{(yuan)}} &  &  &  &  &  \\
\hspace{0.5cm}Avg. total income & 4523 & 4230 & 47108 & 473.3 & 55515 \\
\hspace{0.5cm}Avg. remittance & 325.5 & 893.3 & 46125 & 0 & 13812 \\ \hline\hline
\end{tabular}%
\captionsetup{singlelinecheck=false}
\caption*{\footnotesize \textit{Notes}: This table reports the summary statistics for all households in 24 provinces during 1995 - 2008. The income per capita is deflated to their levels in 1995 using provincial rural CPIs from \cite{Brandt2006-bn}. We measure income by the Chinese currency \textit{yuan}, and one \textit{yuan} is approximately 1/8.4 USD in 1995. }
\end{table}

To split the HEE' impact on rural village as a whole and its impact on households who send out more migrants due to the increasing of education opportunity, we also utilize the household level panel in the same period.

Table \ref{tab_hh_su} reports the summary statistics for households in the NFPS data. Aligning with the classification of variables on the village level, we classify the key outcomes into three groups: population, labor force, and income. The number of one household's residents are the number of people who register their \textit{hukou} in this household. The NFPS does not have a direct record on households' migration. To fit our context we use the difference of number of residents between year $t$ and year $t-1$ as an approximation of the number of permanent migration.

Again, for labor force, Table \ref{tab_hh_su} first reports the total number of rural labor force in each household. Then we report the labor force by educational achievement. The results show consistent pattern with the village level survey -- about half of the total rural labor force has only junior high school education and only 1/9 rural labors have senior school degree or above.

The last part of Table \ref{tab_hh_su} presents the outcomes on households' income. We still deflate it to the level in year 1995 using provincial rural consumer price indices. The household level survey does not specify their access to living facilities. Nevertheless, it specifies the amount of income from the migrated household members' remittance.

In our empirical analysis, we transform all outcomes into their logarithmic levels. For variables equal to zero in some villages (e.g., the number of households with tap water before 1999), we use the inverse hyperbolic sine function to do the transformation \citep{Burbidge1988-sj}.\footnote{The IHS function is similar to a logarithmic transformation but is well deﬁned for values of zero. Thus, we use it for continuous outcomes whose distribution includes a preponderance of zeros and a long right tail. We use the logarithmic transformation for continuous variables without any zeros.} In the paper, the top and bottom 1 percent of values are dropped to deal with outliers. However, even if we don't drop outliers, the results are very similar.

To account for region-specific characteristics, we also collect provincial characteristics including population, aggregate economic outputs, and employment from the Provincial Statistical Yearbooks.

\section{Empirical Strategy}\label{sec_EmpiricalStrategy}

Our empirical strategy consists of three parts. We first propose our preferred measure for the magnitude of the HEE across provinces and check its correlation with other reasonable measures. Second, we examine how the HEE affects people's educational attainment in colleges and senior high schools by exploiting individual characteristics from the 2010 population census.\footnote{As you will see, the magnitude of the HEE is a posterior term, the effect on college enrollment could be mechanical. Nevertheless, the HEE cannot directly affect people's enrollment in senior high schools, so it represents their willingness to pursue education as more post-secondary education opportunities emerge, which we regard as the spillover effects of the HEE.} In addition to educational attainment, we also explore other outcomes associated with education (e.g., income and working industry). Finally, we switch to the panel data on rural villages and households from NFPS. We analyze how rural people respond to the HEE regarding their agricultural activities, work decisions, and migration decisions, assessing the consequences of the HEE in rural areas.

\subsection{The Magnitude of the HEE}\label{sec_EmpiricalStrategy_treatment}
The HEE increased the number of admitted students directly, which makes it easier for students in less populated provinces to attend colleges than those in more populous provinces, so the change in the number of admitted students per capita should be an appropriate candidate to measure the magnitude of the HEE. It captures the policy change \textit{per se} and the competition effect due to population density. Thus, we utilize changes in the number of admitted students per capita after the HEE to measure its magnitudes.\footnote{Here we calculate changes in the share of admitted students from the population. We do not restrict to the population only aged around the standard NCEE-taking age, 18 (for example, from 15 to 21), for two reasons: (1) we cannot observe the exact age distribution in each province across all years; (2) the 2000 Census indicates that age distributions are similar across provinces.} Specifically, we define the level changes or the ratio changes, then take the average across years,
\begin{equation}\label{eq_treatment_def}
    \begin{split}
    Diff_{p} = \frac{1}{5} \sum_{t=1999}^{2003} Diff_{p,t} &= \frac{1}{5} \sum_{t=1999}^{2003} (AdmPC_{p, t} - AdmPC_{p, 1998}), \\
    Ratio_{p} = \frac{1}{5} \sum_{t=1999}^{2003} Ratio_{p,t} &= \frac{1}{5} \sum_{t=1999}^{2003} (AdmPC_{p, t}/AdmPC_{p, 1998}),
    \end{split}
\end{equation}
where $AdmPC_{p,t}$ is the number of admitted students per capita (for exposition, we use 10,000 population) in province $p$ in year $t$. We include only the year 1998 as the pre-HEE period due to two considerations. First, the admission information is missing for many provinces before 1998. Second, the number of college students in all provinces had been stable for years before 1998, so the number of admitted students didn't vary a lot until 1998, consistent with the national pattern in Figure \ref{fig_nation_adm}.

In the two measures, $Diff_{p}$ quantifies the magnitude of the HEE in terms of level changes while $Ratio_{p}$ measures the percentage changes.\footnote{The key difference between the two measures is related to the average number of admitted students in 1998 (i.e., $AdmPC_{p, 1998}$). When we rank all provinces by $Diff_{p}$ or $Ratio_{p}$, a higher (lower) $AdmPC_{p, 1998}$ would lead to a lower (higher) rank in $Ratio_{p}$ comparing to the rank in $Diff_{p}$ since $Ratio_{p} = Diff_{p}/AdmPC_{p, 1998} + 1$.}
Our preferred measure is the 5-year average of the level changes ($Diff_{p}$). For robustness, we also calculate the 5-year average of the percentage changes ($Ratio_{p}$) and the 10-year averages of both measures from 1999 to 2008. Table \ref{tab_treatment_measure} presents the values of each measure and their normalized values for all the 24 provinces in our sample.  Table \ref{tab_treatment_corr} shows the correlation between our preferred measure and other measures of the HEE magnitudes. All of them are highly correlated, indicating that the variation of the HEE magnitudes is consistent across different measures.

\begin{figure}[h!]
    \centering
    \includegraphics[width = 0.8\textwidth]{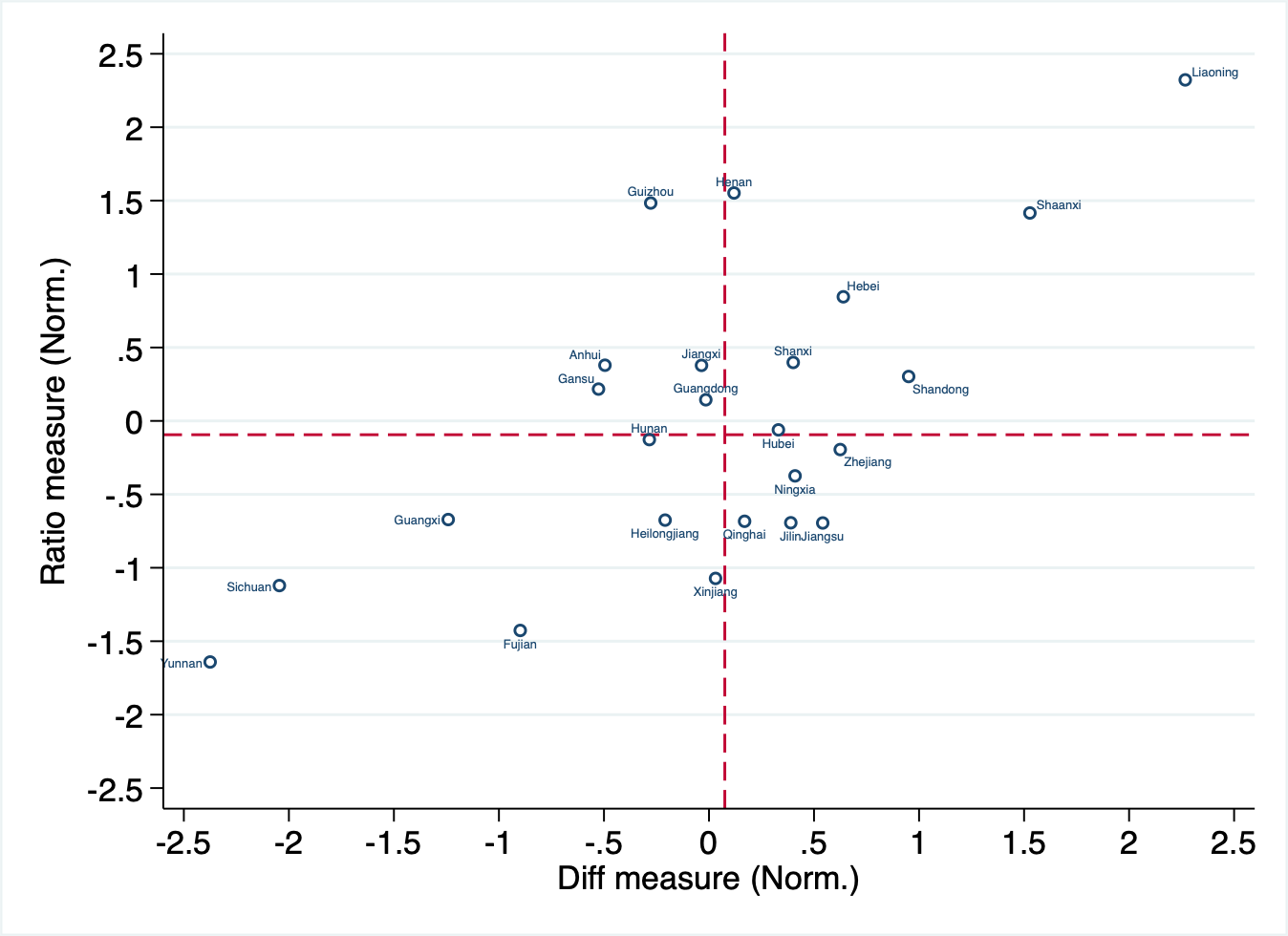}
    \caption{Measures of HEE Magnitudes (5-year average)}
    \label{fig_treatment_avg5}
    \captionsetup{singlelinecheck=false, width = \textwidth}
    \caption*{\footnotesize \textit{Notes}: This figure shows how we measure the magnitudes of HEE across 24 provinces (excluding Hainan, Tibet, Inner Mongolia, and four provincial-level municipalities Beijing, Tianjin, Shanghai, and Chongqing). The $Diff$ and $Ratio$ measures are defined in Equation \eqref{eq_treatment_def}, their normalized 5-year average values are presented in x- and y-axis, respectively. The two dashed lines correspond to the median value of each measure.}
\end{figure}

We classify the 24 provinces into two groups based on their positions relative to their median value. Specifically, we define $HighDiff_{p} = 1$ if $Diff_p$ is above the median of $Diff$ across all provinces and $0$ otherwise.\footnote{In robustness checks, we define $HighRatio_{p} = 1$ if $Ratio_p$ is above the median of $Ratio$ across all provinces and $0$ otherwise.} Figure \ref{fig_treatment_avg5} demonstrates the relationships between $Diff_{p}$ and $Ratio_{p}$ under the 5-year average.\footnote{Figure \ref{fig_treatment_avg10} shows the relationship between $Diff_{p}$ and $Ratio_{p}$ under the 10-year average.} Most provinces are located above or below the medians of both measures, which further reveals the consistency across different measures.


\subsection{Balance Tests}\label{sec_EmpiricalStrategy_balancetest}

To examine whether any provincial-specific characteristics drive the HEE magnitude, we conduct balance tests on a set of provincial economical and education characteristics documented prior to the HEE. Specifically, we use provincial economic and education characteristics in all years prior to the HEE to predict the magnitude of HEE across provinces. We include GDP per capita and employment to capture provincial economic status. We use employment shares in the agricultural and manufacturing sectors and GDP share in the agricultural sector to account for the gap in economic structure across provinces. We also consider population and the number of colleges per capita, which could affect the difficulties of being admitted through NCEE due to its place-based feature.

Table \ref{tab_balancetest} shows the regression results of predicting the two measures defined in Equation \eqref{eq_treatment_def} using province-level annual characteristics from 1995 to 1998. Most economical and education characteristics cannot predict the HEE magnitudes, except for the GDP per capita. The coefficients on GDP per capita indicate that poorer provinces are more likely to have higher magnitudes of the HEE, which means the impact could be stronger for more impoverished rural areas. Thus, to account for plausibly confounding effects caused by the economic gap across provinces, We control for the provincial GDP per capita in 1998.

\begin{table}[h!]
\centering
\caption{Balance Tests of the HEE Magnitude}\label{tab_balancetest}
\scalebox{0.9}{
\begin{tabular}{lcccccccc}
\hline \hline
 & (1) & (2) & (3) & (4) & (5) & (6) & (7) & (8) \\
 & $Diff$ & $Diff$ & $Diff$ & $Diff$ & $Ratio$ & $Ratio$ & $Ratio$ & $Ratio$ \\
\hline
ln(Employment) & 2.347 & 2.258 & 2.362* & 1.854 & 3.035 & 3.454* & 3.702* & 2.628 \\
 & (1.655) & (1.565) & (1.139) & (1.232) & (2.055) & (1.909) & (2.014) & (2.115) \\
Emp. share (agri.) & 0.0110 & 0.00280 & -0.0107 & -0.00787 & -0.0178 & -0.0347 & -0.0590 & -0.0411 \\
 & (0.0470) & (0.0488) & (0.0453) & (0.0451) & (0.0699) & (0.0740) & (0.0677) & (0.0633) \\
Emp. share (manu.) & 0.124 & 0.104 & 0.0843 & 0.103 & 0.0907 & 0.0690 & 0.0349 & 0.106 \\
 & (0.0907) & (0.0940) & (0.0826) & (0.0748) & (0.126) & (0.133) & (0.122) & (0.119) \\
ln(Population) & -1.895 & -1.733 & -1.771 & -1.433 & -2.613 & -2.913 & -3.015 & -2.165 \\
 & (1.708) & (1.661) & (1.166) & (1.300) & (2.151) & (2.012) & (2.037) & (2.153) \\
GDP share (agri.) & 0.0284 & 0.00908 & 0.00390 & -0.0176 & -0.00270 & -0.0237 & -0.0186 & -0.0173 \\
 & (0.0444) & (0.0480) & (0.0431) & (0.0455) & (0.0538) & (0.0606) & (0.0631) & (0.0714) \\
ln (GDP per capita) & -1.080** & -1.462** & -1.837*** & -2.202*** & -2.537*** & -3.042*** & -3.123*** & -3.269*** \\
 & (0.459) & (0.563) & (0.416) & (0.444) & (0.638) & (0.687) & (0.543) & (0.646) \\
\# Colleges per capita & 54.40 & 64.55 & 97.36 & 169.7** & 19.72 & 23.85 & 49.20 & 109.1 \\
 & (97.08) & (91.33) & (67.23) & (69.10) & (137.9) & (137.1) & (115.9) & (120.6) \\
\hline
Year & 1995 & 1996 & 1997 & 1998 & 1995 & 1996 & 1997 & 1998 \\
Observations & 24 & 24 & 24 & 24 & 24 & 24 & 24 & 24 \\
Adjusted R$^2$ & 0.538 & 0.592 & 0.667 & 0.662 & 0.0981 & 0.203 & 0.299 & 0.279 \\
\hline \hline
\end{tabular}
}
\captionsetup{singlelinecheck=false, width=\textwidth}
\caption*{\footnotesize \textit{Notes}: This table presents the balance test of the HEE magnitudes on all provincial economic and education characteristics prior to HEE. Each column corresponds to these characteristics in a specific year, as indicated by Year in the bottom part. For exposition, the unit for employment and population is 10,000, the unit for GDP per capita is \textit{yuan}, the number of colleges per capita is measured in terms of 10,000 population. The outcomes are normalized values of $Diff$ and $Ratio$, which are defined in Equation \eqref{eq_treatment_def}. Robust standard errors are in parentheses, *, **, and *** denote significance at the 10\%, 5\% and 1\% levels, respectively.}
\end{table}

Our causal estimation of the HEE's impact would be biased if other province-level shocks were correlated with the HEE magnitudes and occurred near the time of the HEE. Around 1999, the most significant shock was China's accession to the World Trade Organization (WTO) in 2001.\footnote{Another significant province-level policy change is the Rural Land Contracting Law, which gave farmers legal rights to lease their land while reiterating existing protections for the security of land rights after 2003 \citep{Chari2021-lt}. Exploiting the province-level implementation of RLCL, we could include province-specific indicators in our regression to check whether the main results change after accounting for the impact of RLCL.} This nationwide shock could generate differential impacts across provinces because the degree that a province is affected by the induced tariff reduction depends on the province's initial industrial structure \citep{Tian_2020-ol}. If a province more exposed to the HEE also experienced more tariff reduction, our estimation of the HEE's impact would be invalid. For example, a lower tariff could lead to firms' expansion and hire more workers, prompting people to drop out of school. Also, tariff reduction may connect firms to high-tech firms outside, increasing the demand for high-skill workers and encouraging people to get more education. Thus, the WTO accession could confound the impact of the HEE on educational attainment, and a similar argument can be applied to the confounding issue of WTO accession on rural households' response to the HEE. To mitigate this concern, we examine whether province-specific trade shock correlates with the HEE magnitudes. We use every province's trade volume (in cash) per capita (defined as $TradePC_p$) change after 2001 to capture the WTO accession shock,
\begin{equation}\label{eq_trade}
    \begin{split}
    DiffTrade_p &= \frac{1}{5}\sum_{t=2002}^{2006}TradePC_{p,t}-\frac{1}{5}\sum_{t=1997}^{2001}TradePC_{p,t}, \\
    RatioTrade_p &= \frac{1}{5}\sum_{t=2002}^{2006}TradePC_{p,t}/\left(\frac{1}{5}\sum_{t=1997}^{2001}TradePC_{p,t}\right).
    \end{split}
\end{equation}

Table \ref{tab_trade_hee} reports the regression of HEE magnitudes defined in Equation \eqref{eq_treatment_def} on the magnitudes of trade shock induced by the WTO accession. There is no evidence that the HEE magnitudes correlate with the WTO accession shock. The results are consistent when we change our measure for the HEE magnitudes from the 5-year averages to the 10-year ones or control for provincial characteristics. Therefore, we can exclude the concern that the WTO accession may confound the HEE's impacts.

\begin{table}[h!]
\centering
\caption{Trade shock and HEE magnitudes}\label{tab_trade_hee}
\begin{tabular}{lcccccc}
\hline \hline
  & (1) & (2)  & (3)  & (4)  & (5) & (6) \\
 & $Diff$ & $Diff$ & $Diff$ & $Ratio$ & $Ratio$ & $Ratio$ \\
\hline
$DiffTrade$ & 0.189 & -0.184 & -0.0720 & &  &  \\
 & (0.128) & (0.211) & (0.162) &  &  &  \\
$RatioTrade$ & & & & -0.0961 & 0.0440 & -0.0170 \\
  & & & & (0.124) & (0.312) & (0.247)  \\
\hline
Controls & & GDP pc & All &  & GDP pc & All \\
Observations & 24 & 24 & 24 & 24 & 24 & 24 \\
Adjusted R$^2$ & -0.00798   & 0.0575 & 0.407 & -0.0358 & -0.0693 & 0.295 \\
\hline \hline
\end{tabular}
\captionsetup{singlelinecheck=false}
\caption*{\footnotesize \textit{Notes}: The dependent variables are the 5-year average $Diff$ and $Ratio$ measures described in Equation \eqref{eq_treatment_def}. Columns (1) and (4) don't include any control variable; Columns (2) and (5) control for the logarithmic GDP per capita; Columns (3) and (6) include as controls all the variables listed in Table \ref{tab_balancetest}. All variables are collected from Provincial Statistical Yearbooks.  Robust standard errors are reported in parentheses, *, **, and *** denote significance at the 10\%, 5\% and 1\% levels, respectively.}
\end{table}

\subsection{The Impact on Educational Attainment}\label{sec_EmpiricalStrategy_eduattainment}
To identify the effect of the HEE on the attainment of secondary and post-secondary education, we conduct a cohort-based difference-in-differences regression by exploiting the variation of the HEE magnitudes across provinces and the variation of cohorts' exposure to the HEE. A typical student in China starts her primary school at age six, spends six years in primary school, three years in junior high school, and three years in senior high school, so she will be age 18 when completing senior high school and taking the NCEE, though this age may be different if some people started their education earlier or later or they may skip or repeat some grades. Usually, a higher education institution (university or college) in China requires four years before graduation. Table \ref{tab_ageincensus} illustrates the different cohorts' ages in the 2010 population census and the year when they entered colleges or senior high schools.

\begin{table}[h!]
\centering
\caption{Ages of Different Birth-Cohorts in Census}\label{tab_ageincensus}
\begin{tabular}{cccc}
\hline \hline
\multicolumn{1}{c}{\multirow{2}{*}{Birth year/cohort}} & \multirow{2}{*}{Age in 2010} & \multicolumn{2}{c}{Year entering} \\
\cline{3-4}
 & & College & Senior high \\
\hline
1972 & 38 & 1990 & 1987 \\
1973 & 37 & 1991 & 1988 \\
1974 & 36 & 1992 & 1989 \\
1975 & 35 & 1993 & 1990 \\
1976 & 34 & 1994 & 1991 \\
1977 & 33 & 1995 & 1992 \\
1978 & 32 & 1996 & 1993 \\
1979 & 31 & 1997 & 1994 \\
1980 & 30 & 1998 & 1995 \\
1981 & 29 & \cellcolor[rgb]{0.973, 0.796, 0.678}1999 & 1996 \\
1982 & 28 & \cellcolor[rgb]{0.973, 0.796, 0.678}2000 & 1997 \\
1983 & 27 & \cellcolor[rgb]{0.973, 0.796, 0.678}2001 & 1998 \\
1984 & 26 & \cellcolor[rgb]{0.973, 0.796, 0.678}2002 & \cellcolor[rgb]{0.973, 0.796, 0.678}1999 \\
1985 & 25 & \cellcolor[rgb]{0.973, 0.796, 0.678}2003 & \cellcolor[rgb]{0.973, 0.796, 0.678}2000 \\
1986 & 24 & \cellcolor[rgb]{0.973, 0.796, 0.678}2004 & \cellcolor[rgb]{0.973, 0.796, 0.678}2001 \\
1987 & 23 & \cellcolor[rgb]{0.973, 0.796, 0.678}2005 & \cellcolor[rgb]{0.973, 0.796, 0.678}2002 \\
1988 & 22 & \cellcolor[rgb]{0.973, 0.796, 0.678}2006 & \cellcolor[rgb]{0.973, 0.796, 0.678}2003 \\
1989 & 21 & \cellcolor[rgb]{0.973, 0.796, 0.678}2007 & \cellcolor[rgb]{0.973, 0.796, 0.678}2004 \\
1990 & 20 & \cellcolor[rgb]{0.973, 0.796, 0.678}2008 & \cellcolor[rgb]{0.973, 0.796, 0.678}2005 \\
1991 & 19 & \cellcolor[rgb]{0.973, 0.796, 0.678}2009 & \cellcolor[rgb]{0.973, 0.796, 0.678}2006 \\
1992 & 18 & \cellcolor[rgb]{0.973, 0.796, 0.678}2010 & \cellcolor[rgb]{0.973, 0.796, 0.678}2007 \\
\hline \hline
\end{tabular}
\captionsetup{singlelinecheck=false}
\caption*{\footnotesize \textit{Notes}: A typical student starts primary school at age six, spends six years in primary school, three years in junior high school and three years in senior high school, so she enters senior high school and college at age 15 and 18, respectively. All the highlighted cells indicate people affected by HEE when making education decisions on entering colleges or senior high schools.}
\end{table}

All the highlighted cells in Table \ref{tab_ageincensus} indicate people affected by the HEE when making education decisions on attending colleges or senior high schools. To be specific, the 1981 cohort (i.e., people born in 1981) took their NCEE in 1999, so they are the most senior cohort that has experienced the HEE when making decisions on post-secondary education, though people from other cohorts (e.g., cohort 1980 or 1982) may also take their NCEE in 1999. Similarly, the 1984 cohort is the most senior one affected by the HEE when deciding whether to enter senior high schools. 


By utilizing the difference of each cohort in provinces with higher and lower HEE magnitudes and the difference across cohorts in whether or not being affected by the HEE when choosing their educational attainment, we use the following cohort-based difference-in-differences specification to estimate the effect of HEE on attending colleges or senior high schools,
\begin{equation}\label{eq_did}
    y_{i,c,p} =  \eta_c + \gamma_p + \sum_{k=1972}^{1991} \beta_k \times Cohort_{i,k} \times HighDiff_{p} + X_{i,c,p} + \varepsilon_{i,c,p}.
\end{equation}
Here $y_{i,c,p}$ is the indicator of attending colleges or senior high schools for individual $i$ of cohort $c$ in province $p$.\footnote{Note that we use every individual's \textit{hukou} province if the province where she is surveyed in the population census is different from her \textit{hukou} province. In the 2010 population census, 9.48\% of the population are surveyed in provinces different from their \textit{hukou} provinces.} $HighDiff_{p}$ is the indicator of province $p$ for being more exposed to the HEE, as defined in Section \ref{sec_EmpiricalStrategy_treatment}. We also use binary measure $HighRatio_p$ and non-binary measures ($Diff_p$ and $Ratio_p$) to replace $HighDiff_p$ for robustness. Standard errors are clustered at the province level to allow for arbitrary dependence of the error term $\varepsilon_{i,c,p}$ across cohorts within each province.

We allow for ``buffer'' cohorts when classifying which cohorts are affected by the HEE. In Table \ref{tab_ageincensus}, cohort 1981 is the first one affected by the HEE when making college attendance decision, as they were 18 in 1999. However, people can attend school earlier or later or skip or repeat some grades, so the HEE could also affect cohorts 1980 or 1982's higher education decisions. If we regarded cohort 1981 as the first being affected by the HEE and any earlier cohort as unaffected, our estimation of the HEE's impact on people's educational attainment would be downwards biased since the HEE could also influence those earlier cohorts. To eliminate such concerns, we use cohort 1978 as the reference group and compare it to other cohorts' college attendance. Similarly, we use cohort 1981 as the reference group when estimating the HEE's impact on attending senior high schools.

We control for provincial and individual characteristics in $X_{i,c,p}$. Table \ref{tab_balancetest} indicates that province-level GDP per capita is correlated with the HEE magnitude across provinces, so we include into $X_{i,c,p}$ the provincial GDP per capita interacted with cohort indicators, which flexibly allows for heterogeneous impacts of economic status on all the cohorts.
For individual controls, we include both gender and ethnicity indicators to account for the gender gap of access to higher education \citep{wang2019labor} and minority students' bonuses in China's higher education admission process. Specifically, the gender indicator is zero for a male student, and the ethnicity indicator is one for a Han student (the major ethnicity in China) and zero for other minority students.

In Equation \eqref{eq_did}, $\eta_c$ are a set of cohort fixed effects to control for policy shocks that affect each cohort uniformly across locations, for example, the 1986 compulsory education law, which requires all people to complete at least nine years of schooling \citep{fang2012returns}, and the abolition of the national agricultural tax in 2004 \citep{Wang2014-gw}. $\gamma_p$ are a set of province fixed effects to account for the underlying time-invariant gaps among provinces, such as the gaps of economic status and industrial structure. Thus, our identifying assumption is valid to any nationwide policy change or underlying provincial gap.

The coefficients of interest are $\beta_k$'s. The underlying assumption of our identification is that conditional on observed provincial and individual characteristics and unobserved province- and cohort-specific common shocks, the HEE magnitudes are randomly assigned across provinces. To be specific, let $y^{0}_{i,c,p}$ be the potential educational attainment of individual $i$ in cohort $c$ in province $p$ if province $p$ had lower magnitude of the HEE. We assume that
\begin{equation*}
    E[y^{0}_{i,c,p}|\eta_c, \gamma_p, X_{i,c,p}, HighDiff_{p} = 1] = E[y^{0}_{i,c,p}|\eta_c, \gamma_p, X_{i,c,p}].
\end{equation*}
That is, in the absence of the HEE, the educational attainment of any cohort in provinces more exposed to the HEE would be similar to that of the same cohort in other provinces.

Given the specific urban-rural economy dualism in China and that we want to explore the impact of education opportunities on rural areas, it is essential to disentangle the impacts of HEE on urban people from that on rural people. Thus, we examine how the HEE affects the educational attainment of people with rural \textit{hukou} and those with urban \textit{hukou}, respectively. We use the same regression in Equation \eqref{eq_did} on the sample with only the rural or urban population defined by their \textit{hukou} type.


Besides the direct impact on people's educational attainment, we also dig into the subsequent labor market outcomes related to education. Due to the data availability, we examine two variables in particular: people's earnings (which is only available in the 2005 population census) and the industries in which people were working in 2010.

Since the outcome variables are binary, we also report the logistic specification in addition to the linear regression in Equation \eqref{eq_did} as a robustness check.

\subsection{The Impact on Rural Villages}\label{sec_EmpiricalStrategy_didvillage}

After examining the HEE's impact on people's educational attainment, we focus on how the HEE affects rural villages. In line with the brain gain story, the HEE opened up more opportunities for rural people to acquire higher education, accelerating human capital accumulation and spurring the economic growth of those rural areas. 

However, the HEE could also hinder the development of rural areas due to the brain drain impact of education opportunities. Taking those higher education opportunities enables individuals to acquire skills, be competitive in urban labor markets, and earn higher salaries, attracting them to move into urban areas, thus reducing the human capital in rural areas. This phenomenon is pervasive in many developing countries where there are rural-urban migration barriers, and to some extent, skills acquired from education could alleviate the barriers. For example, high living costs in urban areas would impede rural-urban migration, but high salaries induced by high skills would lead to relatively lower living costs, thus facilitating rural-urban migration.\footnote{China's \textit{hukou} system places fewer migration restrictions on well-educated people, allowing them to migrate into urban areas and live there permanently (\cite{Zhao1997-sw}; \cite{meng2012labor}; \cite{ngai2019china}).} Therefore, whether the HEE helps or hinders rural areas' development is ambiguous and remains as an empirical question to be answered.

To answer this question and get causal estimates of the HEE's impacts on rural villages, we utilize the HEE magnitude variations across provinces and the panel structure of the village-level data from NFPS. Specifically, we estimate the following difference-in-differences equation, 
\begin{equation}\label{eq_did_village}
y_{v, p, t}=\alpha + \beta HighDiff_p \times Post_{t} + X_{v, p,1998} \times Post_{t} \Gamma + \delta_{t} + \theta_{v} + \varepsilon_{v, p, t}.
\end{equation}
Here $y_{v,p,t}$ represents the outcomes of interest for village $v$ in province $p$ in year $t$. $HighDiff_{p}$ is the indicator equal to one if province $p$ is above the median of the $Diff$ measure we defined for the HEE magnitudes in Equation \eqref{eq_treatment_def}, and zero otherwise. $Post_{t}$ is the time indicator that equals one when $t$ is later than the year 1998 and zero otherwise. $\delta_{t}$ are a set of year fixed effects to adjust for any national policy that affects all the villages uniformly in any specific year. $\theta_{v}$ are a set of village fixed effects to control for those underlying village gaps such as geographical location and weather, which are time-invariant but can affect outcomes of interest. We cluster standard errors at the province level to allow for arbitrary dependence of the error term $\varepsilon_{v,p,t}$ across time within each province.

We include province and village characteristics in 1998 in $X_{v, p, 1998}$, which are then interacted with the time indicator $Post_{t}$ to allow for their differential impacts after 1999. All the province characteristics are the same as those in Equation \eqref{eq_did} (and Table \ref{tab_balancetest}) to account for the results of balance test in Section \ref{sec_EmpiricalStrategy_balancetest} and the impact caused by the inherent provincial gaps in economics status and industrial structures. For village-level characteristics, we exploit the indicator from the survey that measures the economic status of each village: whether the village is an impoverished village according to national standards. In those poor villages, it might be harder for people to take those education opportunities due to a lack of educational resources. We will also implement heterogeneity analysis in this regard.

\cite{benjamin2005evolution} mentioned that the sampling in NFPS is not proportional to the provincial population. An example is that the number of villages surveyed in Sichuan and Gansu is around 15 and 9, respectively, but Sichuan has a rural population that is nearly ﬁve times larger.\footnote{There is a more extreme example. The number of surveyed villages is 5 in Yunnan and 11 in Liaoning, but the rural population is very close in these two provinces.} Therefore, we use provincial rural population (by year) to weight our regression where
\begin{equation}
    \text{weight} = \frac{\text{Provincial rural population}}{\text{Number of villages sampled in that province}} \times \text{Village population size}.
\end{equation}
That is, we allocate more weight to a village with larger population in a province with more rural people.

Our identification for the HEE's causal impact on rural villages relies on the assumption that conditional on observed provincial characteristics, unobserved time-invariant village gaps ($\theta_v$) and unobserved uniform shock to all villages ($\delta_t$), the HEE magnitudes are randomly assigned across provinces. 
That is, in the absence of the HEE, outcomes of villages in provinces with higher HEE magnitudes would evolve similarly to those in other provinces, which means they have parallel trends.

If other factors caused villages in provinces with higher HEE magnitudes to show different patterns from those in provinces less exposed to HEE, our estimates would be biased. For example, more roads to rural villages could help rural people move out of agriculture \citep{asher2020rural} and lead to more educational investment \citep{adukia2020educational}, then our results of how the HEE impacts agricultural activities and education investment in rural communities would be confounded. To validate the parallel trends across villages and rule out possible confounding factors, we implement an event study based on the specification in Equation \eqref{eq_did_village}
\begin{equation}\label{eq_eventstudy}
     y_{v,p,t} = \alpha + \sum_{k = 1995, k \neq 1998}^{2008}\beta_{k}  HighDiff_{p} \times \mathbf{1}[t = k] + X_{v,p,t} \Gamma +\delta_t + \eta_v + \varepsilon_{v,p,t},
\end{equation}
where $\mathbf{1}[t = k]$ is the year indicator. We use the year 1998 as the reference category (the year right before HEE). All control variables included in $X_{v,p,t}$ are time-varying. Standard errors are clustered at the province level.

A set of coefficients $\beta_k$'s tests the outcome differences between villages in provinces with higher and lower HEE magnitudes from 1995 to 2008 (relative to the difference in 1998). If the parallel trends assumption holds, all the $\beta_k$'s for years before 1999 should be statistically insignificant, indicating similar trends for those villages without the HEE. We can also examine the dynamic effect of the HEE on rural villages by looking at the estimates $\beta_k$'s with $k$ larger than 1998.

Other caveats to our identification are similar to what we have discussed in Section \ref{sec_EmpiricalStrategy_eduattainment}. All the arguments apply to this part since both identifications rely on the province-level variation of HEE magnitudes.

Our main empirical results utilize the 5-year level-change dummy measure (i.e., $HighDiff$) and a balanced panel including 192 villages in 24 provinces and covering all years from 1995 to 2008. We leave the results from other measures and unbalanced panel in the appendix for robustness checks.

\subsection{The Impact on Rural Households}\label{sec_EmpiricalStrategy_didHH}
Finally, we switch to the household panel data to further disentangle the impact on different households. We do this for two reasons. First, the impact of education opportunities might vary across households since some households may not be able to take those opportunities due to the selectivity of the college entrance examination. Second, probing the household-level impact would help us further distinguish the direct effect of education opportunities from its spillover effect.

Since the household panel has fewer outcomes than the village panel, we will focus on three primary outcomes: household population (and the number of migrants), household labor force (classified by educational attainment), and household income (including remittance). These outcomes are essential to test the impacts of education opportunities on rural development and household welfare. Due to data limitations, we cannot observe why a household member migrated out, but we can check whether more people migrated out in areas more exposed to the HEE and what happened to the welfare of households with migrants after the HEE.

More importantly, by comparing the income of households without migrants in different provinces, we can infer the spillover effect of education opportunities from other households since the individual is not likely to be counted as a rural household member after attending colleges in urban areas. As opposed to the overall impact on rural villages, this approach will uncover the more granular impact of education opportunities on rural areas and meaningful heterogeneities across households.

We use a similar empirical design to identify the impact of education opportunities on rural households,
\begin{equation}\label{eq_did_hh}
    y_{i, p, t}=\alpha + \beta HighDiff_p \times Post_{t} + X_{i, p,1998} \times Post_{t} \Gamma + \delta_{t} + \theta_{i} + \varepsilon_{i, p, t}.
\end{equation}
Here $y_{i,p,t}$ represents the outcomes of interest for household $i$ in province $p$ in year $t$. $HighDiff_{p}$ is the indicator equal to one if province $p$ is above the median of the $Diff$ measure we defined for the HEE magnitudes in Equation \eqref{eq_treatment_def}, and zero otherwise. $Post_{t}$ is the time indicator that equals one when $t$ is later than the year 1998 and zero otherwise. $\delta_{t}$ are a set of year fixed effects to adjust for any national policy that affects all the households uniformly. $\theta_{i}$ are a set of household fixed effects to control for those underlying household gaps such as geographical location and weather, which are time-invariant but can affect outcomes of interest. We cluster standard errors at the province level to allow for arbitrary dependence of the error term $\varepsilon_{v,p,t}$ across time within each province. For robustness check, we report the results using all households in the survey, in addition to the results based on only households in the balanced panel.

\section{Empirical Results}\label{sec_EmpiricalResults}
In this section, we display the empirical results in several steps. Firstly, we present the results of the HEE's impact on people's education attainment and subsequent labor market outcomes related to educational attainment. Then we report the heterogeneous effects of HEE on people with different types of \textit{hukou} status. After that, we examine the impact of HEE on rural villages' population, labor force, agricultural activities, and income and life quality. Finally, we show the results on rural households, exploring the heterogeneous impacts on households with different number of migrants.

\subsection{The Impact on Education Attainment}\label{sec_EmpiricalResults_eduattainment}
We first show the HEE's overall effect on educational attainment and the labor market outcomes related to education. Then we break down to rural people and urban people based on their \textit{hukou} type to explore the heterogeneous effects.

\subsubsection{Baseline}
\begin{figure}[h!]
    \centering
	\subfloat[College and above]{
		\includegraphics[width=.48\textwidth]{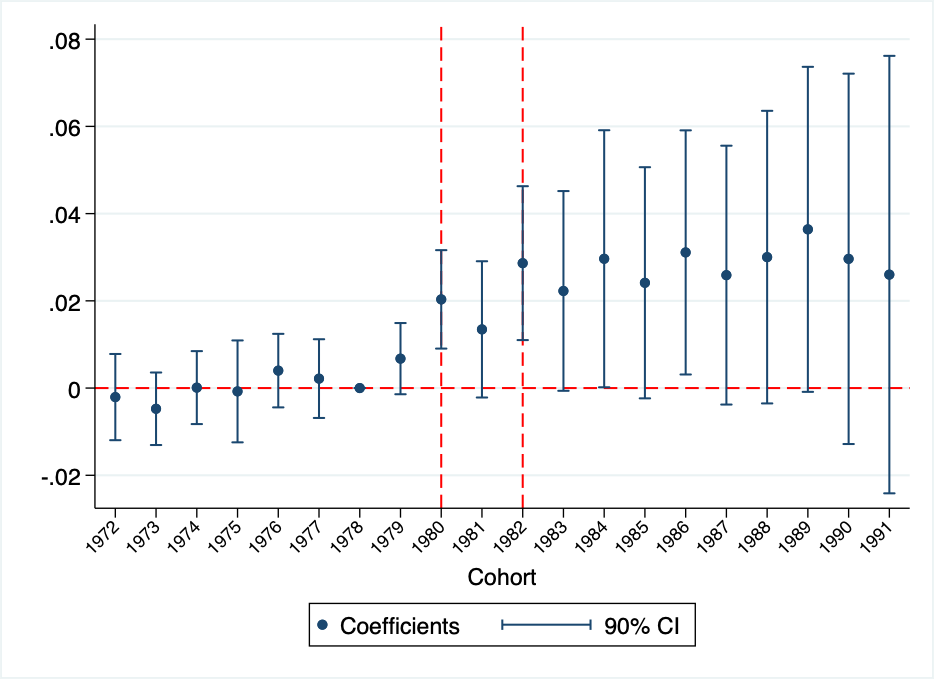}\label{fig_baseline_col}}
	\subfloat[Senior high school and above]{
		\includegraphics[width=.48\textwidth]{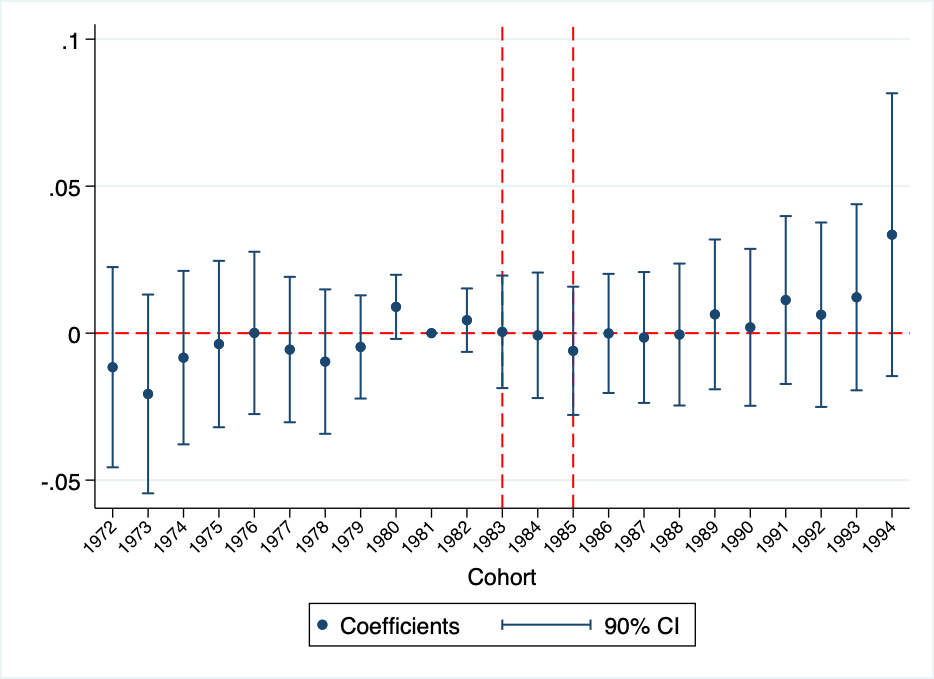}\label{fig_baseline_hs}}
	\caption{The Impact of HEE on Education Attainment}\label{fig_baseline_col_hs}
	\captionsetup{singlelinecheck=false, width = \textwidth}
    \caption*{\footnotesize \textit{Notes}: The estimated coefficients in Equation \eqref{eq_did} and associated 90\% confidence intervals are reported. We use the 5-year averages of the binary $Diff$ measure for the magnitudes of HEE across provinces. The left panel shows the impact on college attendance and the right panel presents the effect on high school enrollment. All the data are from the 2010 population census. Following the discussion in Section \ref{sec_EmpiricalStrategy_eduattainment}, we use cohorts 1978 and 1981 as the reference cohorts for the two outcomes, respectively.}
\end{figure}

We begin by considering the HEE's impact on the educational attainment of all cohorts shown in Table \ref{tab_ageincensus}. Figure \ref{fig_baseline_col_hs} presents the estimated coefficients of Equation \eqref{eq_did} using the 2010 population census. As has been discussed in Section \ref{sec_EmpiricalStrategy_eduattainment}, we use cohorts 1978 and 1981 as the reference cohorts for the two outcomes to account for the fact that people in some cohorts may skip or repeat some grades or start schooling earlier or later. Figure \ref{fig_baseline_col} indicates that before the HEE there is no difference of college attendance between any cohort in more exposed provinces and the same cohort in less exposed provinces. After the HEE, cohorts in provinces more exposed to the HEE are more likely to attend colleges than their peers in provinces less exposed to the HEE, and this effect persists for all the subsequent cohorts. The magnitude of the impact is around 3\%, which is considerable given that only 10\% of the population in China has diplomas from colleges or above in the 2010 population census.

Figure \ref{fig_baseline_hs} presents the impact on people attending senior high school and above. The probability of attending junior high schools and above is not statistically different for cohorts in provinces more exposed to the HEE comparing to cohorts in less exposed provinces. Combined with Figure \ref{fig_baseline_col}, the results imply that more people switched from high schools to colleges in province more exposed to the HEE. That is, the HEE increases local people's education attainment.


The results shown in Figure \ref{fig_baseline_col_hs} are based on the 5-year average of the binary $Diff$ measure defined in Equation \eqref{eq_treatment_def}, we also use other measures to investigate the same outcomes as robustness checks. Figures \ref{fig_baseline_cont_col_hs} changes the binary measures of HEE magnitudes to the continuous ones, the results are similar in sign, though the significance decrease slightly for the impact on college attendance. Figure \ref{fig_baseline_avg10_col_hs} are the results by using the 10-year averages to measure the magnitudes of HEE across provinces. The impact of HEE on college attendance is 1\% stronger than that in Figure \ref{fig_baseline_col_hs} and inference power is also stronger. 

We also control for individual characteristics (ethnicity and gender), but the results barely change. Overall, we find that HEE significantly increased the education attainment of people in provinces that are more exposed to the HEE. The results are robust to different measures of the HEE magnitudes or controlling for provincial or individual characteristics.

\subsubsection{Labor Market Outcomes}

We have found that the HEE induced more people attending colleges, in this part we examine the HEE's impact on labor market outcomes associated to education attainments. Specifically, we investigate two variables: monthly earnings and industry information of jobs. We use the 2005 population census to explore the impact on labor market outcomes because this is the only population census that provides information on individual earnings. The disadvantage of using this census is that we can only observe very a few cohorts who were affected by the HEE and already graduated from colleges (i.e., aged more than 22). For the sector choice of jobs, we use the 2010 population census.

\begin{figure}[h!]
    \centering
	\subfloat[ln(Earnings)]{
		\includegraphics[width=.48\textwidth]{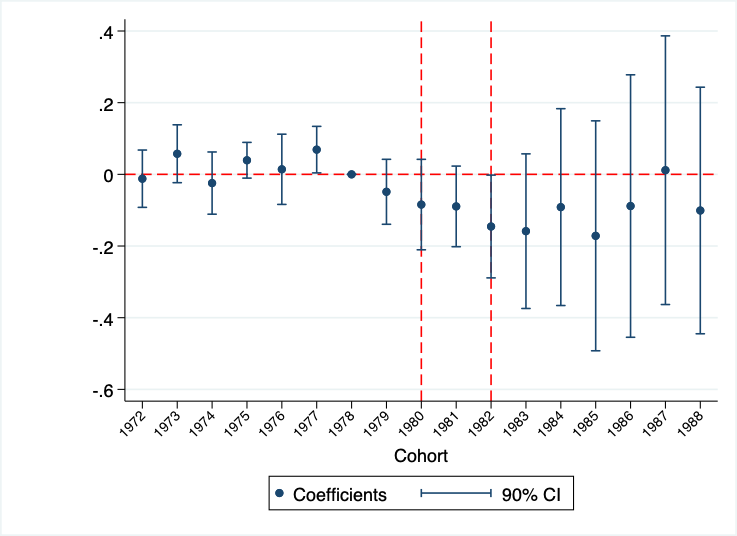}\label{fig_labor_salary}}
	\subfloat[Working in the agricultural sector]{
		\includegraphics[width=.48\textwidth]{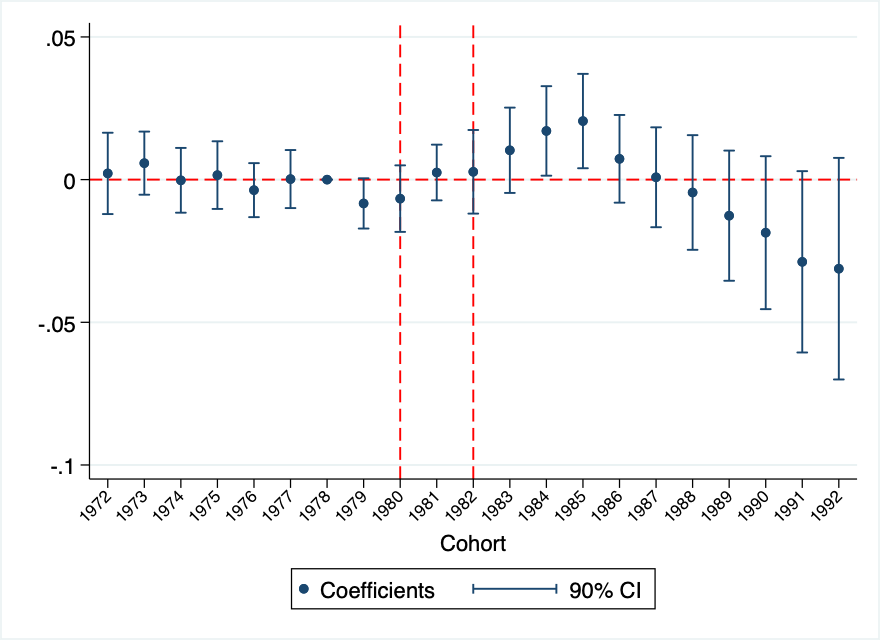}\label{fig_labor_sector}}
	\caption{The Impact of HEE on Labor Market Outcomes}\label{fig_labor}
	\captionsetup{singlelinecheck=false, width = \textwidth}
    \caption*{\footnotesize \textit{Notes}: Figures report estimated coefficients in Equation \eqref{eq_did} and associated 90\% confidence intervals. We use the 5-year averages of the binary $Diff$ measure for the magnitudes of HEE across provinces. The left panel shows the impact on log-earnings and the right panel presents the effect on the probability of working in the agricultural sector. Panel (a) is based on the 2005 population census while Panel (b) uses the 2010 population census. Following the discussion in Section \ref{sec_EmpiricalStrategy_eduattainment}, we use cohorts 1978 as the reference group.}
\end{figure}

Figure \ref{fig_labor_salary} shows that cohorts that were potentially affected by the HEE (i.e., cohorts 1979-1983) on average experienced a drop in their monthly earnings, possibly because more people went to colleges and did not earn much right after their graduation. We do not see significant reduction in wages for cohorts after 1983, partly due to the fact that they had not graduated from colleges in 2005, when they were surveyed.

Figure \ref{fig_labor_sector} indicates that people are more likely to work in the agricultural sector after being affected by the HEE but this pattern changed immediately. Later cohorts in provinces more exposed to the HEE were more prone to work in the non-agricultural sector, and this impact persisted over time.

We also report the results using the continuous measure of the 5-year average level change in HEE magnitudes, as opposite to the binary one in Figure \ref{fig_labor}. The results are reported in Figure \ref{fig_labor_cont_rural}, indicating that people experienced earnings drop in provinces more exposed to the HEE and that more people are working in the non-agricultural sector after being more affected by the HEE. The results are consistent with Figure \ref{fig_labor}.

\subsubsection{Heterogeneous Effects}

As mentioned in Section \ref{sec_Background_ncee}, the national admission pattern diﬀers for people with urban \textit{hukou} and those with rural \textit{hukou}, students in rural areas face more ﬁerce competition to attend colleges due to the larger number of the rural population and relatively scarce education resources in rural areas. 
We then reproduce the results above but separately restrict the sample to include only people with rural \textit{hukou} or urban \textit{hukou}.

\begin{figure}[h!]
    \centering
	\subfloat[College and above]{
		\includegraphics[width=.48\textwidth]{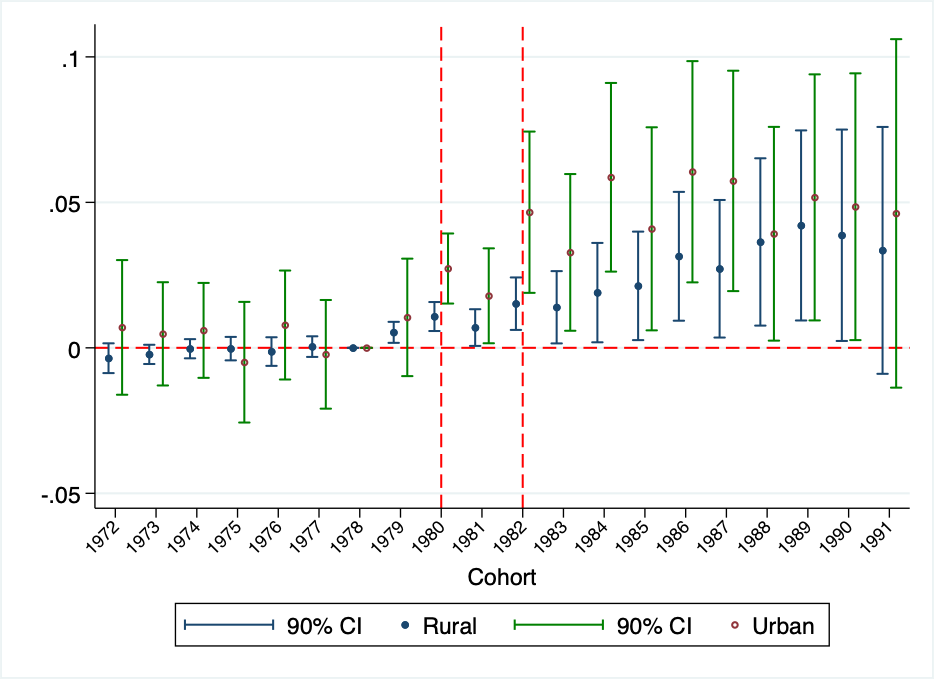}\label{fig_hetero_col}}
	\subfloat[Senior high school and above]{
		\includegraphics[width=.48\textwidth]{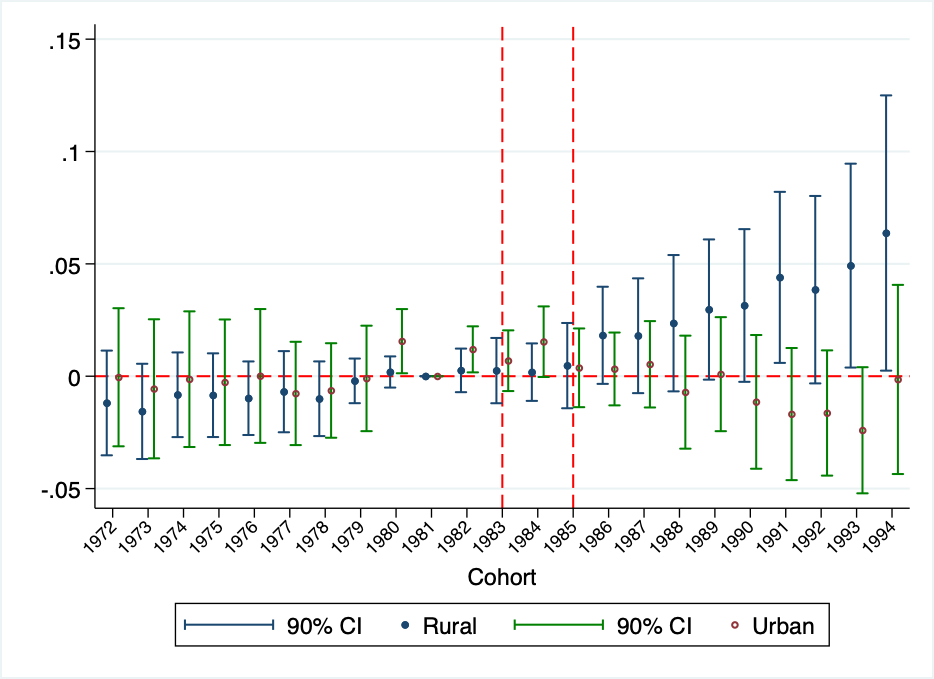}\label{fig_hetero_hs}}
	\caption{The Impact of HEE on Education Attainment}\label{fig_hetero_col_hs}
	\captionsetup{singlelinecheck=false, width = \textwidth}
    \caption*{\footnotesize \textit{Notes}: Figures report estimated coefficients in Equation \eqref{eq_did} and associated 90\% confidence intervals for the sample including only people with rural or urban \textit{hukou}. We use the 5-year averages of the binary $Diff$ measure for the magnitudes of HEE across provinces. The left panel shows the impact on college attendance and the right panel presents the effect on high school enrollment. All the data are from the 2010 population census. Following the discussion in Section \ref{sec_EmpiricalStrategy_eduattainment}, we use cohorts 1978 and 1981 as the reference cohorts for the two outcomes, respectively.}
\end{figure}

Figure \ref{fig_hetero_col} depicts how differently the HEE impacts the educational attainment of rural and urban people. Compared to earlier cohorts, later cohorts with rural \textit{hukou} in provinces more exposed to the HEE experienced around 3\% more college enrollment than their peers in other provinces. This number is larger for people with urban \textit{hukou} at around 5\%, indicating the education disparity between rural and urban areas in China. Nonetheless, the impact on rural people is very substantial. In the 2000 population census, just below 1\% of the rural population have college degrees or above, about 15\% of the urban population have attended colleges or above.\footnote{In the 2010 population census, the two numbers are 4\% and 30\%, respectively.} Our results indicate that the HEE almost tripled the probability of attending colleges for rural people and that the HEE's impact on urban people's educational attainment in colleges and above is around 30\%.

Figure \ref{fig_hetero_hs} demonstrates that the HEE has no significant impact on urban people's education attainment in senior high schools and above, but it leads more rural people to attend senior high schools. On average, the HEE induced around 3\% more rural people to attend senior high schools and colleges. Results in the two figures imply that HEE has both direct impact on rural people's college attendance and positive spillover effect on rural people's education attainment in high schools. 

\begin{figure}[h!]
    \centering
	\subfloat[ln(Income)]{
		\includegraphics[width=.48\textwidth]{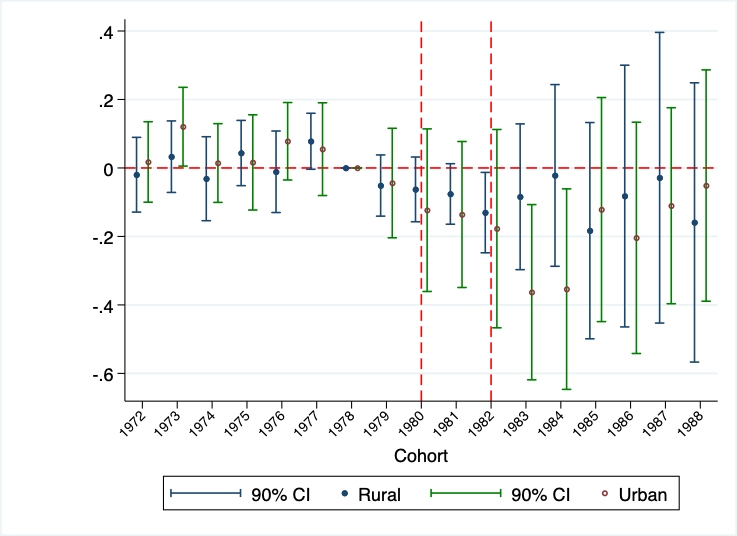}\label{fig_labor_salary_rural}}
	\subfloat[Working in the agricultural sector]{
		\includegraphics[width=.48\textwidth]{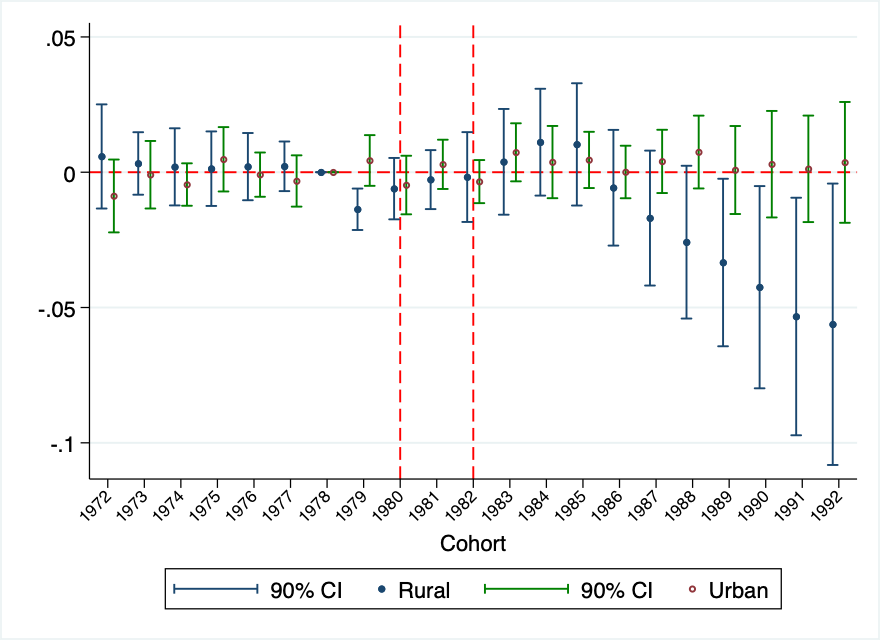}\label{fig_labor_sector_rural}}
	\caption{The Impact of HEE on Labor Market Outcomes}\label{fig_labor_rural}
	\captionsetup{singlelinecheck=false, width = \textwidth}
    \caption*{\footnotesize \textit{Notes}: Figures report estimated coefficients in Equation \eqref{eq_did} and associated 90\% confidence intervals for the sample including only people with rural or urban \textit{hukou}. We use the 5-year averages of the binary $Diff$ measure for the magnitudes of HEE across provinces. The left panel shows the impact on log-earnings and the right panel presents the effect on the probability of working in the agricultural sector. All the data are from the 2010 population census. Following the discussion in Section \ref{sec_EmpiricalStrategy_eduattainment}, we use cohorts 1978 as the reference group.}
\end{figure}

Figure \ref{fig_labor_salary_rural} reveals that rural people in provinces more exposed to the HEE on average experienced a more reduction in monthly earnings, possibly because of attending schools and being absent from work. The HEE's impact on earnings is similar for urban people, but the magnitudes are smaller (20\% relative to 30\% of rural people).

More importantly, Figure \ref{fig_labor_sector_rural} indicates that the HEE has no impact on urban people's working choices on the agricultural or non-agricultural sector. However, the HEE induces around 5\% more people with rural \textit{hukou} moved out of the agricultural sector in provinces more exposed to the HEE. This impact is sizable given that the fraction of rural population working in the agricultural sector is 48\% in the 2000 population.

We also change our measures on the HEE magnitudes and report the results in Figures \ref{fig_baseline_cont_col_hs_rural} and \ref{fig_baseline_avg10_col_hs_rural}. We have similar findings with controlling for provincial characteristics specified in Table \ref{tab_balancetest}. They are consistent with the results in Figures \ref{fig_hetero_col_hs} and \ref{fig_labor_rural} under different specifications, indicating the robustness of our findings.

Overall, the results reveal that the HEE leads to significant increases in the attainment of higher education for both rural and urban people, and encourages the attendance in senior high schools for people in rural areas. The HEE also caused a short-term reduction in earnings due to college attendance. Importantly, the HEE causes salient structural transformation in the sense that more rural people entered the non-agricultural sector in provinces more exposed to the HEE than their rural peers in other provinces. All the results provide evidence that the HEE significantly impacts people's education attainment in rural areas and has pronounced implications on the earnings and working choices of people from rural areas.

\subsection{The Impact on Rural Villages}

The results in Section \ref{sec_EmpiricalResults_eduattainment} indicate that the HEE leads more education attainment of rural people, but this could either help or hinder the development of rural villages. On the one hand, the HEE can contribute to the brain gain for the rural area since more educated population creates more human capital, which enables faster economic development. On the other hand, well-educated people may move out of rural areas to get more earnings and enjoy better urban amenities, which is particularly significant in China's context due to the migration barrier from the \textit{hukou} policy. Then the HEE might drive those educated people out of rural areas, causing a brain drain and inhibiting the economic development in rural villages.

To examine the HEE's effect on rural villages, we first categorize all village-level outcomes in the NFPS data into four groups: population, labor force, agricultural activities, and income and life quality. We employ Equation \eqref{eq_did_village} to identify the overall impact of the HEE, and use Equation \eqref{eq_eventstudy} to explore the dynamic effects and evaluate the parallel trends assumption. We present the results using the balanced village panel spanning from 1995 to 2008, including 192 villages in 24 provinces. 

\subsubsection{Population}\label{sec_vpanel_pop}

We start with the impact of HEE on population and the number of households in rural villages, as shown in Table \ref{tab_v_reg_population}. The population is defined by the number of residents whose \textit{hukou} are registered in the village. The \textit{hukou} system records individuals within one family together as a household. We use end-of-year population and the number of households to approximate migration flows, as a way to tackle the impact of education on migration.\footnote{A more direct way to examine this out-migration effect is to see the change in the number of people moving out of villages. The questionnaire of NFPS asks ``the number of people moving out of their \textit{hukou} village within the year" and ``the number of people who moved into their \textit{hukou} village within the year". However, there are a large number of missing records for these variables. For example, half of the observations are missing for the number of people who moved out of villages. Thus, we use the number of people at the end of each year as a substitution.}

All the dependent variables are in the form of inverse hyperbolic sine to deal with the incompatibility issue of log-transformation and zero value of outcomes. We report the results based on the difference-based measurement of HEE magnitudes as defined in Equation \eqref{eq_treatment_def}. The results in Table \ref{tab_v_reg_population} show that the HEE induced an out-migration in rural villages. Villages in provinces with higher magnitudes of HEE experienced a higher fraction of out-migration due to the fact that it is easier for well-educated rural people to find jobs in cities and live there permanently. Columns (2) and (4) show that population and the number of households of villages decreases by around 1.1\% and 1.6\% in provinces that experienced one unit average increase in the magnitude of HEE (i.e., the number of admitted students per capita increased by 1\%). Columns (6) and (8) indicate that villages in provinces with larger HEE magnitudes (i.e., above the median across all provinces) tended to have around 9\% population and 15\% households fewer than villages in other provinces. To eliminate the concern that the impact on population and households are driven by the HEE through migration rather than people's birth or death, we also check the impact of HEE on the birth and death rates across villages and find negligible effects. Thus, the results confirm that HEE leads to more people moving out of rural villages. 
As a robustness check, in Table \ref{tab_app_v_reg_population} we report the results with the ratio-based treatment, and the effects are pretty consistent.

\begin{table}[h!]
\centering
\caption{The Effect of HEE on the Population in Rural Villages}\label{tab_v_reg_population}
\scalebox{0.9}{
\begin{tabular}{lcccccccc}
\hline \hline
 & (1) & (2) & (3) & (4) & (5) & (6) & (7) & (8) \\
 & Pop. & Pop. & HHs & HHs & Pop. & Pop. & HHs & HHs \\
\hline
$HighDiff \times Post$ & -0.00546* & -0.0114* & -0.00341 & -0.0164** & -0.0652* & -0.0981 & -0.0638 & -0.149** \\
 & (0.00307) & (0.00572) & (0.00304) & (0.00648) & (0.0359) & (0.0678) & (0.0398) & (0.0684) \\
 \hline
Observations & 2,656 & 2,656 & 2,657 & 2,657 & 2,656 & 2,656 & 2,657 & 2,657 \\
Number of villages & 192 & 192 & 192 & 192 & 192 & 192 & 192 & 192 \\
Treatment & Cont. & Cont. & Cont. & Cont. & Binary & Binary & Binary & Binary \\
Provincial Controls &  & Y &  & Y &  & Y &  & Y \\ \hline \hline
\end{tabular}
}
\captionsetup{singlelinecheck=false, width = 0.9\textwidth}
\caption*{\footnotesize \textit{Notes}: All the dependent variables are transformed by inverse hyperbolic sine. The continuous treatment is defined by Equation \eqref{eq_treatment_def}  The binary treatment defined by if the province has a continuous HEE measure above the medium. Provincial controls include the GDP share of the agricultural sector, ln(Population), ln(GDP per capita), ln(total employment), the employment shares in the agricultural sector and the manufacturing sector, and the share of rural employment in each province in 1998. All the provincial controls are interacted with $Post_{98}$. Standard errors in parentheses are clustered at the province level, *, **, and *** denote significance at the 10\%, 5\% and 1\% levels, respectively.}
\end{table}

\begin{figure}[h!]
    \centering
	\subfloat[ln(Population)]{
		\includegraphics[width=.48\textwidth]{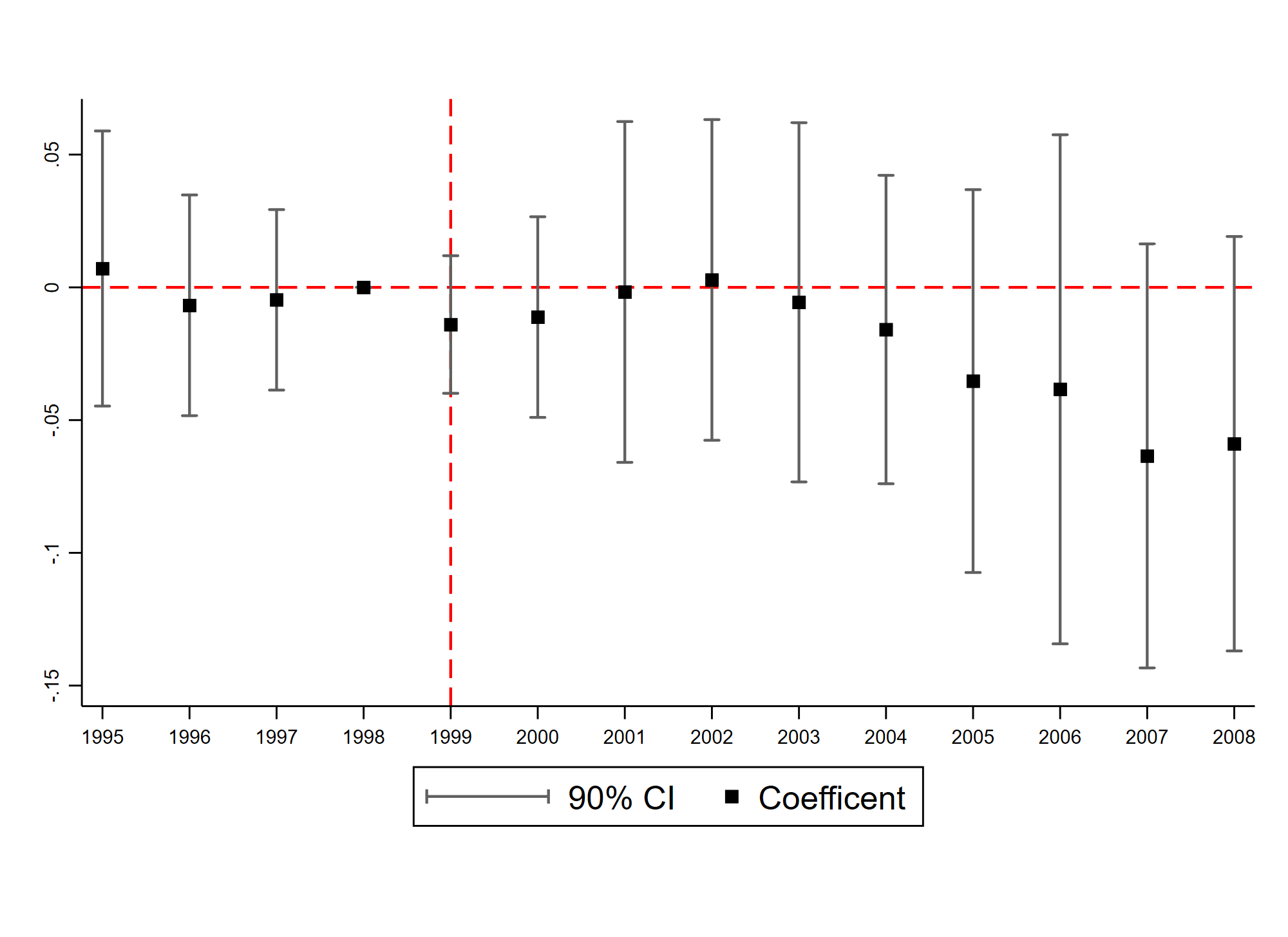}\label{fig_v_p14_pop_yrend}}
	\subfloat[ln(Number of households)]{
		\includegraphics[width=.48\textwidth]{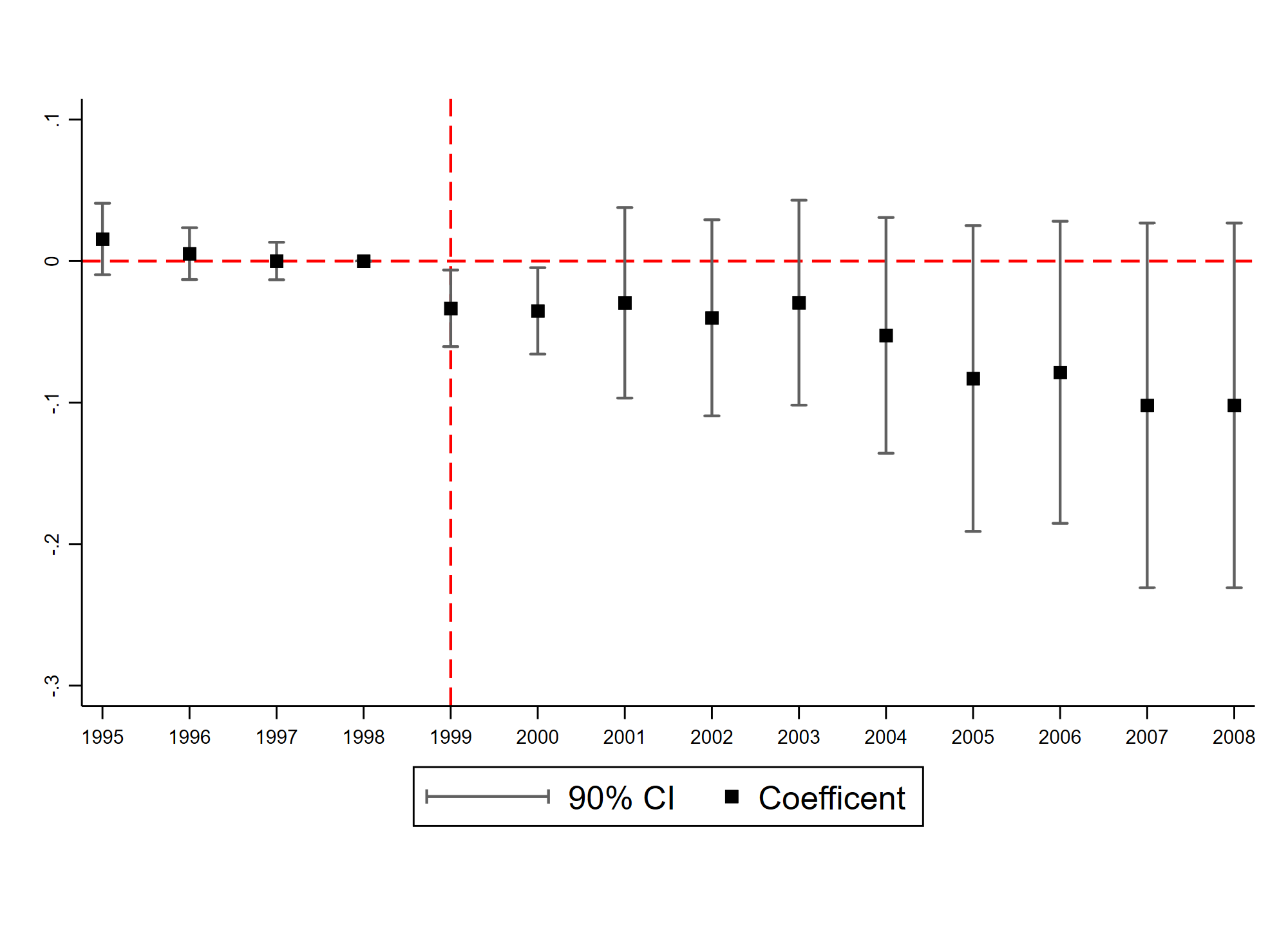}\label{fig_v_p14_hhnum}}
	\caption{The Dynamic Effects of HEE on the Population in Rural Villages}\label{fig_v_p14_population}
	\captionsetup{singlelinecheck=false, width = 0.9\textwidth}
    \caption*{\footnotesize \textit{Notes}: Both figures show the coefficients of the interaction term of between $HighDiff$ and year indicators as presented in Equation \eqref{eq_eventstudy}. The results include 24 provinces as mentioned in Section \ref{sec_EmpiricalStrategy_treatment}. The year 1998 serves as the reference category, which is omitted in the regression. The vertical line indicates the start year of HEE.}
\end{figure}

To see the dynamic effects of HEE on population and households, we conducted an event study as specified in Equation \eqref{eq_eventstudy}. The results are shown in Figure \ref{fig_v_p14_population}. Both figures indicate that before 1999 there was no difference in the end-of-year population and the number of households between villages in provinces with higher magnitudes of HEE and those in provinces with lower magnitudes of HEE, which validates the parallel trends assumption. After 1999, both population and the number of households decreased, which is consistent with the out-migration shown in Table \ref{tab_v_reg_population}, though the magnitudes are not statistically different from zero. The downward trending in the two figures indicates that the out-migration impacts on population and households are persistent. 

\subsubsection{Education Attainment of the ``Left-behind" Labor Force}\label{sec_vpanel_labor}
From the census data, we have learned that the HEE caused rural people to get better education attainment and find jobs in the non-agricultural sector. Given the HEE's effect on the out-migration as shown in Table \ref{tab_v_reg_population}, we then dig into the education attainment of people who stayed behind. It is likely that the human capital increment led by the HEE was not preserved in the rural area due to migration. The NFPS data record the education attainment of any labor force who has stayed in the village for more than six months in the year, which provides us with the opportunity to discuss the HEE's effect on the human capital accumulation in the rural areas and on the outcomes of people who stayed behind.

The outcomes in our regressions are the number of laborers who have attained senior high school degrees or above, junior high school degrees or above, primary school degree or above, and the number of laborers without any education, respectively. To address the concern that the decrease in the population could mechanically drive the results on the number of educated laborers, we then examine the HEE's effect on the total number of remaining labor in the village and add village population as additional controls.

Table \ref{tab_v_reg_laborforce} presents the impact of HEE on the labor force in rural areas. Column (1) examines the change in the total labor force in rural villages due to a higher magnitude of HEE. It slightly affects the total labor force, and the negative impact is not significant from zero. These results indicate that the total labor force did not decrease in those villages experiencing a higher magnitude of HEE, although there were out-migrations. The insignificance in total remaining labor is because only school-aged people can be affected by HEE, but NFPS does not include school-aged children or students as part of the labor force.\footnote{As mentioned in Section \ref{sec_Data}, the labor force in NFPS includes all rural residents between 15 and 64 years of age who have lived in villages for more than six months in a year and are not employed outside the village, except for students, military personnel, and the disabled.}

\begin{table}[h!]
\centering
\caption{The Impact on the "Left-behind" Labor Force}\label{tab_v_reg_laborforce}
\scalebox{0.8}{
\begin{tabular}{lcccccc}
\hline \hline
 & (1) & (2) & (3) & (4) & (5) & (6) \\ 
 & \multirow{2}{*}{Total} & \multicolumn{5}{c}{By education attainment} \\ \cline{3-7}
 &  & senior sch. & junior sch. & primary sch. & illiteracy & \% senior sch.\\
\hline
\multicolumn{7}{l}{\textit{\textbf{Panel A: No controls}}} \\
$HighDiff \times Post$ & -0.00888 & -0.0642*** & -0.0220** & -0.00458 & -0.0425* & -0.00374** \\
 & (0.00598) & (0.0103) & (0.0104) & (0.00596) & (0.021) & (0.00166) \\
Observations & 2440 & 2454 & 2457 & 2451 & 2154 & 2427 \\
Number of villages & 191 & 192 & 192 & 192 & 192 & 192 \\ \hline
\multicolumn{7}{l}{\textit{\textbf{Panel B: Control for provincial characteristics}}} \\
$HighDiff \times Post$ & -0.00373 & -0.0688*** & -0.00951 & -0.00493 & -0.0256 & -0.00655*** \\
 & (0.00701 & (0.0194) & (0.0149) & (0.00823) & (0.022) & (0.00189) \\
Observations & 2440 & 2454 & 2457 & 2451 & 2154 & 2427 \\
Number of villages & 191 & 192 & 192 & 192 & 192 & 192 \\ \hline
\multicolumn{7}{l}{\textit{\textbf{Panel C: Control for provincial characteristics and village population}}} \\
$HighDiff \times Post$ & 0.00741 & -0.0557*** & 0.00256 & 0.0062 & -0.0152 & -0.00634*** \\
 & (0.0077 & (0.0141) & (0.0168) & (0.00956) & (0.0203) & (0.00183) \\
Observations & 2439 & 2453 & 2456 & 2450 & 2153 & 2426 \\
Number of villages & 191 & 192 & 192 & 192 & 192 & 192 \\
\hline \hline
\end{tabular}
}
\captionsetup{singlelinecheck=false}
\caption*{\footnotesize \textit{Notes}: Except for the percentage of senior high school or above degree holders among left-behind rural labors in Column (6), all the dependent variables are transformed by inverse hyperbolic sine. Except for illiteracy, the education achievement is defined by the number of labors who get the degree and above. E.g., \textit{senior sch.} refers to the number of labors with senior high school degree and above. $HighDiff$ is defined by Equation \eqref{eq_treatment_def}. Provincial controls include the GDP share of the agricultural sector, ln(Population), ln(GDP per capita), ln(total employment), the employment shares in the agricultural sector and the manufacturing sector, and the share of rural employment in each province in 1998. All the provincial controls are interacted with $Post_{98}$. Village population is the logged value in year $t$. Standard errors in parentheses are clustered at the province level, *, **, and *** denote significance at the 10\%, 5\% and 1\% levels, respectively.}
\end{table}

However, the educational attainment of the left-behind labor force may change because of HEE, although the total labor force remained the same. Columns (2)-(4) report the HEE's impact on the logged number of rural remaining labors earning senior high school degrees or above, junior high school degrees or above, and primary degrees or above. For completeness, we also report the impact on the logged number of illiteracy among remaining rural laborers in Column (5). 

In Column (2), the coefficients range from -5.8\% to -6.8\%, with significance levels at 99\% under all three specifications. Column (2) in Panel C shows that, after controlling for the number of people with their \textit{hukou} registering at the village, HEE still has a significant negative influence on the stock of the senior high school educated labor. In other words, though some well-educated rural workers do not change their \textit{hukou}, they still manage to work outside their rural hometowns. The comparison between Panel B and Panel C presents that the population size can explain about 19\% of the HEE's impact on the stock of senior high school degree holders in the village's labor force. It indicates that some of the well-educated people move their \textit{hukou} successfully. 

Together with the result in Column (1), we have presented that with a higher level of HEE, the village tends to lose more well-educated laborers. At the same time, there is little impact on the total remaining rural labor. The difference is because the absolute number of senior high school or even collage educated left-behind rural laborers are relatively small compared to the total amount. As Table \ref{tab_v_su} shows, only about 10\% of the total have received this level of education. This result is consistent with the estimation in Columns (3) and (4) -- there is no significant impact if we expand the education level to junior high school or primary school. Both are required by the compulsory education law and account for a large proportion of the remaining laborers.\footnote{For example, in Table \ref{tab_v_su}, junior school degree holders take up about half of the total on average.}  Since these levels of education are mandatory, HEE would not change them dramatically. In conclusion, the HEE particularly influences the stock of the most educated labor in rural areas. Besides, to more explicitly show that HEE would change the education structure among the remaining rural laborers, we also report its effect on the percentage of at least senior high school degree holders in Column (6). After controlling for provincial level variables in 1998 (or controlling for the population additionally), the proportion of senior high school educated ones reduces by 0.63\% - 0.65\% among the village's labor stock due to 1 unit increase in HEE. The average level of this percentage is 9.6\%, among the villages locating in provinces that have HEE magnitude in the bottom half. Thus, it is clear that the HEE would extract particularly more well-educated laborers.

Our results in Section \ref{sec_EmpiricalResults_eduattainment} indicates the positive effect of education opportunities on school attendance, which means that, due to HEE, the education attainment increases among young people holding rural \textit{hukou}. Nevertheless, in the section, from the perspective of the rural village, those people did not stay in rural areas after graduation. Though more at least senior high school degree holders appear, less stock of them the villages would have. Altogether, HEE leads to more education opportunities and improves the educational attainment of the labor force in rural areas. However, at the same time, those well-educated people are not preserved in rural areas. HEE accelerates the human capital accumulation to boost the economic development in China and extract more well-trained men from rural areas.

\begin{figure}[h!]
    \centering
	\subfloat[ln(Senior school educated labor)]{
		\includegraphics[width=.48\textwidth]{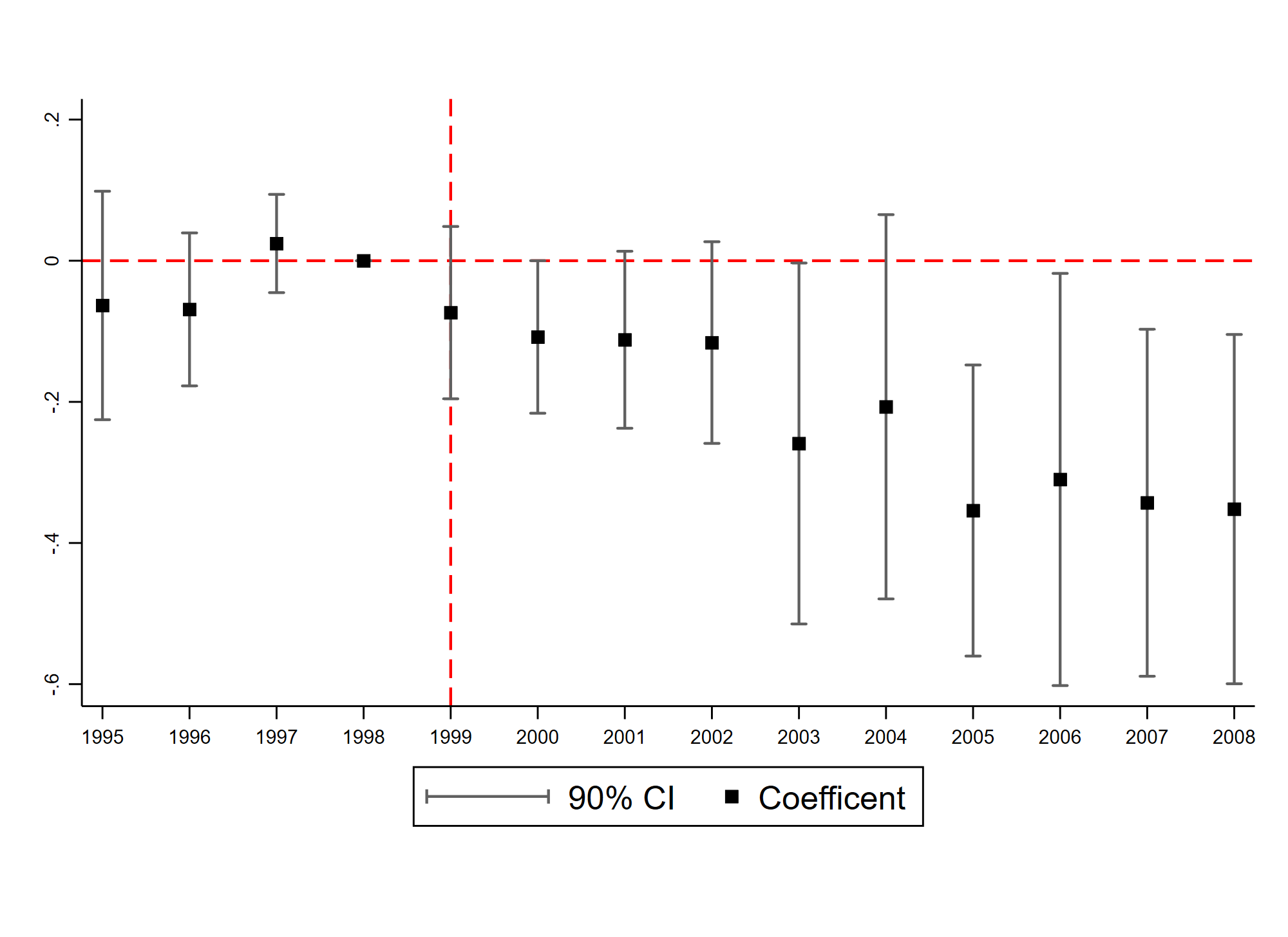}\label{fig_v_p14_labor_hsch}}
	\subfloat[\% Senior school educated labor]{
		\includegraphics[width=.48\textwidth]{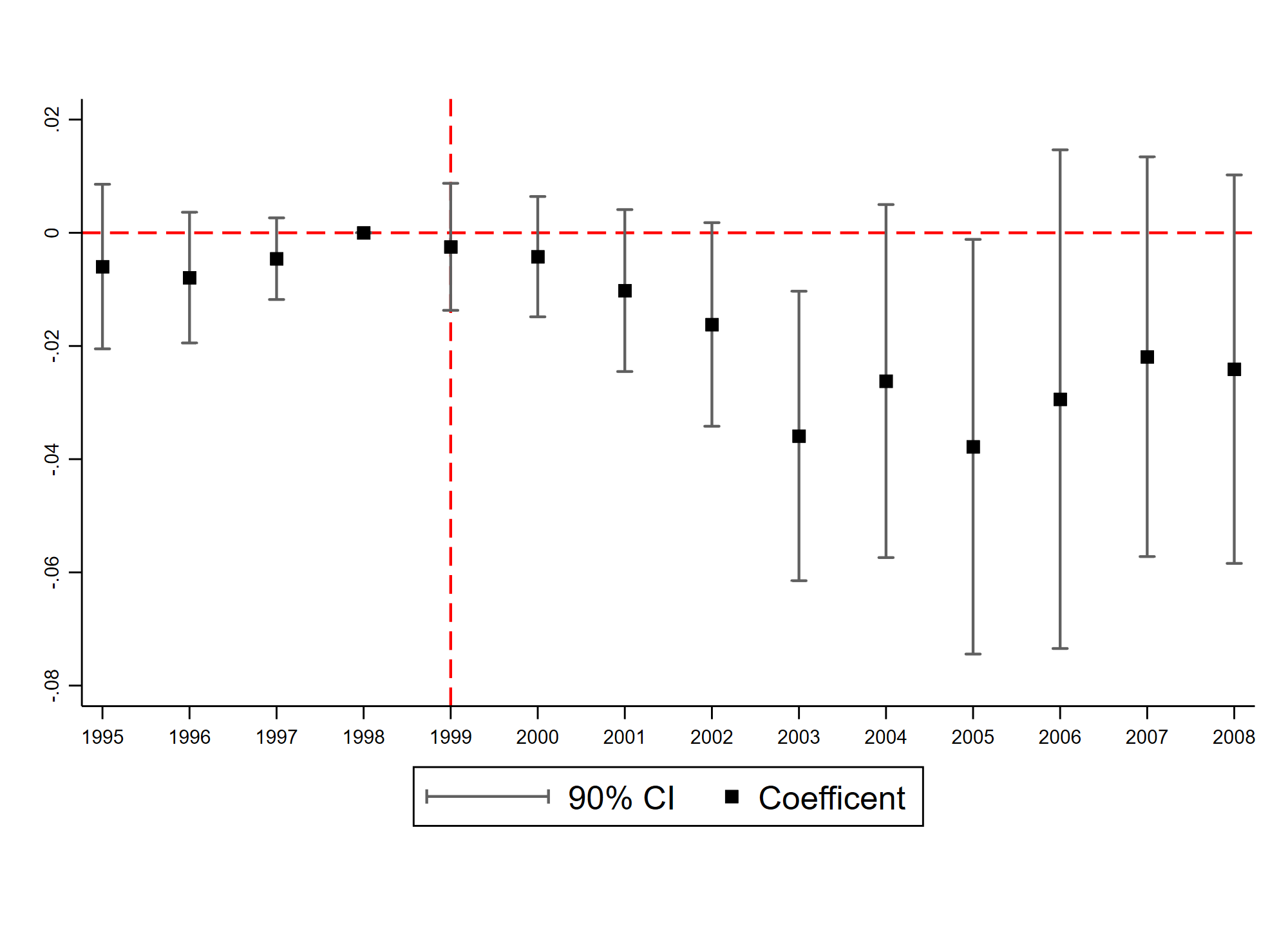}\label{fig_v_p14_labor_nonagri}}
	\caption{The Dynamic Effects of HEE on the Labor Force in Rural Villages}\label{fig_v_p14_labor}
	\captionsetup{singlelinecheck=false, width = 0.9\textwidth}
    \caption*{\footnotesize \textit{Notes}: Both figures show the coefficients of the interaction term of between $HighDiff$ and year indicators as presented in Equation \eqref{eq_eventstudy}. The results include all 24 provinces. The year 1998 serves as the reference category, which is omitted in the regression. The vertical line indicates the start year of HEE.}
\end{figure}

To see the dynamic effects of HEE on the education level of left-behind laborers, we conducted the event study as presented by Equation \eqref{eq_eventstudy}. The results are shown in Figure \ref{fig_v_p14_labor}.  Both figures demonstrate that the differences of both the number and percentage of senior school educated rural laborers are not significant between villages in provinces with higher magnitude of HEE and those in provinces with lower magnitude of HEE before 1999, which verifies the validity of the parallel trends assumption. After 1999, all the coefficients are negative and significantly different from zero at the 90\% level, which is consistent with the selection effect shown in Table \ref{tab_v_reg_laborforce}. 

\subsubsection{Agricultural Activities}\label{sec_vpanel_agract}

Sections \ref{sec_vpanel_pop} and \ref{sec_vpanel_labor} reveal that HEE causes the out-migration of people in rural villages, extracts the labor force with better education attainment. Nevertheless, it is still unclear how HEE would influence the productivity in the rural area and the life quality of remaining rural residents.

In this section, we start with agricultural activity, the main business in the rural villages. We will use the area of cropped land to measure the effect of HEE on agricultural activities. We use this measure to approximate it rather than the output of agricultural products (e.g., corn, cotton) because the output could be affected by the change of geographical conditions and weather, which are unobserved to us. Besides, we also attach the HEE's impact on the participation in agriculture among the left-behind laborers in the villages.

Table \ref{tab_v_reg_agriactivities} reports the results of HEE's impact on the agricultural activities. Here we include the area of cropped land in the sense of total cropped land, grain-crop land, and cash-crop land in Column (1) - (3). Columns (4) and (5) display the impact on the percentage and the logarithmic number of agricultural laborers. To address the concern that the population size could mechanically drive the outcomes, besides the provincial level controls in their 1998 level, we also control the village population in year $t$. The results indicate that the HEE would not influence agricultural activities in rural areas. We observe insignificant effects the HEE has on both the area of cropped land and participants in the agricultural production in all panels.

The results in the area of cropped land and the labor are consistent with each other. From the last section discussing the HEE's impact on the stock of labor across different education levels, we find it has little effect on the stock of relatively low-level educated labor. Together with the results in this section, the picture is more explicit. The HEE encourages more people in rural areas to gain a higher level of education (senior high school degree or above) and helps these people move or work outside. However, since the well-educated people/laborers are still a small part of the rural population/laborers, the HEE does not influence rural places' fundamental business and labor participation.


\begin{table}[h!]
\centering
\caption{The Effect of HEE on the Agricultural Activities in Rural Communities}\label{tab_v_reg_agriactivities}
\begin{tabular}{lccccc}
\hline \hline
 & (1) & (2) & (3) & (4) & (5) \\
 & \multicolumn{3}{c}{ln(Area of cropped land)} & \multirow{2}{*}{\% Agri. Labor} & \multirow{2}{*}{ln(Agri. Labor)} \\ \cline{2-4}
 & total & grain crop & cash crop &  &  \\
 \hline
\multicolumn{5}{l}{\textit{\textbf{Panel A: No controls}}} & \multicolumn{1}{l}{} \\
$HighDiff \times Post$ & -0.00944 & -0.00329 & 0.00458 & -0.00102 & -0.0111 \\
 & (0.00851) & (0.00684) & (0.0141) & (0.00336) & (0.0115) \\
Observations & 2645 & 2620 & 2431 & 2435 & 2435 \\
Number of villages & 191 & 190 & 192 & 192 & 192 \\ \hline
\multicolumn{6}{l}{\textit{\textbf{Panel B: Control for provincial characteristics}}}  \\
$HighDiff \times Post$ & -0.00907 & -0.00297 & 0.00329 & -0.00688 & -0.0193 \\
 & (0.00855) & (0.00607) & (0.0149) & (0.00515) & (0.0128) \\
Observations & 2645 & 2620 & 2431 & 2435 & 2435 \\
Number of villages & 191 & 190 & 192 & 192 & 192 \\ \hline
\multicolumn{6}{l}{\textit{\textbf{Panel C: Control for provincial characteristics and village population}}} \\
$HighDiff \times Post$ & -0.00705 & -0.000146 & 0.00819 & -0.00738 & -0.01 \\
 & (0.00841) & (0.00652) & (0.0144) & (0.00502) & (0.0148) \\
Observations & 2643 & 2619 & 2431 & 2434 & 2434 \\
Number of villages & 191 & 190 & 192 & 192 & 192 \\
\hline \hline
\end{tabular}
\captionsetup{singlelinecheck=false, width = 0.9\textwidth}
\caption*{\footnotesize \textit{Notes}: Column (4) and (5) are still restricted in the village's remaining labor force. Except for the percentage of agricultural laborers among left-behind rural laborers in Column (4), all the dependent variables are transformed by inverse hyperbolic sine. Provincial controls include the GDP share of the agricultural sector, ln(Population), ln(GDP per capita), ln(total employment), the employment shares in the agricultural sector and the manufacturing sector, and the share of rural employment in each province in 1998. All the provincial controls are interacted with $Post_{98}$. Village population is the logged value in year $t$. Standard errors in parentheses are clustered at the province level, *, **, and *** denote significance at the 10\%, 5\% and 1\% levels, respectively.}

\end{table}

\subsubsection{Income and Life Quality}\label{sec_vpanel_lifequality}

Finally, we switch to the HEE's impact on remaining rural residents' life quality. Table \ref{tab_v_reg_lifequality} reports the results. To split the effect of out-flow migration from the HEE's impact on the village per se in some sense, besides provincial level controls, we also control the logarithmic village population in year $t$. Additionally, since some variables are counted by household numbers (e.g., the number of households having a TV), we control the logarithmic household number as a precaution, in case the decrease of households mechanically drives the results. 

Column (1) displays the change of average income in rural areas due to HEE. The denominator of the village level average income covers residents who live in the village for more than six months and migrant workers who may work outside more than six months but bring their income home. The income covers all the sources including the remittances from people outside (\cite{Policy_Research_Office_of_CPC2010-us}; \cite{Moa_datacoll-aj}). We adjust the income to the 1995 level. We observe a significant negative impact the HEE has on the average income with a 10\% confidence level. After controlling the provincial and village level control variables, the village with one more unit in the HEE magnitude suffers about a 6\% decline in the income per capita. Moreover, the effect is slightly smaller when we add the population and household number into the controls. In other words, the HEE has a negative impact on the average income by itself. In Table \ref{tab_app_inc9503}, we report the result with the same specification but just using the data during 1995 - 2003, and there is no significant effect in this relatively short period. It indicates that this effect is not likely to be temporarily driven by the potential decrease in the labor force as some of the labor may switch to education given the HEE. Also, as the migrant workers are part of the average income, it is possible that the HEE affects the average income through people permanently moving out. In other words, HEE helps people who could earn more than the remaining permanently move out. Thus, the average income naturally falls for the left-behind ones.    

\begin{figure}[h!]
    \centering
    \includegraphics[width=.48\textwidth]{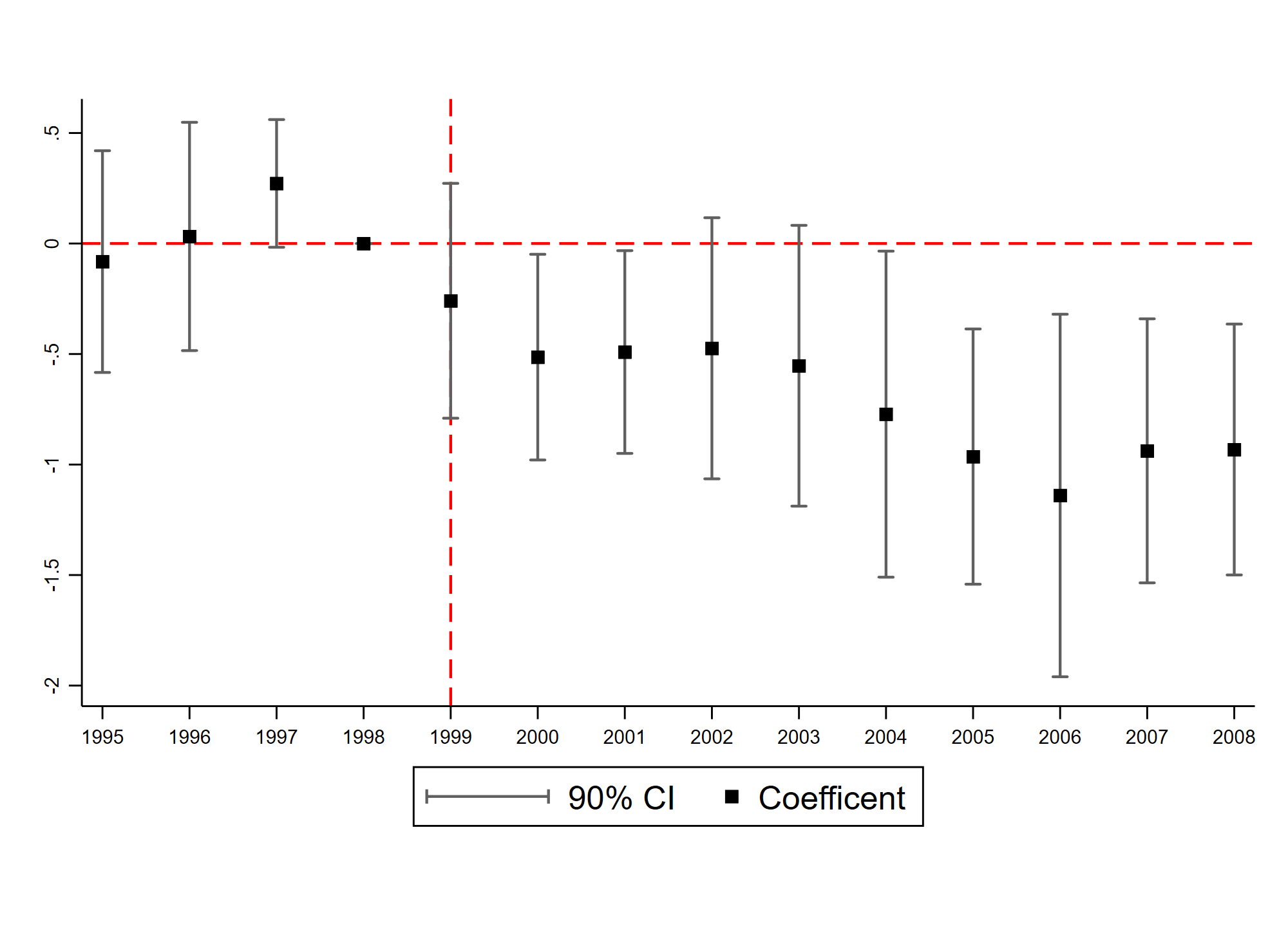}
	\caption{The Dynamic Effect of HEE on the Number of Households Using Tap Water}\label{fig_v_tapwater}
	\captionsetup{singlelinecheck=false}
    \caption*{\footnotesize \textit{Notes}: This figure shows the coefficients of the interaction term of between $HighDiff$ and year indicators as presented in Equation \eqref{eq_eventstudy}. The results include 24 provinces. The year 1998 serves as the reference category, which is omitted in the regression. The vertical line indicates the start year of HEE.}
\end{figure}

Through Columns (2)-(6), we use the number of households with access to better living facilities to measure life quality. Column (7) reports the total number of telephones the village has, and it is still very likely to be the case that one household equips one telephone. As we have argued in Section \ref{sec_Data}, tap water could be a good approximation for both village and household wealth, and safe water is an indicator of village health status. After controlling for the provincial and village level control variables, we have found that with one unit increase in the HEE magnitude, the number of households that invest in a tap water system declines about 2.3\%. We also report the event study result of the access to tap water in Figure \ref{fig_v_tapwater}. Consistently, we also find that the number of households who live in the concrete house drops about 8.5\% to 9.5\% in a village with one unit more HEE magnitude. Though significance not found in other facilities, the HEE generally has a negative effect across all the variables describing the life quality of the remaining households in the villages. By comparing the estimated coefficient between Panel B and C and D, we could conclude that in some sense, the negative effect is brought by the out-flow of migrants and the decrease of household number per se. However, the survival significance in tap water and good housing after controlling the population size and household number implies the similar mechanism shown in Column (1) that the HEE moves the people/households who have a higher potential to have better life qualities out of the village. In other words, the negative effect is not only due to the decline in household/population mechanically, but also the HEE picks up the cherries within the village.       

To more explicitly describe the whole picture of remaining residents' life quality. We also conduct a principal component analysis (PCA) to construct an aggregated life-quality index for rural villages, as presented in Column (9). We include all variables listed in Columns (2)-(7) in PCA. The index can capture about 60\% of the variation across villages. The results indicate that villages exposed to a higher magnitude of HEE are worse off than those exposed to a lower magnitude of HEE. Figure \ref{fig_v_lifeindex} presents the dynamic effects of HEE on the life-quality index, the insignificant coefficients before 1999 indicate the validity of the parallel trends assumption. The significant downward trend reveals the persistently adverse effects of HEE on rural life qualities.

\subsubsection{Heterogeneity Analysis}
In the above sections we have learned that the HEE (i) causes out-flow both population and households in the rural villages; (ii) induces decline in the stock of senior school educated laborers in rural village particularly, while it increase the overall education attainment of rural \textit{hukou} holders; (iii) shows little impact on the agriculture 
activities, which absorb the majority of rural labor; (iv) has negative effect on the remaining residents life quality even after controlling the population and households number.

These results are explained by the possible mechanism that the HEE helps the education system pick more village residents with potential in earnings. This mechanism is consistent with the fact that though more well-educated rural people the HEE produces, less stock of them the rural place has eventually. To provide more consistent evidence of this mechanism, we conduct a heterogeneity analysis on the village level, exploiting variation of villages within the province. 

In every province, one village, if it is very poor or rich, would be categorized as "Poor village" (\textit{Chinese: Pin Kun Cun}, the poor village, the same below) and "Moderately Prosperous village" (\textit{Chinese: Xiao Kang Cun}, for simplicity, the rich village, the same below) by the provincial civil authority based on certain province-made criterion. Namely, though these standard official titles are applied across the country but they are categories within one province. Given the HEE magnitude is the same across villages within one province, this province-made heterogeneity enable us to explore HEE's impact controlling province-level shock in some sense.

To provide more evidence on the mechanisms of how the HEE impacts rural village, we estimate the following equation to detect the HEE's effect on poor (or rich) villages particularly,
\begin{equation}\label{eq_did_village_heter}
y_{v, p, t}=\alpha + \beta_{1} D_{v} \times HighDiff_{p} \times Post_{t} + \beta_{2} HighDiff_{p} \times Post_{t} + X_{v, p,1998} \times Post_{t} \Gamma + \delta_{t} + \theta_{v} + \varepsilon_{v, p, t}
\end{equation}
where all terms are the same as in Equation \eqref{eq_did_village} except for $D_{v}$. Here we have $D_{v} = 1$ if village $v$ is a poor (or rich) village. $\beta_{1}$ is the coefficient of interest.

To be consistent with the possible mechanism that HEE helps education picks more individuals with higher potential, we would expect that the HEE's influence is larger in the rich villages within the same province. This is because comparing to other villages, the province-certificated rich village shall have a higher level of human capital stock and development potential per se. Namely, it has relatively more candidates for education (e.g. more senior high school students who can at least compete for college education) and also residents are more likely to face loose budget constraint facing education decision. 

Table \ref{tab_v_heter_xiaokang} presents the estimations of Equation \eqref{eq_did_village_heter}. We add control variables by the most precautionary specification as we have discussed in the above sections. We do the exercises across outcomes which the HEE has significant impact on in the main result part. As one can observe from the heterogeneity analysis results, the HEE does have stronger impacts among the rich villages, even though villages in one provinces recieve the same level of HEE magnitude. Rich villages tend to suffer more from the out-flow of well-educated laborers. For example, they would lose 2\% more senior school or above degree holders while the average effect is around 5.6\% for the whole group. The heterogeneity analysis supports our argument about the mechanism by which the HEE could affect the rural villages in some sense.    

\newpage
\begin{landscape}
\begin{table}[htbp!]
\centering
\caption{The Effect of HEE on the Income and Life Quality in Rural Villages}\label{tab_v_reg_lifequality}
\begin{tabular}{lcccccccc}
\hline \hline
 & (1) & (2) & (3) & (4) & (5) & (6) & (7) &  (8) \\
 & \multirow{2}{*}{ln(Avg. income)} & \multicolumn{5}{c}{ln(Number of HHs)} & \multirow{2}{*}{ln(Telephone)} &  \multirow{2}{*}{Index} \\
\cline{3-7}
 &  & tap water & safe water & good housing & electricity & TV &   &  \\
\hline
\multicolumn{9}{l}{\textit{\textbf{Panel A: No controls}}} \\
$\Delta \overline{AdmPC} \times Post$ & -0.0260 & -0.0505* & -0.00525 & -0.0714*** & -0.0332 & -0.0203*** & -0.0841** & -0.0635 \\
 & (0.0182) & (0.0281) & (0.0502) & (0.0229) & (0.0247) & (0.00434) & (0.0384) & (0.0756) \\
Observations & 2579 & 2399 & 2363 & 2381 & 2648 & 2638 & 2653 & 1481 \\
Number of villages & 188 & 192 & 192 & 192 & 192 & 192 & 192 & 192 \\ \midrule
\multicolumn{9}{l}{\textit{\textbf{Panel B: Control for provincial characteristics}}} \\
$\Delta \overline{AdmPC} \times Post$ & -0.0612* & -0.0287*** & -0.113* & -0.102*** & -0.0421* & -0.0116 & -0.0814 & -0.215** \\
 & (0.0301) & (0.0120) & (0.0661) & (0.0363) & (0.0213) & (0.0149) & (0.0539)  & (0.0838) \\
Observations & 2579 & 2399 & 2363 & 2381 & 2648 & 2638 & 2653 & 1481 \\
Number of villages & 188 & 192 & 192 & 192 & 192 & 192 & 192  & 192 \\ \midrule
\multicolumn{9}{l}{\textit{\textbf{Panel C: Control for: provincial characteristics + village population}}} \\
$\Delta \overline{AdmPC} \times Post$ & -0.0598* & -0.0228** & -0.107 & -0.0945** & -0.0323 & -0.00460 & -0.0689 & -0.201** \\
 & (0.0304) & (0.0113) & (0.0669) & (0.0343) & (0.0235) & (0.0143) & (0.0500) & (0.0813) \\
Observations & 2577 & 2399 & 2363 & 2380 & 2648 & 2638 & 2653  & 1481 \\
Number of villages & 188 & 192 & 192 & 192 & 192 & 192 & 192  & 192 \\ \midrule
\multicolumn{9}{l}{\textit{\textbf{Panel D: Control for: provincial characteristics + village population + village number of HHs}}} \\
$\Delta \overline{AdmPC} \times Post$ & -0.0578* & -0.0237** & -0.104 & -0.0851** & -0.0286 & 0.00401 & -0.0661  & -0.138* \\
 & (0.0317) & (0.0119) & (0.0659) & (0.0321) & (0.0236) & (0.0141) & (0.0510)  & (0.0723) \\
Observations & 2576 & 2398 & 2362 & 2379 & 2647 & 2637 & 2652 & 1481 \\
Number of villages & 188 & 192 & 192 & 192 & 192 & 192 & 192 & 192 \\
\hline \hline
\end{tabular}
\captionsetup{singlelinecheck=false, width = 1.0\textwidth}
\caption*{\footnotesize \textit{Notes}: Dependent variables in Column (2)-(6) are defined by the number of HHs equipped with them. (e.g., \textit{TV} refers to the number of HHs having a TV.) The life-quality index in Column (8) is generated by the principal component analysis based on variables listed in Columns (2)-(7). Except for the PCA-based index all outcomes are transformed by inverse hyperbolic sine. Provincial controls include the GDP share of the agricultural sector, ln(Population), ln(GDP per capita), ln(total employment), the employment shares in the agricultural sector and the manufacturing sector, and the share of rural employment in each province in 1998. Both village population and households number are in year $t$. Standard errors in parentheses are clustered at the province level, *, **, and *** denote significance at the 10\%, 5\% and 1\% levels, respectively.}
\end{table}
\end{landscape}

\newpage
\begin{landscape}
\begin{table}[h!]
\centering
\caption{Heterogenous effect on the rich village}
\label{tab_v_heter_xiaokang}
\scalebox{0.9}{
\begin{tabular}{lccccccccc}
\hline \hline
 & (1) & (2) &  & (3) & (4) &  & (5) & (6) & (7) \\ 
 & \multicolumn{2}{c}{Population} &  & \multicolumn{2}{c}{Edu of remaining labor} &  & \multicolumn{3}{c}{Income and Life quality} \\ \cline{2-3} \cline{5-6} \cline{8-10} 
 & ln(Population) & ln(HHs) &  & ln(Senior sch.) & \% Senior sch. &  & ln(Avg. income) & ln(HHs: good housing) & ln(HHs: tap water) \\
\hline
$D_{v,t} \times \Delta\overline{AdmPC}_{p} \times Post$ & -0.0101 & -0.0165** &  & -0.0209*** & -0.00394** &  & -0.0288* & -0.0130*** & -0.0105 \\
 & (0.00590) & (0.00689) &  & (0.0042) & (0.00149) &  & (0.0157) & (0.0077) & (0.0080) \\
$ \Delta\overline{AdmPC}_{p} \times Post$ & -0.0112* & -0.0084** &  & -0.0489*** & -0.00327** &  & -0.0485* & -0.0756*** & -0.0315** \\
 & (0.00560) & (0.00327) &  & (0.0075) & (0.00164) &  & (0.0238) & (0.0236) & (0.0159) \\
Observations & 2656 & 2657 &  & 2453 & 2426 &  & 2576 & 2379 & 2398 \\
Number of villages & 192 & 192 &  & 192 & 192 &  & 188 & 192 & 192 \\ \hline
Provincial control & Y & Y &  & Y & Y &  & Y & Y & Y \\
Village population &  &  &  & Y & Y &  & Y & Y & Y \\
Village HHs number &  &  &  &  &  &  & Y & Y & Y \\
\hline \hline
\end{tabular}
}

\captionsetup{singlelinecheck=false, width = 1.0\textwidth}
\caption*{\footnotesize \textit{Notes}: $D_{v,t} = 1$ if village $v$ at year $t$ is a rich village. Provincial controls include the GDP share of the agricultural sector, ln(Population), ln(GDP per capita), ln(total employment), the employment shares in the agricultural sector and the manufacturing sector, and the share of rural employment in each province in 1998. Both village population and households number are in year $t$. Standard errors in parentheses are clustered at the province level, *, **, and *** denote significance at the 10\%, 5\% and 1\% levels, respectively.}

\end{table}
\end{landscape}

\subsection{The Impact on Rural Households}\label{sec_EmpiricalResults_didHH}

In the previous section, we have found that villages in provinces more exposed to the HEE experienced out-migration, drop in the labor force with senior high school education, lower income and worse life qualities. The results indicate that the HEE had side-effects the development of rural villages. However, the overall negative impact on the village level does not indicate the same impact on the households living in the villages living in places that experienced stronger magnitude of the HEE. In this section, we then extend our analysis to the household level.

\subsubsection{Households Population and Education}
We first examine the HEE's impact on households' population and households' members' education attainments. We focus on three primary outcomes: household size, the number of migrants, number of household labors with at least senior high school education.

Table \ref{tab_hh_size} shows the results. Column 1 reports the effect on households' population size, which is defined by the number of residents who register their \textit{hukou} at one household. It shows that households in provinces more exposed to the HEE experienced a 5\% more decline in household size after the HEE. The second column  reports the HEE's effect on the households permanent migrants. The NFPS household panel does not have the direct records of households' permanent migrants (who leave the household forever). Here we approximate the households permanent migrants by the percentage change of residents numbers between the base year 1995 and year $t$. Column 2 indicates that those households had around 5 percentage point increase in the share of household migrants. To address ones' concern that this result may be driven by the HEE's impact on households fertility or mortality, we also examined these two outcomes at the village level since there is no information of these two variables in the household panel. The HEE shows no significant impact on the villages' birth/death rate, which release the concern. The third column demonstrates the impact on the the amount of high-quality left-behind labor (measured by whether finishing senior high school). Households with more people can provide around 10\% more high-quality labor, but the impact could be offset by the HEE, which on average reduced each household's high-quality labor by 9\% in provinces with higher expansion magnitudes. 

In short, on households level, HEE led to more permanent migrants and fewer number of educated labors left in the family. Also, one shall notice these estimations on households level are consistent with the village-level results on the population and left-behind labors' education attainment.

\begin{table}[h!]
\centering
\caption{The HEE's Impact on Household Size and Labor Force}\label{tab_hh_size}
\begin{tabular}{lccc}
\hline \hline
& (1) & (2) & (3) \\
 & ln(HH size) & \%(Migrants) & ln(\# Labor Hsch. Edu) \\
 \hline
$\Delta \overline{AdmPC} \times Post$ & -0.0540*** & 0.0497*** & -0.0892* \\
 & (0.0183) & (0.0155) & (0.0514) \\
ln(HH size) &  &  & 0.109*** \\
 &  &  & (0.0192) \\
\hline
Control Mean & 2.0605 & -.0063 & .1502 \\
Adjusted R$^2$ & 0.517 & 0.476 & 0.425 \\
Observations & 52771 & 52771 & 52771 \\
\hline \hline
\end{tabular}
\captionsetup{singlelinecheck=false, width = 0.8\textwidth}
\caption*{\footnotesize \textit{Notes}: The outcomes in Columns (2) and (3) are the percentage of accumulated migrants and the amount of labor force with at least senior high school education, respectively. Standard errors in parentheses are clustered at the province level, *, **, and *** denote significance at the 10\%, 5\% and 1\% levels, respectively.}
\end{table}

\subsubsection{Households Income through the Migration Channel}
Next, we move forward to HEE's impact on households' income. On the village level, we detect that this policy has some side-effect on the villages' public goods provision and income per capita. However, the overall effect on villages does not indicate the same effect on households affected by the HEE directly (households who sent out more students due to the policy). 

Table \ref{tab_hh_income} reports the HEE's impact on the welfare of households, as measured by total income, remittance sent by migrants, and other income. We calculate both the total income and the per capita income in each household to account for the change in household size due to migration. After controlling for the number of migrants, households in provinces more exposed to the HEE experienced increase in both total income and remittance. The results are opposed to that at the village level, where we see negative impact of the HEE on per capita income. More importantly, the increase in household income is mainly due to remittance but not other income, indicating that the HEE provides a way for out-migration.

Other results from Table \ref{tab_hh_income} further confirm the migration channel through which the HEE affects rural households. Without the HEE, the amount of migrants has a negative impact on household income due to the loss of labor force. After the HEE, the impact is reversed, more migrants lead to higher household income. The reason is as follows. Some household members obtained high education after the HEE, then they can earn higher salaries in urban areas. Those rural-urban migrants may not support their families directly through remittances, but their households on average will have higher income than other households with fewer migrants. The results are robust to other specifications when we change the definition for migrants. The event studies also confirms the parallel trends assumption in our identification.

Overall, we find contrasting impacts on rural households when compared to rural villages. The HEE provides opportunities for some rural residents to receive higher education, migrate to urban areas, and get higher salaries, which ultimately increases their households welfare. That is, we see the positive impact of education opportunities on rural households who can take those opportunities. However, for households that cannot take those high education opportunities, they will experience decreases in income and worse living facilities, as indicated by the HEE's impact on rural villages.

\begin{table}[h!]
\centering
\caption{The HEE's Impact on Household Income and Remittance}\label{tab_hh_income}
\resizebox{\textwidth}{!}{%
\begin{tabular}{@{}lccclccc@{}}
\hline\hline
& (1) & (2) & (3) & & (4) & (5) & (6) \\
 & \multicolumn{3}{c}{Total} &  & \multicolumn{3}{c}{Average (per capita)} \\ \cline{2-4} \cline{6-8}
 & ln(Income) & ln(Remittance) & ln(Other) &  & ln(Income) & ln(Remittance) & ln(Other) \\
 \hline
$HighDiff \times Post$ & 0.0845** & 0.530* & 0.0355 &  & 0.0836** & 0.463** & 0.0504 \\
 & (0.0365) & (0.268) & (0.0359) &  & (0.0358) & (0.212) & (0.0363) \\
$\# Migrants$ & -0.154*** & 0.0175 & -0.179*** &  & 0.0739*** & 0.140*** & 0.0560*** \\
 & (0.00887) & (0.0260) & (0.0103) &  & (0.00573) & (0.0209) & (0.00807) \\
$\# Migrants \times Post$ & 0.0149** & -0.000579 & 0.0270*** &  & 0.00456 & -0.0121 & 0.0144** \\
 & (0.00574) & (0.0298) & (0.00724) &  & (0.00555) & (0.0246) & (0.00614) \\
\hline
Control Mean & 9.9053 & 3.7131 & 9.81 &  & 8.5505 & 2.9786 & 8.4538 \\
Adjusted R$^2$ & 0.692 & 0.367 & 0.680 &  & 0.670 & 0.370 & 0.642 \\
Observations & 47187 & 46138 & 45592 &  & 47106 & 46125 & 45547 \\
\hline \hline
\end{tabular}%
}
\captionsetup{singlelinecheck=false, width = 1.0\textwidth}
\caption*{\footnotesize \textit{Notes}: Columns (2) and (5) show the result for household remittance from migrants. Outcomes in Columns (3) and (6) are all the household income except for remittance. the Standard errors in parentheses are clustered at the province level, *, **, and *** denote significance at the 10\%, 5\% and 1\% levels, respectively.}
\end{table}

\section{Conclusions}\label{sec_Conclusion}

Due to the lack of comprehensive data on rural places, the effect of education opportunities in rural areas remains anecdotal. On the one hand, it could be a ``brain gain'' since more educated people could lead to faster human capital accumulation and economic development of rural areas. On the other hand, the effect could be brain-drain. Taking those higher education opportunities enables individuals to acquire
skills, be competitive in urban labor markets, and earn higher salaries, attracting them to move into
urban areas, thus reducing the human capital in rural areas.

Exploiting the higher education expansion in China and the restricted annual panel data of rural villages and households, this paper causally estimates the impact of education opportunities on human capital and quality of life in rural areas. We find that the HEE induced the educational attainment of rural people, but at the same time, it extracted those well-educated people out of villages, all the people who stayed behind experienced a drop in life quality, which validates the possible brain drain impact of education opportunities. 
However, households sending out migrants after the HEE experienced an increase in their per capita income. The phenomenon where villages experienced ``brain drain'' and households with migrants gained after the HEE is explained by the fact that education could serve as a way to overcome the barrier of rural-urban migration, which is prominent in developing countries.




This paper utilizes the setting in China but can shed light on other situations with migration constraints and low urbanization rates. Our results suggest that policymakers should be cautious about the reallocation effect and the adverse spillover effect on less developed areas when implementing such nationwide policies.

\newpage
\bibliographystyle{jpe}
\bibliography{HigherEduExp.bib}

\begin{thebibliography}{50}
\newcommand{\enquote}[1]{``#1''}
\providecommand{\natexlab}[1]{#1}
\providecommand{\url}[1]{\texttt{#1}}
\providecommand{\urlprefix}{URL }

\bibitem[{Adukia, Asher, and Novosad(2020)}]{adukia2020educational}
Adukia, Anjali, Sam Asher, and Paul Novosad. 2020.
\newblock \enquote{Educational investment responses to economic opportunity:
  evidence from Indian road construction.}
\newblock \emph{American Economic Journal: Applied Economics} 12~(1):348--76.

\bibitem[{Asher and Novosad(2020)}]{asher2020rural}
Asher, Sam and Paul Novosad. 2020.
\newblock \enquote{Rural roads and local economic development.}
\newblock \emph{American Economic Review} 110~(3):797--823.

\bibitem[{Beine, Docquier, and Rapoport(2001)}]{Beine2001-re}
Beine, Michel, Fr{\'e}d{\'e}ric Docquier, and Hillel Rapoport. 2001.
\newblock \enquote{Brain drain and economic growth: theory and evidence.}
\newblock \emph{Journal of Development Economics} 64~(1):275--289.

\bibitem[{Beine, Docquier, and Rapoport(2008)}]{Beine2008-zv}
Beine, Michel, Frederic Docquier, and Hillel Rapoport. 2008.
\newblock \enquote{Brain drain and human capital formation in developing
  countries: winners and losers.}
\newblock \emph{The Economic Journal} 118~(528):631--652.

\bibitem[{Benjamin, Brandt, and Giles(2005)}]{benjamin2005evolution}
Benjamin, Dwayne, Loren Brandt, and John Giles. 2005.
\newblock \enquote{The evolution of income inequality in rural China.}
\newblock \emph{Economic Development and Cultural Change} 53~(4):769--824.

\bibitem[{Bloom, Canning, and Chan(2006)}]{bloom2006higher}
Bloom, David~Elliot, David Canning, and Kevin Chan. 2006.
\newblock \emph{Higher education and economic development in Africa}, vol. 102.
\newblock World Bank Washington, DC.

\bibitem[{Blossfeld et~al.(2016)Blossfeld, Buchholz, Skopek, and
  Triventi}]{blossfeld2016models}
Blossfeld, Hans-Peter, Sandra Buchholz, Jan Skopek, and Moris Triventi. 2016.
\newblock \emph{Models of secondary education and social inequality: An
  international comparison}.
\newblock Edward Elgar Publishing.

\bibitem[{Brandt and Holz(2006)}]{Brandt2006-bn}
Brandt, Loren and Carsten~A Holz. 2006.
\newblock \enquote{Spatial price differences in China: Estimates and
  implications.}
\newblock \emph{Economic development and cultural change} 55~(1):43--86.

\bibitem[{Burbidge, Magee, and Robb(1988)}]{Burbidge1988-sj}
Burbidge, John~B, Lonnie Magee, and A~Leslie Robb. 1988.
\newblock \enquote{Alternative transformations to handle extreme values of the
  dependent variable.}
\newblock \emph{Journal of the American statistical Association}
  83~(401):123--127.

\bibitem[{Chari et~al.(2021)Chari, Liu, Wang, and Wang}]{Chari2021-lt}
Chari, Amalavoyal, Elaine~M Liu, Shing-Yi Wang, and Yongxiang Wang. 2021.
\newblock \enquote{Property rights, land misallocation, and agricultural
  efficiency in China.}
\newblock \emph{The Review of Economic Studies} 88~(4):1831--1862.

\bibitem[{Chatterton and Goddard(2000)}]{chatterton2000response}
Chatterton, Paul and John Goddard. 2000.
\newblock \enquote{The response of higher education institutions to regional
  needs.}
\newblock \emph{European Journal of Education} 35~(4):475--496.

\bibitem[{{Congressional-Executive Commission on
  China}(2005)}]{Congressional-Executive_Commission_on_China2005-oq}
{Congressional-Executive Commission on China}. 2005.
\newblock \enquote{Recent Chinese Hukou Reforms.}
\newblock \url{https://www.cecc.gov/recent-chinese-hukou-reforms}.
\newblock Accessed: 2020-12-11.

\bibitem[{D\'{e}murger, Hanushek, and Zhang(2023)}]{Demurger2019-st}
D\'{e}murger, Sylvie, Eric~A. Hanushek, and Lei Zhang. 2023.
\newblock \enquote{Employer Learning and the Dynamics of Returns to
  Universities: Evidence from Chinese Elite Education during University
  Expansion.}
\newblock \emph{Economic Development and Cultural Change} .

\bibitem[{Docquier and Rapoport(2012)}]{Docquier2012-kj}
Docquier, Fr{\'e}d{\'e}ric and Hillel Rapoport. 2012.
\newblock \enquote{Globalization, brain drain, and development.}
\newblock \emph{Journal of Economic Literature} 50~(3):681--730.

\bibitem[{Duflo(2001)}]{Duflo2001-qb}
Duflo, Esther. 2001.
\newblock \enquote{Schooling and labor market consequences of school
  construction in Indonesia: Evidence from an unusual policy experiment.}
\newblock \emph{American Economic Review} 91~(4):795--813.

\bibitem[{Fan(2007)}]{fan2007china}
Fan, C~Cindy. 2007.
\newblock \emph{China on the Move: Migration, the State, and the Household}.
\newblock Routledge.

\bibitem[{Fang et~al.(2012)Fang, Eggleston, Rizzo, Rozelle, and
  Zeckhauser}]{fang2012returns}
Fang, Hai, Karen~N Eggleston, John~A Rizzo, Scott Rozelle, and Richard~J
  Zeckhauser. 2012.
\newblock \enquote{The returns to education in China: Evidence from the 1986
  compulsory education law.}
\newblock Tech. rep., National Bureau of Economic Research.

\bibitem[{Feng(2010)}]{Feng2010Fiscal}
Feng, Lin. 2010.
\newblock \emph{Fiscal Expenditure of Rural Infrastructure}.
\newblock Ph.D. thesis, Shangdong Agriculture University.

\bibitem[{Gollin, Lagakos, and Waugh(2014)}]{gollin2014agricultural}
Gollin, Douglas, David Lagakos, and Michael~E Waugh. 2014.
\newblock \enquote{The agricultural productivity gap.}
\newblock \emph{The Quarterly Journal of Economics} 129~(2):939--993.

\bibitem[{Ji(2009)}]{Ji2009-cg}
Ji, Baocheng. 2009.
\newblock \enquote{The 1999 {HEE} leads higher education to a passing lane (in
  Chinese).}
\newblock \emph{China Education Daily} .

\bibitem[{Kang(2000)}]{Kang2000-uw}
Kang, Ning. 2000.
\newblock \enquote{Decision-making in Education Policy and Institutional
  Innovation: A case study of the higher education expansion in 1999 (in
  Chinses).}
\newblock \emph{Journal of Higher Education} ~(8):31--38.

\bibitem[{Kang(2005)}]{kang2005institutional}
---{}---{}---. 2005.
\newblock \enquote{Institutional Innovation in Higher Education Resource
  Distribution in a Transitional Economy (in Chinese).}

\bibitem[{Kang and Min(2005)}]{Kang2005-kt}
Kang, Ning and Weifang Min. 2005.
\newblock \enquote{Institutional Innovation in Higher Education Resource
  Distribution in a Transitional Economy (in Chinese).}
\newblock \emph{Journal of Higher Education} 26~(6):6--6.

\bibitem[{Kinnan, Wang, and Wang(2018)}]{Kinnan2018-bd}
Kinnan, Cynthia, Shing-Yi Wang, and Yongxiang Wang. 2018.
\newblock \enquote{Access to migration for rural households.}
\newblock \emph{American Economic Journal: Applied Economics} 10~(4):79--119.

\bibitem[{Lagakos(2020)}]{Lagakos2020-gc}
Lagakos, David. 2020.
\newblock \enquote{Urban-rural gaps in the developing world: Does internal
  migration offer opportunities?}
\newblock \emph{Journal of Economic perspectives} 34~(3):174--192.

\bibitem[{Li, Whalley, and Xing(2014)}]{Li2014-ip}
Li, Shi, John Whalley, and Chunbing Xing. 2014.
\newblock \enquote{China's higher education expansion and unemployment of
  college graduates.}
\newblock \emph{China Economic Review} 30:567--582.

\bibitem[{Lucas(1988)}]{lucas1988mechanics}
Lucas, Robert~E. 1988.
\newblock \enquote{On the mechanics of economic development.}
\newblock \emph{Journal of Monetary Economics} 22~(1):3--42.

\bibitem[{Ma(2024)}]{Ma_undated-gk}
Ma, Xiao. 2024.
\newblock \enquote{College expansion, trade, and innovation: Evidence from
  China.}
\newblock \emph{International Economic Review} 65~(1):315--351.

\bibitem[{McCulloch and Yellen(1977)}]{McCulloch1977-wz}
McCulloch, Rachel and Janet~L Yellen. 1977.
\newblock \enquote{Factor mobility, regional development, and the distribution
  of income.}
\newblock \emph{Journal of Political Economy} 85~(1):79--96.

\bibitem[{Meng(2012)}]{meng2012labor}
Meng, Xin. 2012.
\newblock \enquote{Labor market outcomes and reforms in China.}
\newblock \emph{Journal of Economic Perspectives} 26~(4):75--102.

\bibitem[{Ministry~of Education(2016)}]{Moe2016-ke}
Ministry~of Education, People's Republic of~China. 2016.
\newblock \enquote{How to decide the admission plan of college?}
\newblock
  \url{http://www.moe.gov.cn/jyb_xwfb/moe_2082/zl_2016n/2016_zl30/201606/t20160606_248332.html}.
\newblock Accessed: 2021-6-7.

\bibitem[{Mountford(1997)}]{Mountford1997-al}
Mountford, Andrew. 1997.
\newblock \enquote{Can a brain drain be good for growth in the source economy?}
\newblock \emph{Journal of Development Economics} 53~(2):287--303.

\bibitem[{National Bureau~of Statistics(2019)}]{Moa_datacoll-aj}
National Bureau~of Statistics, People's Republic of~China. 2019.
\newblock \enquote{Rural comprehensive survey system (Nong cun zong he tong ji
  diao cha zhi du).}
\newblock
  \url{http://www.stats.gov.cn/tjfw/bmdcxmsp/bmzd/201902/t20190221_1650141.html}.
\newblock Accessed: 2021-8-27.

\bibitem[{Ngai, Pissarides, and Wang(2019)}]{ngai2019china}
Ngai, L~Rachel, Christopher~A Pissarides, and Jin Wang. 2019.
\newblock \enquote{China’s mobility barriers and employment allocations.}
\newblock \emph{Journal of the European Economic Association}
  17~(5):1617--1653.

\bibitem[{{Policy Research Office of CPC} and
  {MOA}(2010)}]{Policy_Research_Office_of_CPC2010-us}
{Policy Research Office of CPC} and {MOA}. 2010.
\newblock \emph{Summary of National Fixed Point Survey (2000-2009)}.
\newblock China Agriculture Press.

\bibitem[{Psacharopoulos and Patrinos(2004)}]{Psacharopoulos2004-bf}
Psacharopoulos, George and Harry~Anthony Patrinos. 2004.
\newblock \enquote{Returns to investment in education: a further update.}
\newblock \emph{Education Economics} 12~(2):111--134.

\bibitem[{Rong and Wu(2020)}]{Rong2020-ad}
Rong, Zhao and Binzhen Wu. 2020.
\newblock \enquote{Scientific personnel reallocation and firm innovation:
  Evidence from China’s college expansion.}
\newblock \emph{Journal of Comparative Economics} 48~(3):709--728.

\bibitem[{sen University(2017)}]{Suyu2017-dt}
sen University, Sun~Yat. 2017.
\newblock \enquote{Sun Yat-sen University - Admission Plan (2017).}
\newblock \url{http://admission.sysu.edu.cn/zs01/zs01d/index5.htm}.
\newblock Accessed: 2021-6-7.

\bibitem[{Tian, Xia, and Yang(2020)}]{Tian_2020-ol}
Tian, Yuan, Junjie Xia, and Rudai Yang. 2020.
\newblock \enquote{Trade-induced urbanization and the making of modern
  agriculture.}
\newblock \emph{Working Paper} .

\bibitem[{Todaro and Smith(2021)}]{Todaro2021-bd}
Todaro, Michael~P and Stephen~C Smith. 2021.
\newblock \emph{Economic development}.
\newblock Addison-Wesley.

\bibitem[{Wan(2006)}]{Wan2006-wd}
Wan, Yinmei. 2006.
\newblock \enquote{Expansion of Chinese higher education since 1998: Its causes
  and outcomes.}
\newblock \emph{Asia Pacific Education Review} 7:19--32.

\bibitem[{Wang(2019)}]{wang2019labor}
Wang, Shing-Yi. 2019.
\newblock \enquote{The Labor Supply Consequences of Having a Boy in China.}
\newblock Tech. rep., National Bureau of Economic Research.

\bibitem[{Wang and Shen(2014)}]{Wang2014-gw}
Wang, Xiaxin and Yan Shen. 2014.
\newblock \enquote{The effect of China's agricultural tax abolition on rural
  families' incomes and production.}
\newblock \emph{China Economic Review} 29:185--199.

\bibitem[{Wu and You(2020)}]{wu2020welfare}
Wu, Wenbin and Wei You. 2020.
\newblock \enquote{The Welfare Implications of Internal Migration Restrictions:
  Evidence from China.}
\newblock \emph{SSRN Electronic Journal} .

\bibitem[{Xing(2014)}]{Xing2014-yp}
Xing, Chunbing. 2014.
\newblock \enquote{Education {Expansion,Migration} and {Rural-Urban} Education
  {Gap: A} Case Study on the Effect of University Expansion (in Chinese).}
\newblock \emph{China Economic Quarterly} .

\bibitem[{Yang, Hu, and Wu(2008)}]{Yang2008-tvprice}
Yang, Ruidong, Song Hu, and Shuyuan Wu. 2008.
\newblock \enquote{Relations of brand price level and specifications to demand
  price elasticity of China's color {TV} products (in Chinese).}
\newblock \emph{Journal of Tsinghua University (Science and Technology)}
  48~(12):2141--2144.

\bibitem[{Yang(2021)}]{yang2021place}
Yang, Yu. 2021.
\newblock \enquote{Place-Based College Admission, Migration and the Spatial
  Distribution of Human Capital: Evidence from China.}
\newblock \emph{Working Paper} .

\bibitem[{Yao(2019)}]{Yao2019-vg}
Yao, Yao. 2019.
\newblock \enquote{Does higher education expansion enhance productivity?}
\newblock \emph{Journal of Macroeconomics} 59:169--194.

\bibitem[{Zhao(1997{\natexlab{a}})}]{Zhao1997-sw}
Zhao, Yaohui. 1997{\natexlab{a}}.
\newblock \enquote{China's rural labor mobility and the role of education (in
  Chinese).}
\newblock \emph{Economic Research Journal} 2:37--42.

\bibitem[{Zhao(1997{\natexlab{b}})}]{Zhao1997-rs}
---{}---{}---. 1997{\natexlab{b}}.
\newblock \enquote{Labor migration and returns to rural education in China.}
\newblock \emph{American Journal of Agricultural Economics} 79~(4):1278--1287.

\end{thebibliography}

\newpage
\begin{appendices}

\renewcommand\thetable{\thesection.\arabic{table}}
\setcounter{table}{0}
\setcounter{section}{0}
\renewcommand\thefigure{\thesection.\arabic{figure}}
\setcounter{figure}{0}
\setcounter{section}{0}
\renewcommand{\theequation}{\Alph{section}.\arabic{equation}}

\section{Tables}\label{sec_Tables}

\begin{table}[h!]
\centering
\caption{Measures of the Magnitude of HEE}\label{tab_treatment_measure}
\begin{tabular}{lcccccccccc}
\hline \hline
\multicolumn{1}{c}{\multirow{2}{*}{Province name}} & \multicolumn{1}{c}{\multirow{2}{*}{Code}} & \multicolumn{4}{c}{5-year average} & & \multicolumn{4}{c}{10-year average} \\
\cline{3-6}\cline{8-11}
& & $Diff$ & Norm. & $Ratio$ & Norm. & & $Diff$ & Norm. & $Ratio$ & Norm. \\
\hline
Hebei & 13 & 1.50 & 0.64 & 2.78 & 0.85 &  & 2.59 & 0.39 & 4.07 & 0.64 \\
Shanxi & 14 & 1.42 & 0.40 & 2.60 & 0.40 &  & 2.63 & 0.48 & 3.96 & 0.45 \\
Liaoning & 21 & 2.04 & 2.27 & 3.38 & 2.32 &  & 2.97 & 1.18 & 4.47 & 1.28 \\
Jilin & 22 & 1.41 & 0.39 & 2.16 & -0.69 &  & 2.25 & -0.30 & 2.84 & -1.39 \\
Heilongjiang & 23 & 1.22 & -0.21 & 2.16 & -0.68 &  & 2.37 & -0.05 & 3.27 & -0.68 \\
Jiangsu & 32 & 1.46 & 0.54 & 2.15 & -0.69 &  & 2.52 & 0.26 & 2.99 & -1.14 \\
Zhejiang & 33 & 1.49 & 0.62 & 2.36 & -0.19 &  & 2.65 & 0.52 & 3.41 & -0.46 \\
Anhui & 34 & 1.12 & -0.50 & 2.59 & 0.38 &  & 2.44 & 0.09 & 4.46 & 1.28 \\
Fujian & 35 & 0.99 & -0.90 & 1.86 & -1.43 &  & 2.65 & 0.52 & 3.31 & -0.63 \\
Jiangxi & 36 & 1.27 & -0.04 & 2.59 & 0.38 &  & 2.74 & 0.71 & 4.42 & 1.21 \\
Shandong & 37 & 1.60 & 0.95 & 2.56 & 0.30 &  & 2.77 & 0.77 & 3.70 & 0.02 \\
Henan & 41 & 1.32 & 0.12 & 3.06 & 1.55 &  & 2.44 & 0.09 & 4.80 & 1.84 \\
Hubei & 42 & 1.39 & 0.33 & 2.41 & -0.06 &  & 2.87 & 0.98 & 3.91 & 0.37 \\
Hunan & 43 & 1.19 & -0.28 & 2.38 & -0.13 &  & 2.25 & -0.30 & 3.62 & -0.11 \\
Guangdong & 44 & 1.28 & -0.02 & 2.49 & 0.14 &  & 2.27 & -0.27 & 3.65 & -0.06 \\
Guangxi & 45 & 0.87 & -1.24 & 2.16 & -0.67 &  & 1.72 & -1.40 & 3.29 & -0.65 \\
Sichuan & 51 & 0.61 & -2.05 & 1.98 & -1.12 &  & 1.72 & -1.39 & 3.79 & 0.17 \\
Guizhou & 52 & 1.19 & -0.28 & 3.04 & 1.48 &  & 1.54 & -1.76 & 3.64 & -0.08 \\
Yunnan & 53 & 0.50 & -2.37 & 1.77 & -1.64 &  & 1.03 & -2.82 & 2.60 & -1.78 \\
Shaanxi & 61 & 1.79 & 1.53 & 3.01 & 1.42 &  & 3.15 & 1.55 & 4.53 & 1.39 \\
Gansu & 62 & 1.11 & -0.53 & 2.52 & 0.22 &  & 2.39 & -0.01 & 4.28 & 0.98 \\
Qinghai & 63 & 1.34 & 0.17 & 2.16 & -0.68 &  & 2.61 & 0.44 & 3.26 & -0.70 \\
Ningxia & 64 & 1.42 & 0.41 & 2.28 & -0.37 &  & 2.76 & 0.75 & 3.50 & -0.31 \\
Xinjiang & 65 & 1.29 & 0.03 & 2.00 & -1.07 &  & 2.19 & -0.44 & 2.69 & -1.64 \\
\hline
\multicolumn{2}{c}{Mean} & 1.28 & 0.00 & 2.44 & 0.00 & & 2.40 & 0.00 & 3.69 & 0.00 \\
\multicolumn{2}{c}{S.D.} & 0.33 & 1.00 & 0.40 & 1.00 & & 0.48 & 1.00 & 0.61 & 1.00 \\
\multicolumn{2}{c}{Median} & 1.31 & 0.08 & 2.40 & -0.09 & & 2.48 & 0.17 & 3.64 & -0.07 \\
\hline \hline
\end{tabular}
\captionsetup{singlelinecheck=false, width=\textwidth}
\caption*{\footnotesize \textit{Notes}: This table presents measures of the HEE magnitudes across provinces. The number of admitted students per capita is applied to quantify difficulties of the NCEE admission. For exposition, we present the number of admitted students per thousand people in the table. We employ both the level change ($Diff$) and the percentage change ($Ratio$) and then take the average across years to capture the magnitude of HEE, as defined in Equation \eqref{eq_treatment_def}. Both 5-year (1999-2003) and 10-year (1999-2008) averages are shown. The normalization columns (Norm.) after each $Diff$ or $Ratio$ column are the corresponding normalized values across provinces.}
\end{table}

\newpage
\begin{table}[h!]
\centering
\caption{Correlations Between Different Measures of HEE Magnitude}\label{tab_treatment_corr}
\begin{tabular}{clc}
\hline \hline
&  & $\Delta \overline{AdmPC}$ ($Diff$, 5-year average, all people) \\
(1) & $\Delta \overline{AdmPC}$ ($Ratio$, 5-year average, all people) & 0.664*** \\
(2) & $\Delta \overline{AdmPC}$ ($Diff$, 10-year average, all people) & 0.809*** \\
(3) & $\Delta \overline{AdmPC}$ ($Ratio$, 10-year average, all people) & 0.370* \\
(4) & $\Delta \overline{AdmPC}$ ($Diff$, 5-year average, age 16-64) & 0.986*** \\
(5) & $\Delta \overline{AdmPC}$ ($Ratio$, 5-year average, age 16-64) & 0.664*** \\
(6) & $\Delta \overline{AdmPC}$ ($Diff$, 5-year average, age 14-22) & 0.941*** \\
(7) & $\Delta \overline{AdmPC}$ ($Ratio$, 5-year average, age 14-22) & 0.664*** \\
\hline \hline
\end{tabular}
\captionsetup{singlelinecheck=false, width=\textwidth}
\caption*{\footnotesize \textit{Notes}: This table presents correlations between different measures of HEE magnitude across provinces. $\Delta \overline{AdmPC}$ is calculated according to Equation \eqref{eq_treatment_def}, based on the number of admitted students per capita. Rows (1)-(3) is calculated relative to the whole population, Rows (4) and (5) include all working-age population, Rows (6) and (7) include only people around the NCEE-taking age (i.e., 18). The significance level of correlation coefficients is specified,  *, **, and *** denote significance at the 10\%, 5\% and 1\% levels, respectively.}
\end{table}

\newpage
\begin{table}[h!]
\centering
\caption{The Effect of HEE on the Population in Rural Villages}\label{tab_app_v_reg_population}
\scalebox{0.77}{
\begin{tabular}{lcccccccc}
\hline \hline
 & (1) & (2) & (3) & (4) & (5) & (6) & (7) & (8) \\
 & Population & Population & HHs & HHs & Population & Population & HHs & HHs \\
\hline
$\Delta \overline{AdmPC} \times Post$ & -0.0544 & -0.122*** & -0.0653 & -0.153*** & -0.0427 & -0.0383 & -0.0463 & -0.0451 \\
 & (0.0347) & (0.0433) & (0.0399) & (0.0495) & (0.0423) & (0.0647) & (0.0450) & (0.0712) \\
 \hline
Observations & 2656 & 2656 & 2657 & 2657 & 2656 & 2656 & 2657 & 2657 \\
Number of villages & 192 & 192 & 192 & 192 & 192 & 192 & 192 & 192 \\
Treatment & Continuous & Continuous & Continuous & Continuous & Binary & Binary & Binary & Binary \\
Provincial Controls &  & Y &  & Y &  & Y &  & Y \\ \hline \hline
\end{tabular}
}
\captionsetup{singlelinecheck=false}
\caption*{\footnotesize \textit{Notes}: All the dependent variables are transformed by inverse hyperbolic sine. The continuous treatment is defined by Equation \eqref{eq_treatment_def} with the ratio form. The binary treatment defined by if the province has a continuous HEE measure above the medium. Provincial controls include the GDP share of the agricultural sector, ln(Population), ln(GDP per capita), ln(total employment), the employment shares in the agricultural sector and the manufacturing sector, and the share of rural employment in each province in 1998. Standard errors in parentheses are clustered at the province level, *, **, and *** denote significance at the 10\%, 5\% and 1\% levels, respectively.}
\end{table}

\begin{table}[h!]
\centering
\caption{HEE's impact on income per capita (1995-2003)}
\label{tab_app_inc9503}
\begin{tabular}{lccc}
\hline \hline
                                           & (1)      & (2)      & (3)      \\ \hline
$\Delta \overline{AdmPC} \times Post$ & -0.0645  & -0.0680  & -0.0663  \\
                                           & (0.0391) & (0.0399) & (0.0400) \\
Observations                               & 2084     & 2083     & 2083     \\
Number of villages                         & 234      & 234      & 234      \\ \hline
Provincial control                         & Y        & Y        & Y        \\
Village population                         &          & Y        & Y        \\
Village HHs number                         &          &          & Y        \\
\hline \hline
\end{tabular}
\captionsetup{singlelinecheck=false}
\caption*{\footnotesize \textit{Notes}: Dependent variables in Column (1) - (3) are logged income per capita. Provincial controls include the GDP share of the agricultural sector, ln(Population), ln(GDP per capita), ln(total employment), the employment shares in the agricultural sector and the manufacturing sector, and the share of rural employment in each province in 1998. Standard errors in parentheses are clustered at the province level, *, **, and *** denote significance at the 10\%, 5\% and 1\% levels, respectively.}

\end{table}

\newpage
\section{Figures}\label{sec_Figures}

\begin{figure}[htbp]
    \centering
    \includegraphics[width=0.8 \linewidth]{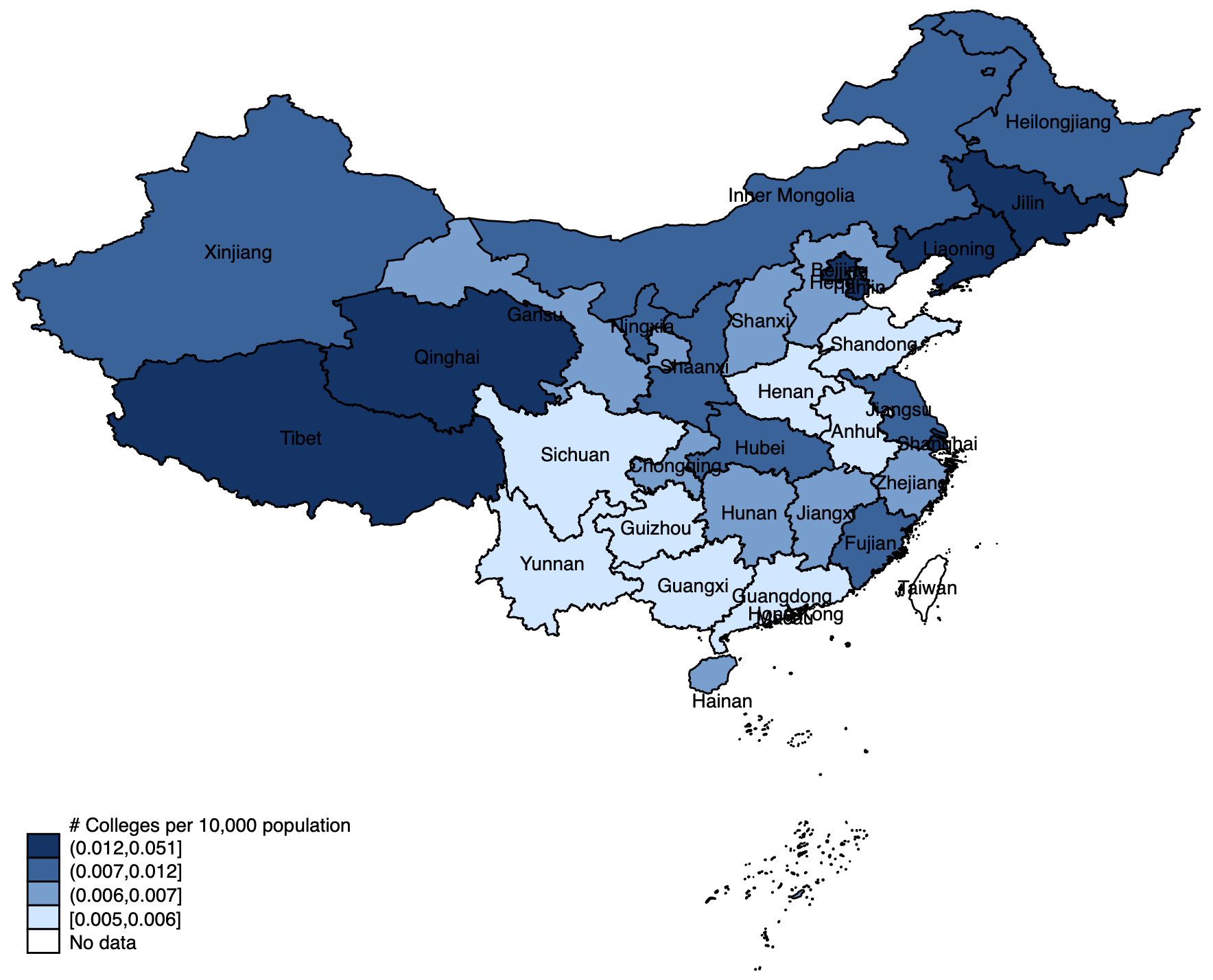}
    \caption{The average number of colleges across provinces in 1998}
    \label{fig_map_ncolpc}
    \captionsetup{singlelinecheck=false, width=\textwidth}
    \caption*{\footnotesize \textit{Notes}: This figure shows the cross-province variations of average number of post-secondary institutions in 1998. It is the ratio between the number of colleges and the population in that province.}
\end{figure}




\newpage
\begin{figure}[h!]
    \centering
	\subfloat[Rural]{
		\includegraphics[width=.48\textwidth]{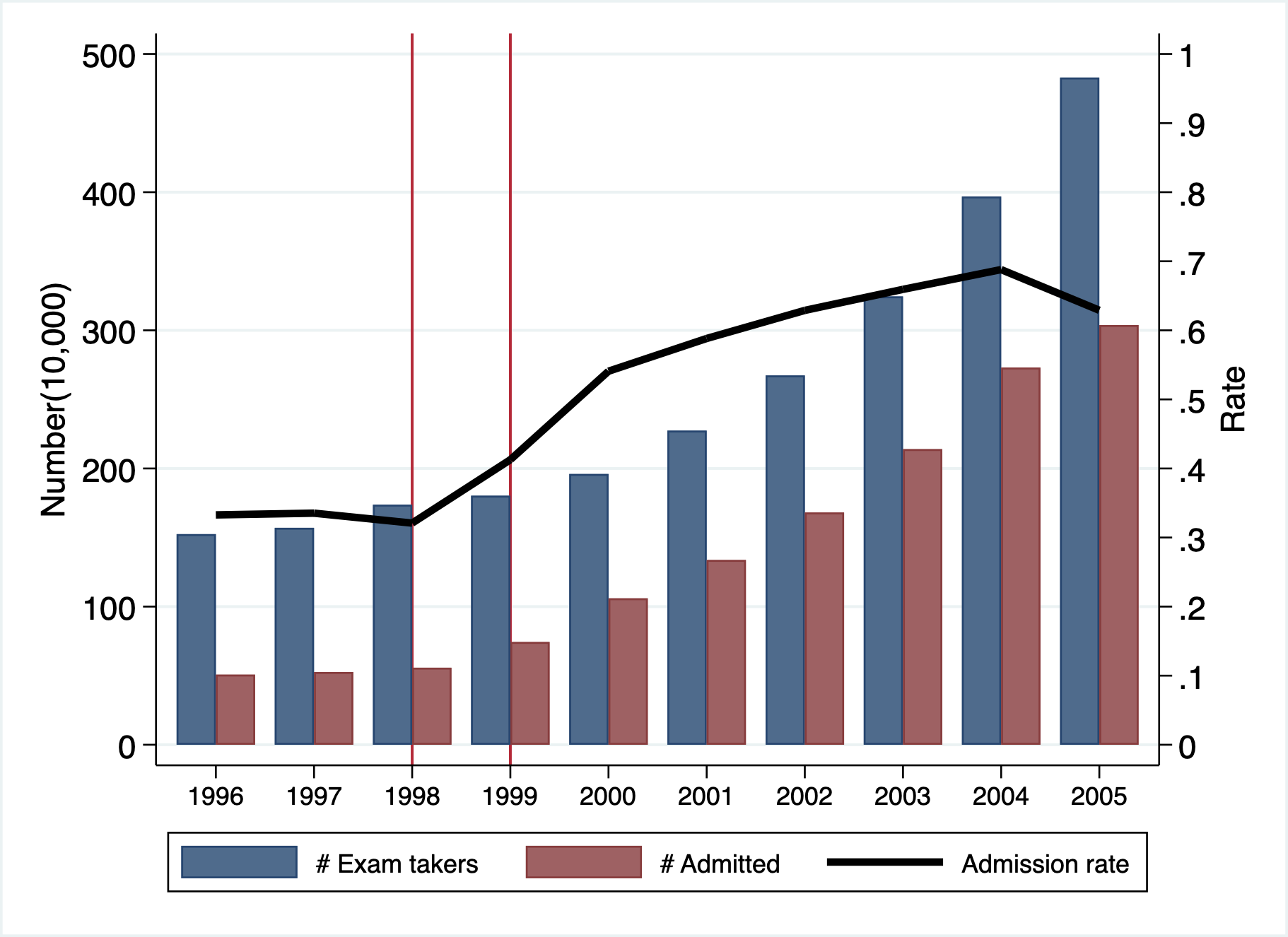}\label{fig_nation_avg_adm_rural}}
	\subfloat[Urban]{
		\includegraphics[width=.48\textwidth]{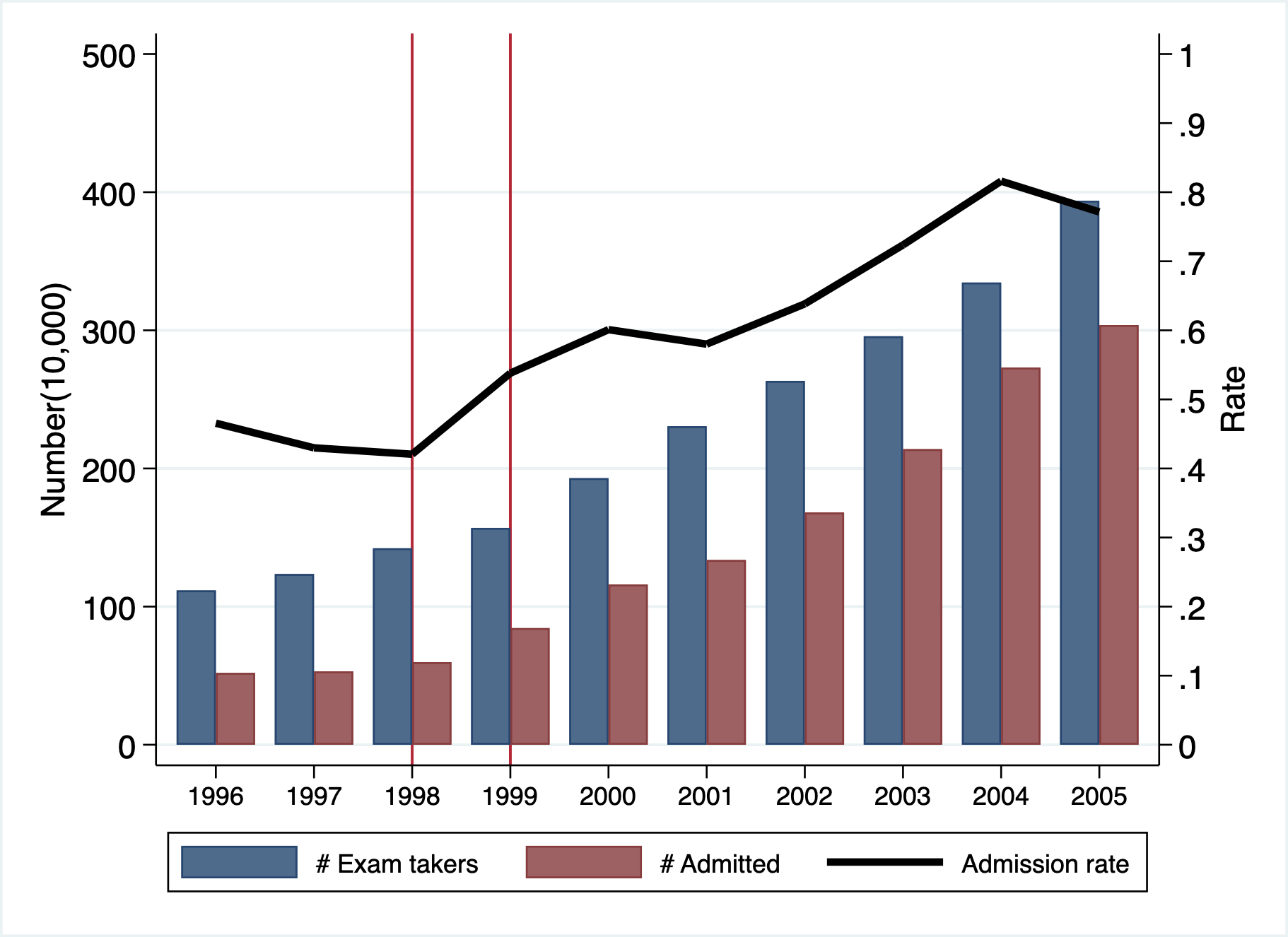}\label{fig_nation_avg_adm_urban}}
	\caption{The national pattern of NCEE admission}\label{fig_nation_adm_ru}
	\captionsetup{singlelinecheck=false, width=\textwidth}
    \caption*{\footnotesize \textit{Notes}: In both panels, the two bars refer to the number of NCEE takers and the number of admitted students from 1996 to 2005. The broken line is the admission rate, equal to the ratio between the two bars. The two vertical lines indicate the higher education expansion from 1998 to 1999. Panel (a) is the admission pattern for people with rural \textit{hukou} and Panel (b) is for thse with urban \textit{hukou}.}
\end{figure}

\begin{figure}[h!]
    \centering
    \includegraphics[width= 0.65 \linewidth]{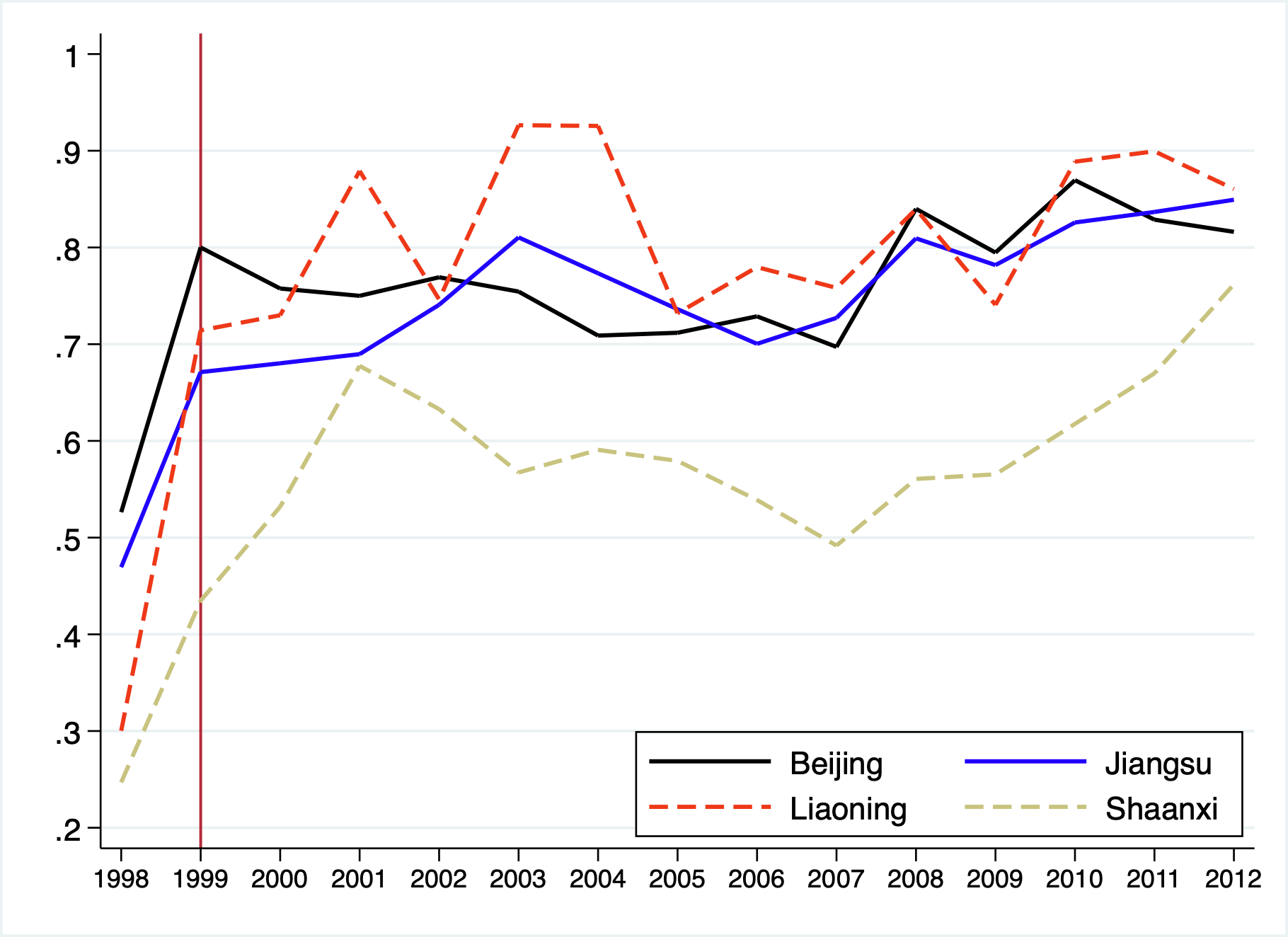}
    \caption{The admission rates over time}\label{fig_province_admrate}
	\captionsetup{singlelinecheck=false, width=\textwidth}
    \caption*{\footnotesize \textit{Notes}: This figure shows the admission rates in four provinces across time. The admission rate is the ratio between the number of students admitted to higher education and the number of NCEE takers. The vertical line indicates the year when HEE started.}
\end{figure}

\newpage
\begin{figure}[h!]
    \centering
    \includegraphics[width = 0.8\textwidth]{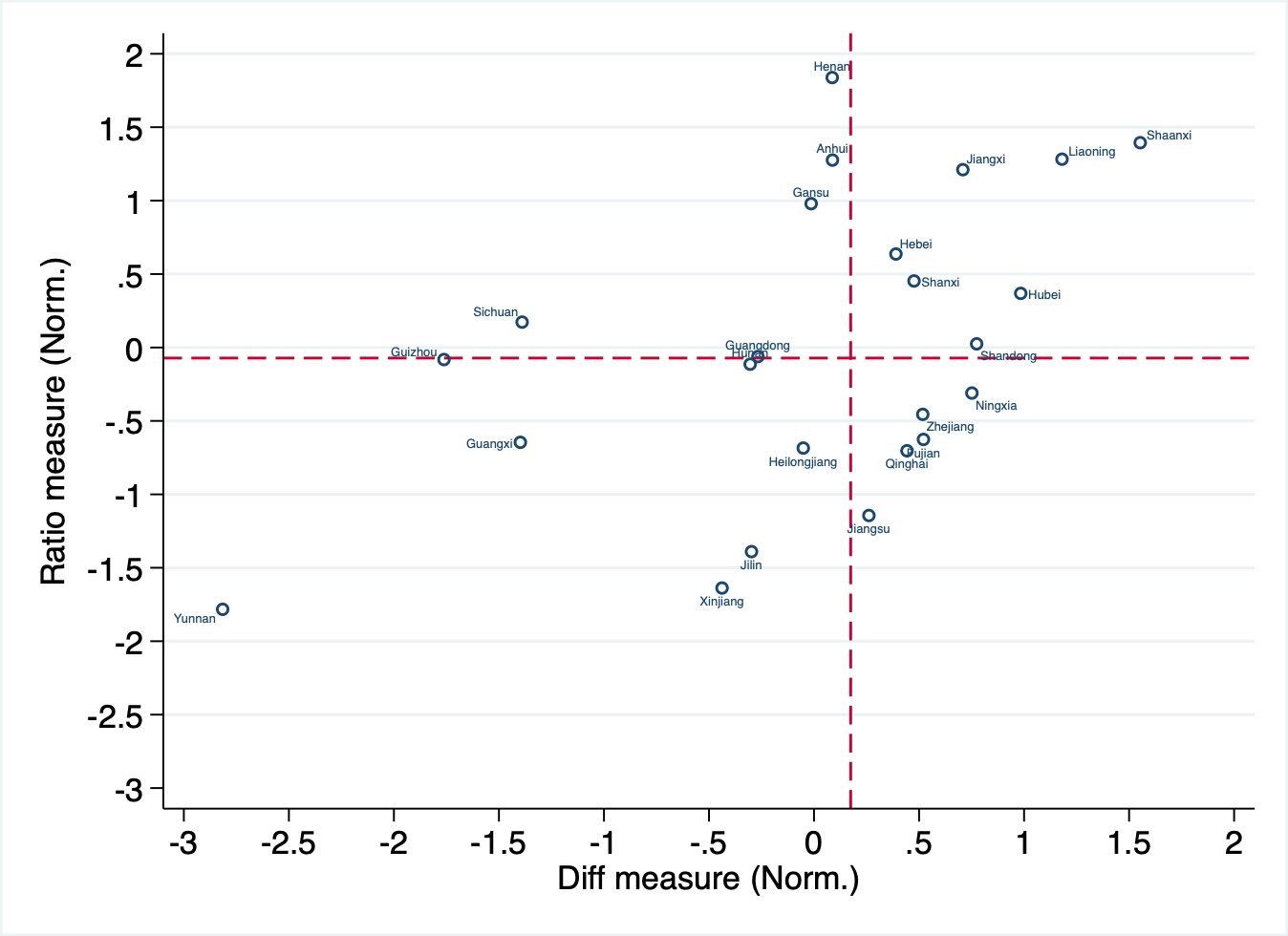}
    \caption{Measures of HEE Magnitudes (10-year average)}
    \label{fig_treatment_avg10}
    \captionsetup{singlelinecheck=false, width = \textwidth}
    \caption*{\footnotesize \textit{Notes}: This figure shows how we measure the magnitudes of HEE across 24 provinces (excluding Hainan, Tibet, Inner Mongolia, and four provincial-level municipalities Beijing, Tianjin, Shanghai, and Chongqing). The $Diff$ and $Ratio$ measures are defined in Equation \eqref{eq_treatment_def}, their normalized 10-year average values are presented in x- and y-axis, respectively. The two dashed lines correspond to the median value of each measure.}
\end{figure}

\newpage

\begin{figure}[h!]
    \centering
	\subfloat[College and above]{
		\includegraphics[width=.48\textwidth]{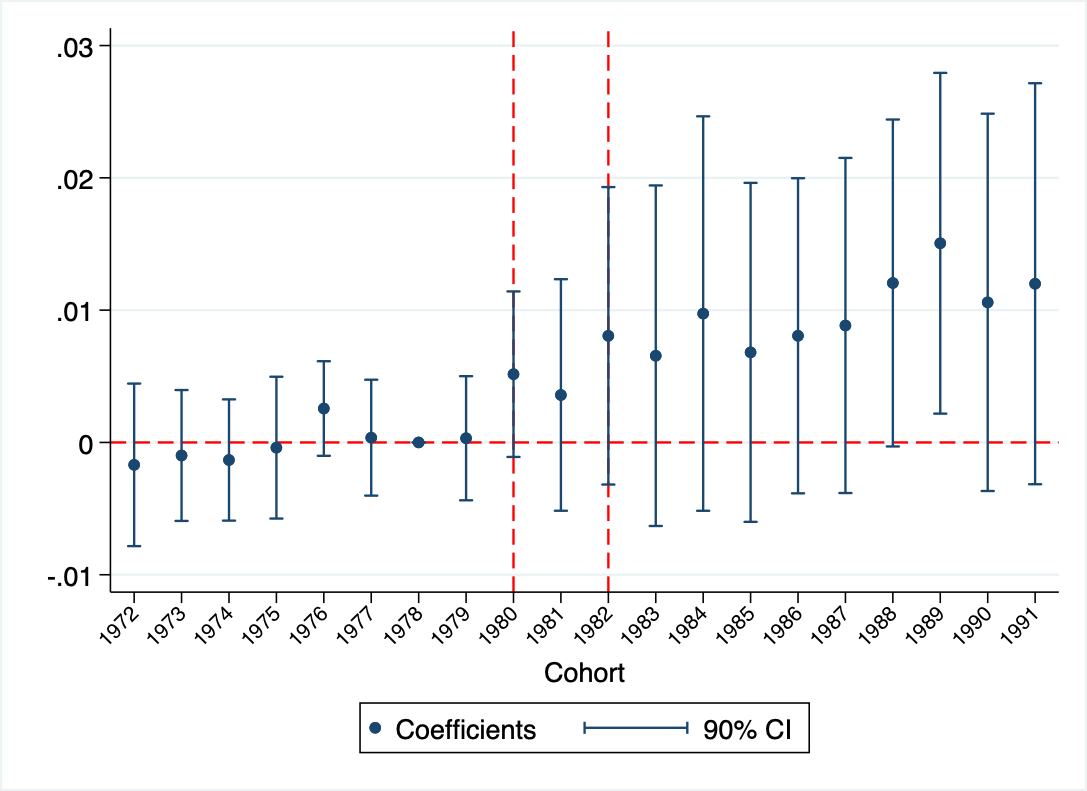}\label{fig_baseline_cont_col}}
	\subfloat[Senior high school and above]{
		\includegraphics[width=.48\textwidth]{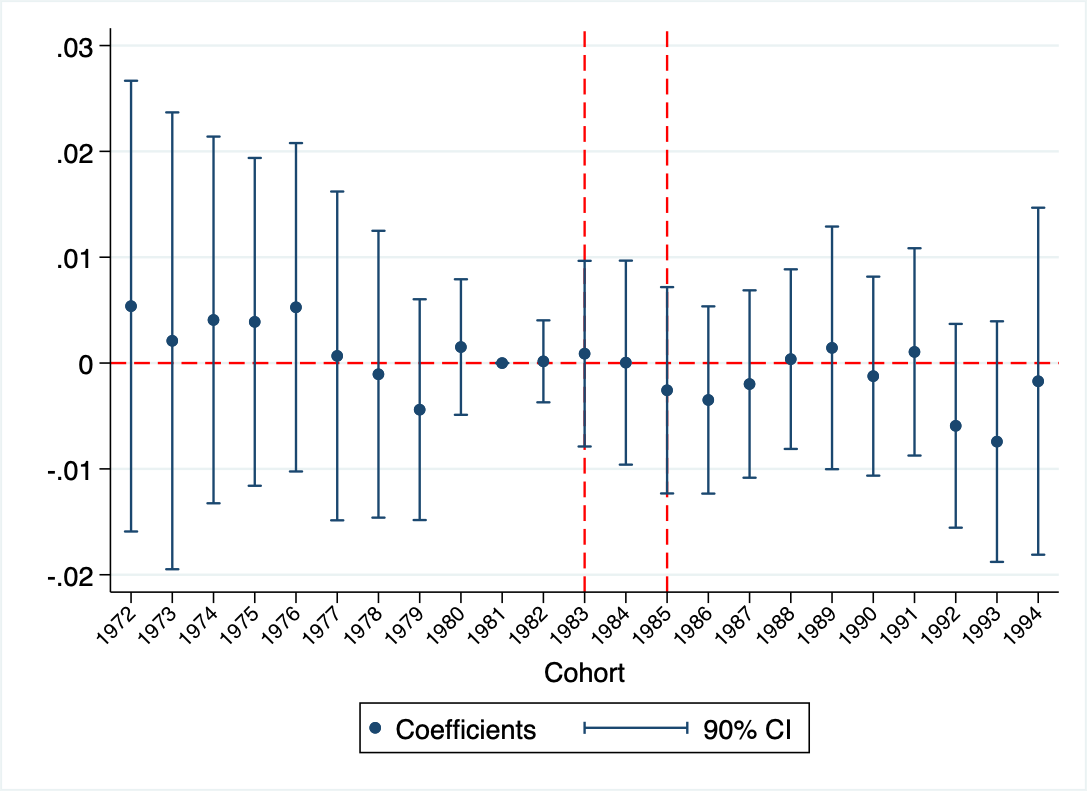}\label{fig_baseline_cont_hs}}
	\caption{The Impact of HEE on Education Attainment (5-year, continuous)}\label{fig_baseline_cont_col_hs}
	\captionsetup{singlelinecheck=false, width = \textwidth}
    \caption*{\footnotesize \textit{Notes}: The estimated coefficients in Equation \eqref{eq_did} and associated 90\% confidence intervals are reported. We use the 5-year averages of the continuous $Diff$ measure for the magnitudes of HEE across provinces. The left panel shows the impact on college attendance and the right panel presents the effect on high school enrollment. All the data are from the 2010 population census. Following the discussion in Section \ref{sec_EmpiricalStrategy_eduattainment}, we use cohorts 1978 and 1981 as the reference cohorts for the two outcomes, respectively.}
\end{figure}

\newpage
\begin{figure}[h!]
    \centering
	\subfloat[College and above]{
		\includegraphics[width=.48\textwidth]{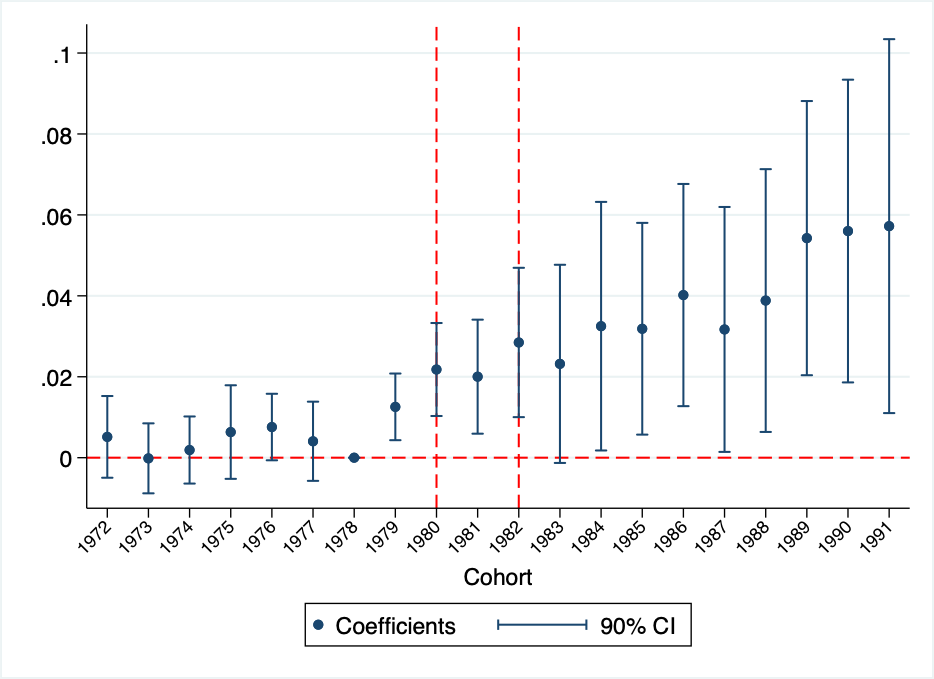}\label{fig_baseline_avg10_col}}
	\subfloat[Senior high school and above]{
		\includegraphics[width=.48\textwidth]{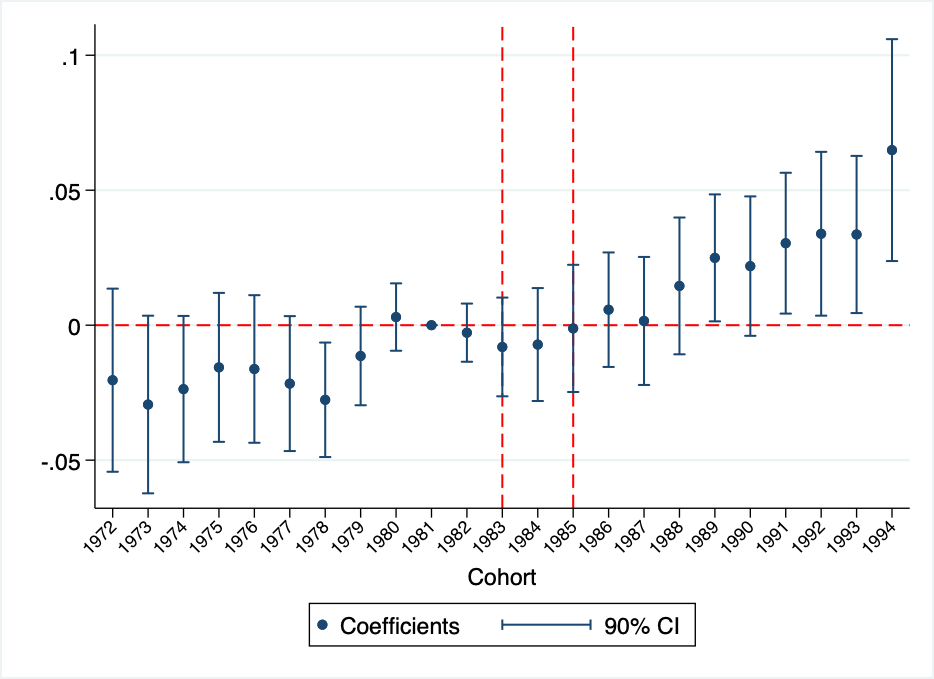}\label{fig_baseline_avg10_hs}}
	\caption{The Impact of HEE on Education Attainment (10-year, binary)}\label{fig_baseline_avg10_col_hs}
	\captionsetup{singlelinecheck=false, width = \textwidth}
    \caption*{\footnotesize \textit{Notes}: The estimated coefficients in Equation \eqref{eq_did} and associated 90\% confidence intervals are reported. We use the 10-year averages of the continuous $Diff$ measure for the magnitudes of HEE across provinces. The left panel shows the impact on college attendance and the right panel presents the effect on high school enrollment. All the data are from the 2010 population census. Following the discussion in Section \ref{sec_EmpiricalStrategy_eduattainment}, we use cohorts 1978 and 1981 as the reference cohorts for the two outcomes, respectively.}
\end{figure}

\begin{figure}[h!]
    \centering
	\subfloat[ln(Earnings)]{
		\includegraphics[width=.48\textwidth]{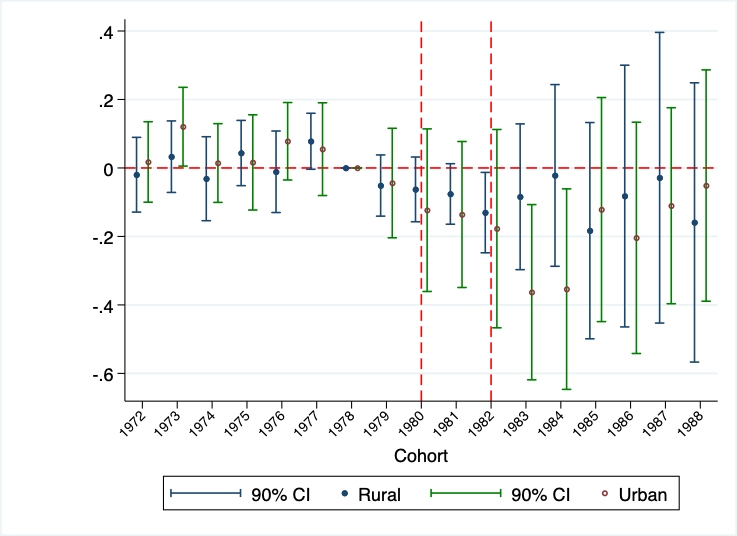}\label{fig_labor_cont_salary_rural}}
	\subfloat[Working in the agricultural sector]{
		\includegraphics[width=.48\textwidth]{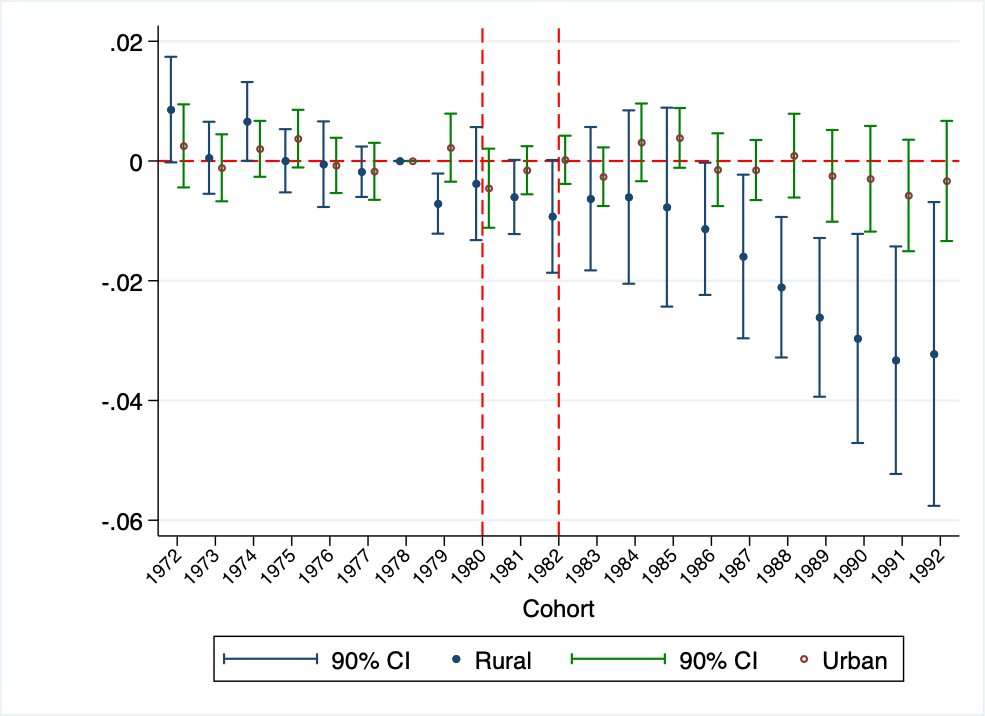}\label{fig_labor_cont_sector_rural}}
	\caption{The Impact of HEE on Labor Market Outcomes}\label{fig_labor_cont_rural}
	\captionsetup{singlelinecheck=false, width = \textwidth}
    \caption*{\footnotesize \textit{Notes}: Figures report estimated coefficients in Equation \eqref{eq_did} and associated 90\% confidence intervals for the sample including only people with rural or urban \textit{hukou}. We use the 5-year averages of the continuous $Diff$ measure for the magnitudes of HEE across provinces. The left panel shows the impact on log-earnings and the right panel presents the effect on the probability of working in the agricultural sector. Panel (a) is based on the 2005 population census while Panel (b) uses the 2010 population census.  Following the discussion in Section \ref{sec_EmpiricalStrategy_eduattainment}, we use cohorts 1978 as the reference group.}
\end{figure}

\newpage
\begin{figure}[h!]
    \centering
	\subfloat[College and above]{
		\includegraphics[width=.48\textwidth]{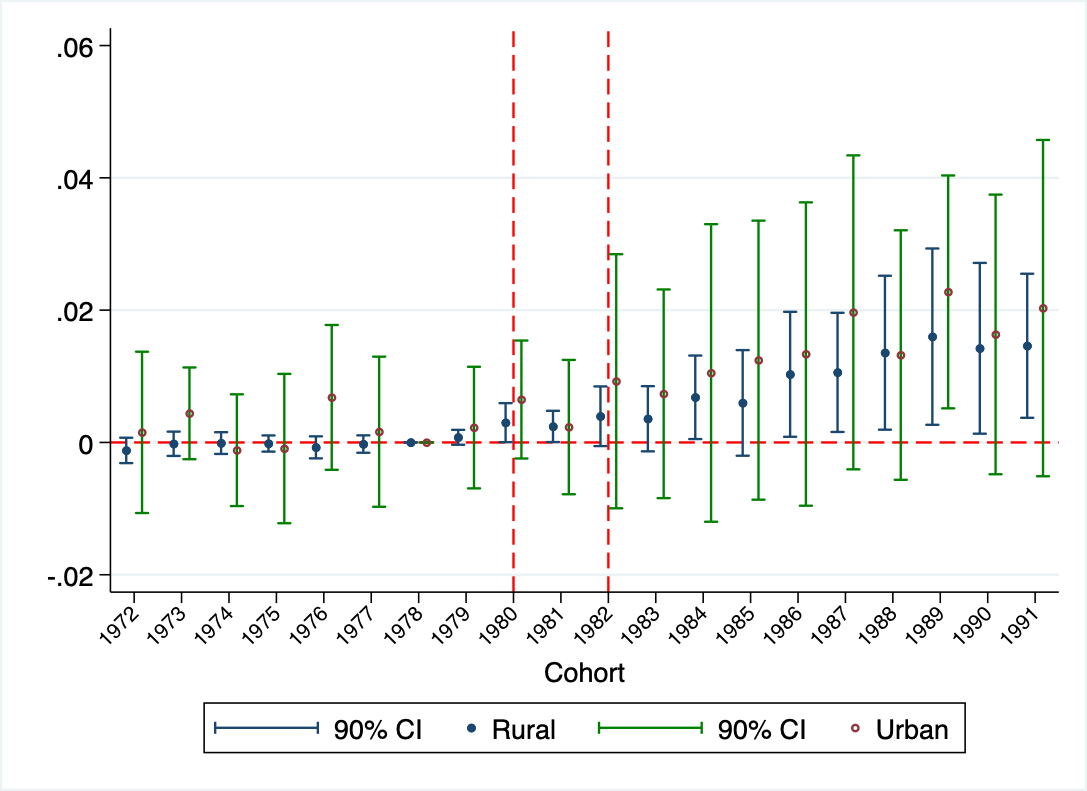}\label{fig_baseline_cont_col_rural}}
	\subfloat[Senior high school and above]{
		\includegraphics[width=.48\textwidth]{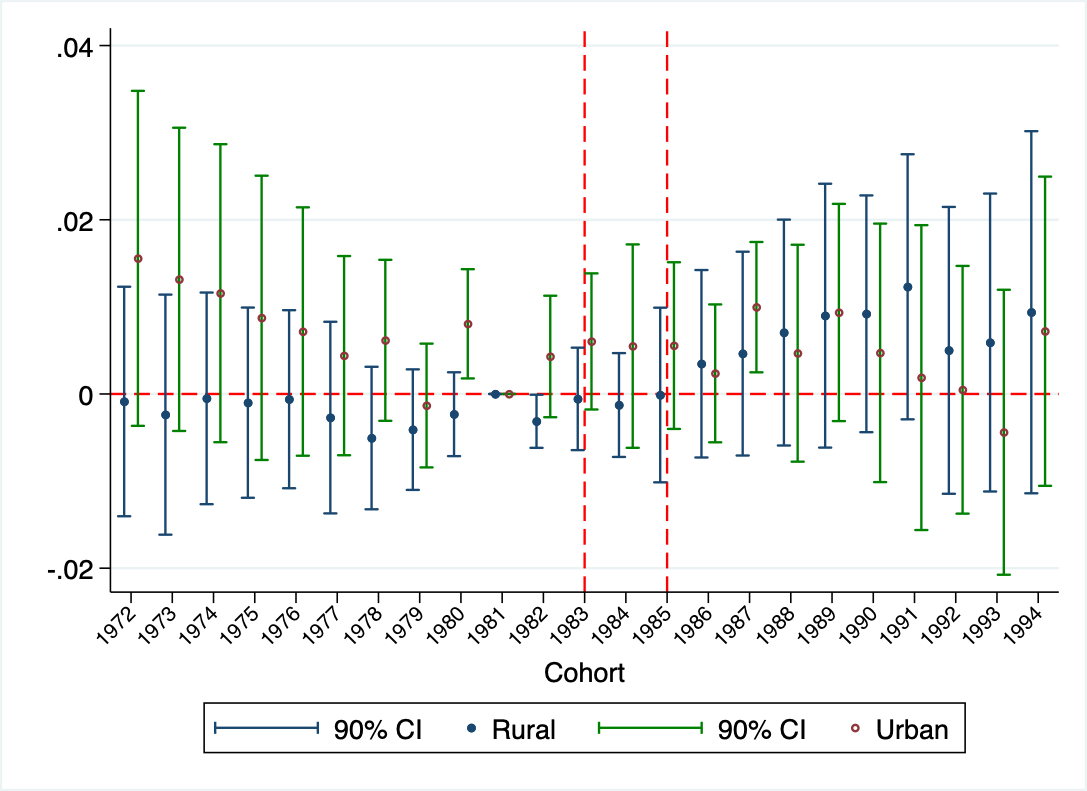}\label{fig_baseline_cont_hs_rural}}
	\caption{The Impact of HEE on Education Attainment (5-year, continuous)}\label{fig_baseline_cont_col_hs_rural}
	\captionsetup{singlelinecheck=false, width = \textwidth}
    \caption*{\footnotesize \textit{Notes}: Figures report estimated coefficients in Equation \eqref{eq_did} and associated 90\% confidence intervals for the sample including only people with rural or urban \textit{hukou}. We use the 5-year averages of the continuous $Diff$ measure for the magnitudes of HEE across provinces. The left panel shows the impact on college attendance and the right panel presents the effect on high school enrollment. All the data are from the 2010 population census. Following the discussion in Section \ref{sec_EmpiricalStrategy_eduattainment}, we use cohorts 1978 and 1981 as the reference cohorts for the two outcomes, respectively.}
\end{figure}

\begin{figure}[h!]
    \centering
	\subfloat[College and above]{
		\includegraphics[width=.48\textwidth]{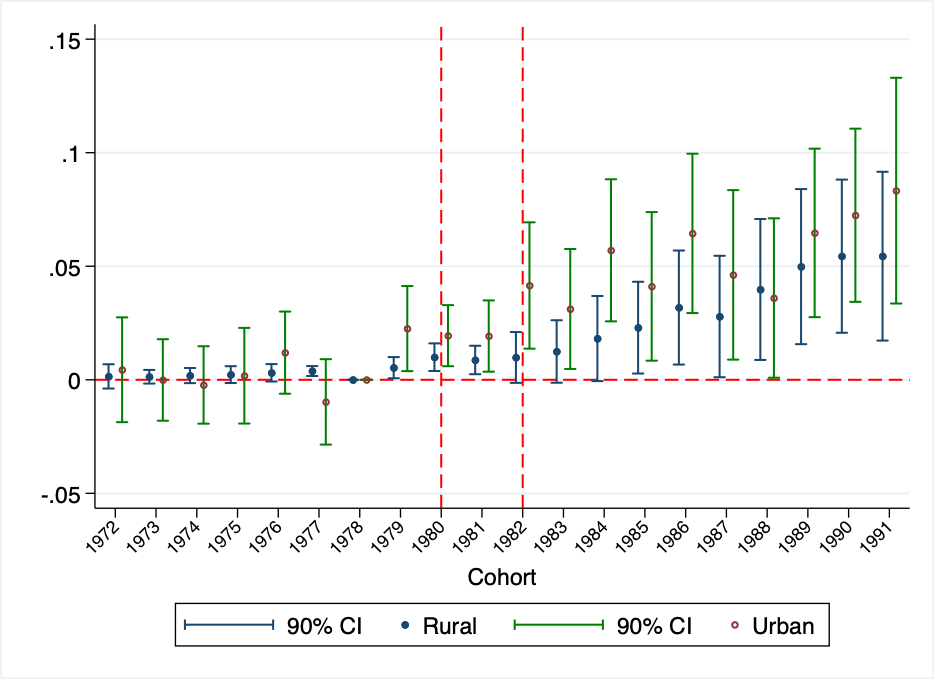}\label{fig_hetero_avg10_col_rural}}
	\subfloat[Senior high school and above]{
		\includegraphics[width=.48\textwidth]{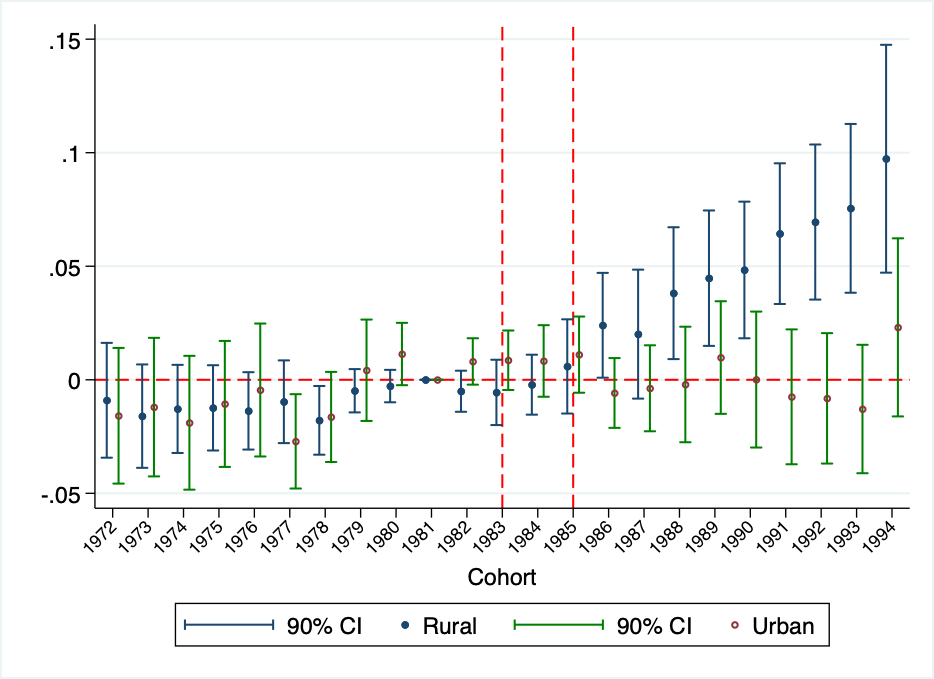}\label{fig_hetero_avg10_hs_rural}}
	\caption{The Impact of HEE on Education Attainment (10-year, binary)}\label{fig_baseline_avg10_col_hs_rural}
	\captionsetup{singlelinecheck=false, width = \textwidth}
    \caption*{\footnotesize \textit{Notes}: Figures report estimated coefficients in Equation \eqref{eq_did} and associated 90\% confidence intervals for the sample including only people with rural or urban \textit{hukou}. We use the 10-year averages of the binary $Diff$ measure for the magnitudes of HEE across provinces. The left panel shows the impact on college attendance and the right panel presents the effect on high school enrollment. All the data are from the 2010 population census. Following the discussion in Section \ref{sec_EmpiricalStrategy_eduattainment}, we use cohorts 1978 and 1981 as the reference cohorts for the two outcomes, respectively.}
\end{figure}

\newpage

\begin{figure}[h!]
    \centering
	\includegraphics[width=.48\textwidth]{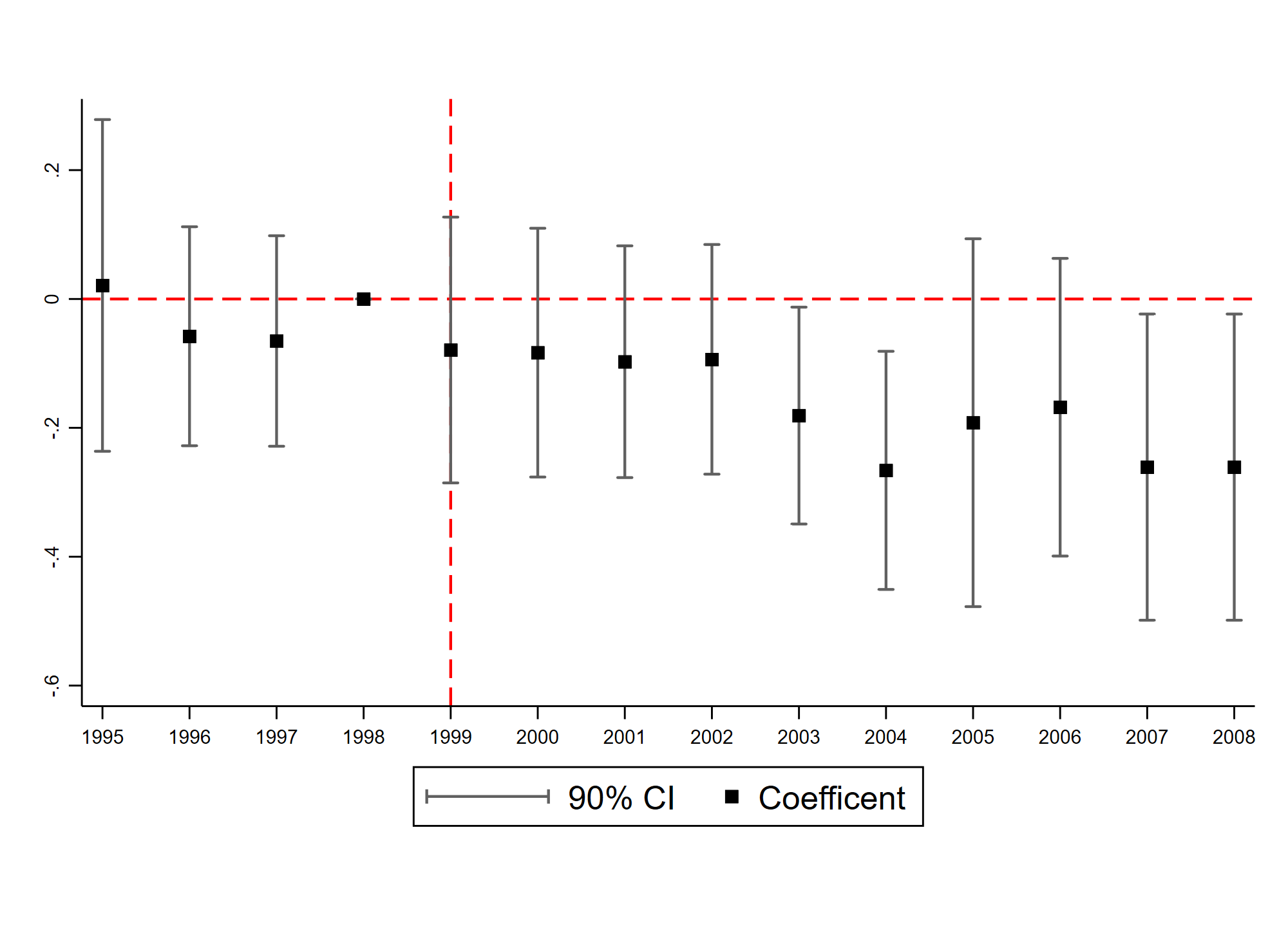}
	\caption{The Dynamic Effect of HEE on the Life Quality Index  in Rural Villages}\label{fig_v_lifeindex}
	\captionsetup{singlelinecheck=false}
    \caption*{\footnotesize \textit{Notes}: This figure shows the coefficients of the interaction term of between $\Delta \overline{AdmPC}$ and year indicators as presented in Equation \eqref{eq_eventstudy}. The results include 24 provinces. The outcome is the life quality index, which is generated by the principal component analysis based on the variables listed in Columns (2)-(7) of Table \ref{tab_v_reg_lifequality}. The year 1998 serves as the reference category, which is omitted in the regression. The vertical line indicates the start year of HEE.}
\end{figure}

\end{appendices}

\end{document}